\documentclass[11pt]{article}

\pdfoutput=1 

\usepackage[textwidth = 460 pt, textheight = 650 pt]{geometry}
\usepackage{hyperref,amsthm,amsmath,amssymb,slashed,graphicx,bbold,fancybox,mathtools,color,tikz-cd,
	tikz,caption,subcaption,geometry,marginnote,verbatim, marginnote, stmaryrd, manfnt,ytableau,xcolor,bm,wasysym} 
\usepackage{wrapfig}
\usepackage{mathrsfs}
\usepackage{simplewick}
\usepackage{cite, mcite}
\usepackage{microtype}
\usepackage{chapterbib}
\usepackage{cancel}
\usepackage[en-US, showisoZ=false, showseconds=false]{datetime2} %datetime is obsolete
\usepackage{tikz-cd,tikz}
\usetikzlibrary{arrows,decorations.markings} % tikz is a package for quiver diagrams and mindmaps

\usepackage[tikz]{mdframed}

\newenvironment{roundbox}[1]{\begin{mdframed}[
	outerlinewidth=1, 
	roundcorner = 10pt,
	userdefinedwidth=2.3cm,
	align=center,
	backgroundcolor = light-yellow,
	outerlinecolor = MyDarkBlue,
	innertopmargin = \topskip, 
	splittopskip = \topskip]}
{\end{mdframed}}

\newcommand{\Comment}[1]{{}}

\usepackage{framed} % shaded box is defined in this package
\usetikzlibrary{mindmap,backgrounds}

\definecolor{MyDarkBlue}{rgb}{0.15,0.15,0.45}
\definecolor{shadecolor}{rgb}{0.85,0.85,0.85}
\definecolor{light-blue}{rgb}{0.15,0.15,0.65}
\definecolor{light-yellow}{rgb}{1,0.98,0.73}
\definecolor{link-blue}{rgb}{0.15,0.15,0.65}
\definecolor{link-red}{rgb}{0.8,0.15,0.1}
\definecolor{link-green}{rgb}{0.15,0.50,0.15}
\definecolor{link}{rgb}{0.45,0.18,0.22}

\usepackage{graphicx}

\def\bpm{\begin{pmatrix}}
\def\epm{\end{pmatrix}}

\def\red{\textcolor{red}}

\newcommand{\be}{\begin{equation}}
\newcommand{\ee}{\end{equation}}
\newcommand{\bse}{\begin{subequations}}
\newcommand{\ese}{\end{subequations}}
\newcommand{\bea}{\begin{eqnarray}}
\newcommand{\eea}{\end{eqnarray}}

\newcommand{\nn}{\nonumber}

\newcommand{\one}[1]{\ydiagram[*(lightgray)]{1}}
\newcommand{\twoa}[1]{\ydiagram[*(lightgray)]{2}}
\newcommand{\twob}[1]{\ydiagram[*(lightgray)]{1,1}}
\newcommand{\twoc}[1]{\ytableaushort[*(lightgray)]{2 }}

\newtheorem*{hypo}{Hypothesis}

\hypersetup{
pdftitle={plane partitions},
pdfsubject={High Energy Physics},
pdfauthor={Thiago Araujo},
pdfkeywords={gauge; susy; strings; fields; cft},
pdfsubject={hep-th},
colorlinks=true,linkcolor=link,citecolor=link,urlcolor=link,linktocpage
}

\begin{document}

\renewcommand{\thefootnote}{\fnsymbol{footnote}}

\makeatletter
\@addtoreset{equation}{section}
\makeatother
\renewcommand{\theequation}{\thesection.\arabic{equation}}

\rightline{}
\rightline{}

\vspace{0.3cm}

\begin{center}
{\Large \bf{Towards Yang-Baxter integrability of quantum crystal melting:\\ \vspace{5pt} 
		From Kagome lattice to vertex models}} \\
\end{center}
 \vspace{0.5cm}
 
\thispagestyle{empty} 

\centerline{{\large {\bf Thiago Araujo}}}
\vspace{0.5cm}

\centerline{{\it
Albert Einstein Center for Fundamental Physics,}}
\centerline{{\it
Institute for Theoretical Physics, University of Bern}}
\centerline{{\it
Sidlerstrasse 5, ch-3012, Bern, Switzerland}}

\vspace{0.5cm}

\centerline{\small \texttt{\href{thgr.araujo@gmail.com}{thgr.araujo@gmail.com} } }

\vspace{0.5cm}

\thispagestyle{empty}

\centerline{\bf Abstract}

\begin{center}
\begin{minipage}[c]{390pt}
{\noindent This paper considers aspects of a Kagome lattice system with states classified by plane partitions. Using two sets of free fermions, we rewrite the lattice in terms of two families of spin chains. In this formalism, the quantum crystals Hamiltonian becomes more transparent, and we determine expressions for the first 3 levels of plane partition states. A classical statistical model associated with local configurations in the Kagome lattice is also defined, and we show that a reduction of this classical system gives two Yang-Baxter integrable subsystems analogous to a descendant of the 6-vertex model.}
\end{minipage}
\end{center}

\setcounter{page}{1}

\renewcommand{\thefootnote}{\arabic{footnote}}

\setcounter{footnote}{0}

\setcounter{tocdepth}{1}

\tableofcontents

%%%%%%%%%%%%%%%%%%%%%%%%%%%%%%%%%%%%%
%%%%%%%%%%%%%%%%%%%%%%%%%%%%%%%%%%%%%
%%%%%%%%%%%%%%%%%%%%%%%%%%%%%%%%%%%%%
%%%%%%%%%%%%%%%%%%%%%%%%%%%%%%%%%%%%%
%%%%%%%%%%%%%%%%%%%%%%%%%%%%%%%%%%%%%

\section{Introduction}

Realistic exactly solvable models are incredibly rare, and that explains why the few available examples are so celebrated. The Ising model, unarguably the best example, was once regarded as too simple; nowadays, universality shows how it dictates the behavior of a vast collection of systems~\cite{Niss2008}. As a matter of fact, it is often believed that each university class has at least one integrable model where the critical exponents can be calculated. 

From the mathematical perspective, it is known that exact solvability generally means that there are conservation laws, or symmetry groups, underlying the dynamics. The wealth of mathematical structures in a physical system constrained by symmetry principles might be less surprising, but these conservation laws generally imply a bidirectional interaction between physics and mathematics, see for example~\cite{Hitchin1999, Miwa2000}. 

Motivated by recent developments in supersymmetric gauge theories, black holes and in the AdS/Higher Spins correspondence~\cite{Okounkov:2003sp, Nekrasov:2003rj,  Heckman:2006sk, Okounkov:2006, Gaberdiel:2017dbk}, we aim to study quantum integrable aspects of systems with states labeled by \emph{plane partitions}. These are pervasive structures in theoretical physics, and have a rich and long tradition in mathematics~\cite{macmahon1915,  *macmahon1916}. More specifically, we study a simple \emph{quantum crystal melting} Hamiltonian with  dynamics akin to the growth of plane partition configurations~\cite{Dijkgraaf:2008ua}.

\begin{figure}[h!]
		\centering
		\includegraphics[height=1.5in]{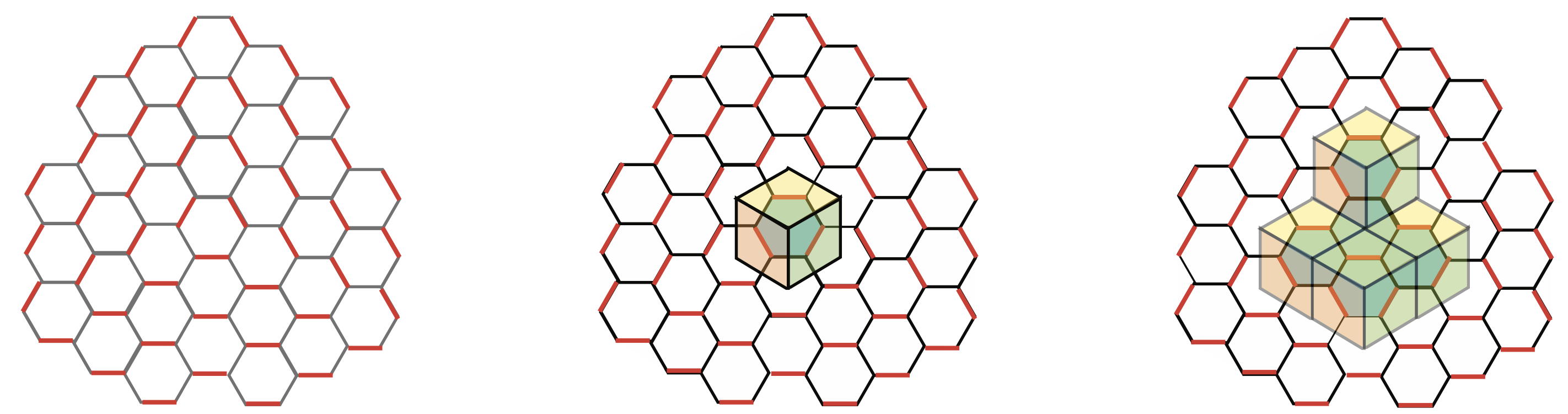}
		\caption{Plane partition configurations in terms of dimers.}
		\label{fig:dimers-1box}
\end{figure}

The first, and fundamental, observation is that plane partitions can be rephrased in terms of dimers living in a hexagonal lattice\footnote{I thank Susanne Reffert and Domenico Orlando for allowing me to use some of their figures.} as in figure~\ref{fig:dimers-1box}. Although the exact solvability of this problem and its relation to physical systems become more evident in the latter description~\cite{Kenyon2003, Okounkov:2006}, a definite understanding of how the Yang-Baxter equation, a hallmark for integrability, emerges in the quantum crystal melting problem is still lacking.

In~\cite{Dijkgraaf:2008ua}, the following quantum crystal melting Hamiltonian has been found
\be 
\label{eq:qtcrystal}
H = - J \sum |\blacksquare \rangle \langle \square |  +  |\square \rangle \langle \blacksquare | +
V \sum \sqrt{q}  |\square \rangle \langle \square | + \frac{1}{\sqrt{q}}  |\blacksquare \rangle \langle \blacksquare |\; .
\ee
The first two operators in the Hamiltonian above, the kinetic terms, describe the creation and annihilation of boxes in a given plane partition configuration. The diagonal operators, the potential terms, define an asymmetric diffusion process, where \(|\square \rangle \langle \square |\) gives the number of places where we can consistently add a box, and \(|\blacksquare \rangle \langle \blacksquare | \) gives the number of boxes that can be removed from a given configuration. We refer to these terms as \emph{growth operators}, although this terminology is unarguably very imprecise since just one of the four operators promotes the actual growth of the plane partitions.

\begin{figure}[h!]
	\centering
	\includegraphics[height=1.5in]{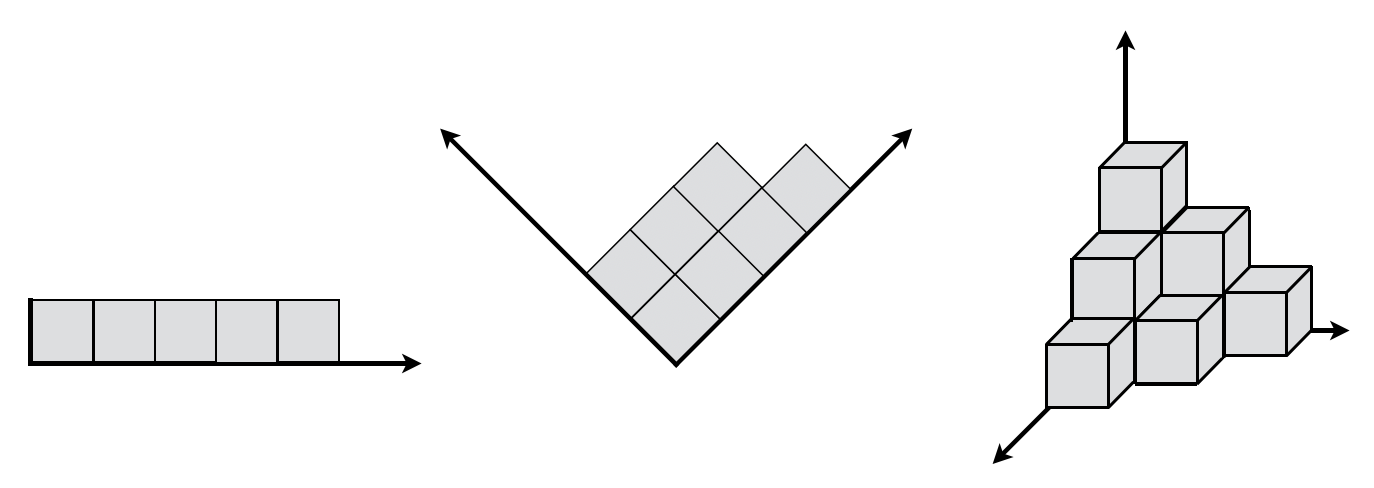}
	\caption{The 1D, 2D and 3D problems.}
	\label{fig:123d}
\end{figure}

The Hamiltonian~(\ref{eq:qtcrystal}) describes a system living in the first octant \(\mathbb{R}^3_{+++}\), which naturally means that one can add and remove boxes along in the three positive directions, we say it is the 3D problem. The form~(\ref{eq:qtcrystal}), on the other hand, does not dependent on the dimensionality we are interested in; and it can also describe the crystals growth dynamics confined to the first quadrant \(\mathbb{R}^2_{++}\), the 2D problem, or along the positive axis \(\mathbb{R}_{+}\), see figure~\ref{fig:123d}.

Quite remarkably, the 1D and 2D problems are exact solvable; in particular, it has been shown that the 2D system is just the XXZ model in disguise~\cite{Dijkgraaf:2008ua}, and it naturally implies the Bethe solvability of the 2D quantum crystal melting problem~\cite{Orlando:2009kd}. Additionally, the 1D and 2D have the same mass gap. These results (and further numerical analysis) led the authors of~\cite{Dijkgraaf:2008ua} to conjecture that the 3 dimensional quantum crystal melting is also integrable, and has the mass gap of its lower dimensional cousins.

Several questions on the 3D Hamiltonian have been addressed by D. Orlando, S. Reffert and the author in~\cite{Araujo:2020opn}. Among our findings, we have defined a novel fermion-boson correspondence for plane partitions that generalizes the usual two dimensional duality. We have also shown that the bosonized partition states are closely related to the MacMahon representation of the affine Yangian \({\cal Y}[\hat{\mathfrak{gl}}(1) ]\), see~\cite{maulik2012, Prochazka:2015deb, Gaberdiel:2017dbk}. Although the presence of the \({\cal Y}[\hat{\mathfrak{gl}}(1) ]\) algebra denotes an underlying integrable structure in this problem, a more pragmatic relation between the Yangian algebra above and the quantum crystal melting still eludes us.

\begin{figure}[h!]	
	\centering
	\begin{subfigure}[h!]{0.4\textwidth}
		\centering
		\includegraphics[height=2.5cm]{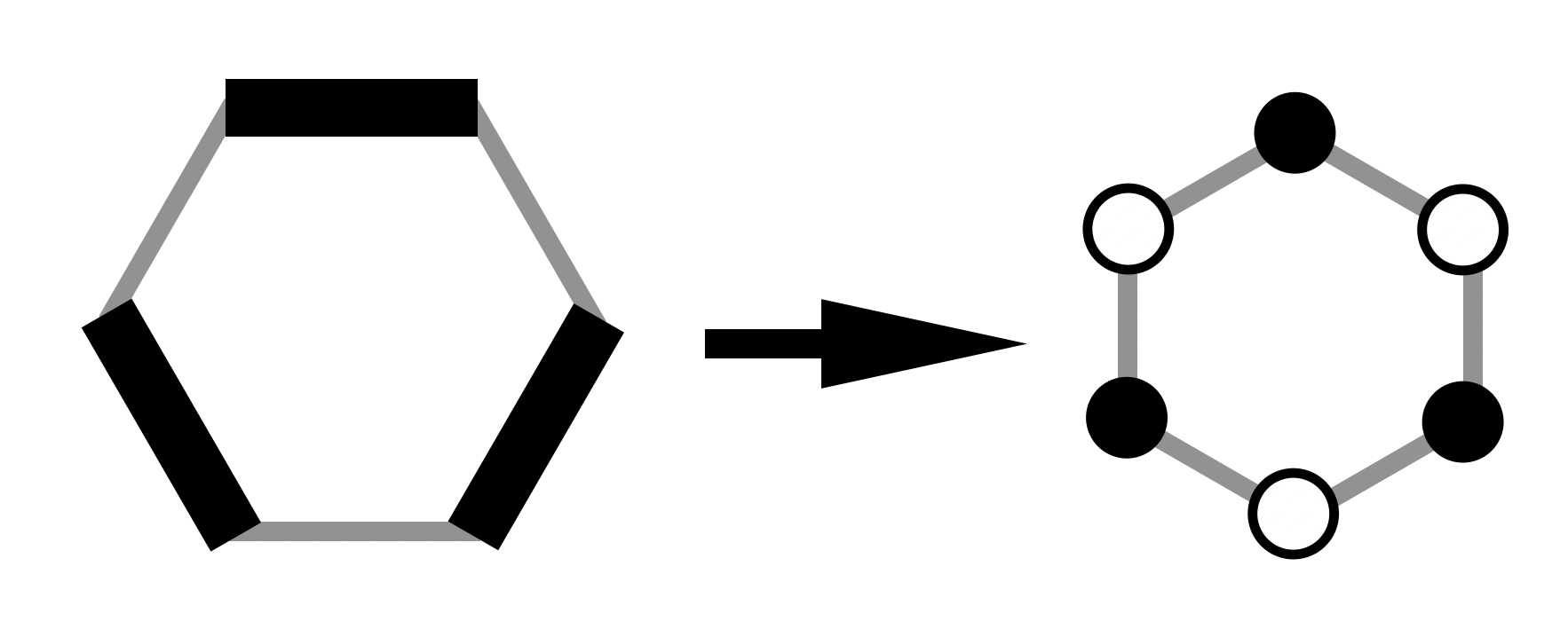}
		\caption{Dimers-particles transformation.}
		\label{fig:dimerparticle}
	\end{subfigure}
	~
	\begin{subfigure}[h!]{0.4\textwidth}
		\centering
		\includegraphics[height=2.5cm]{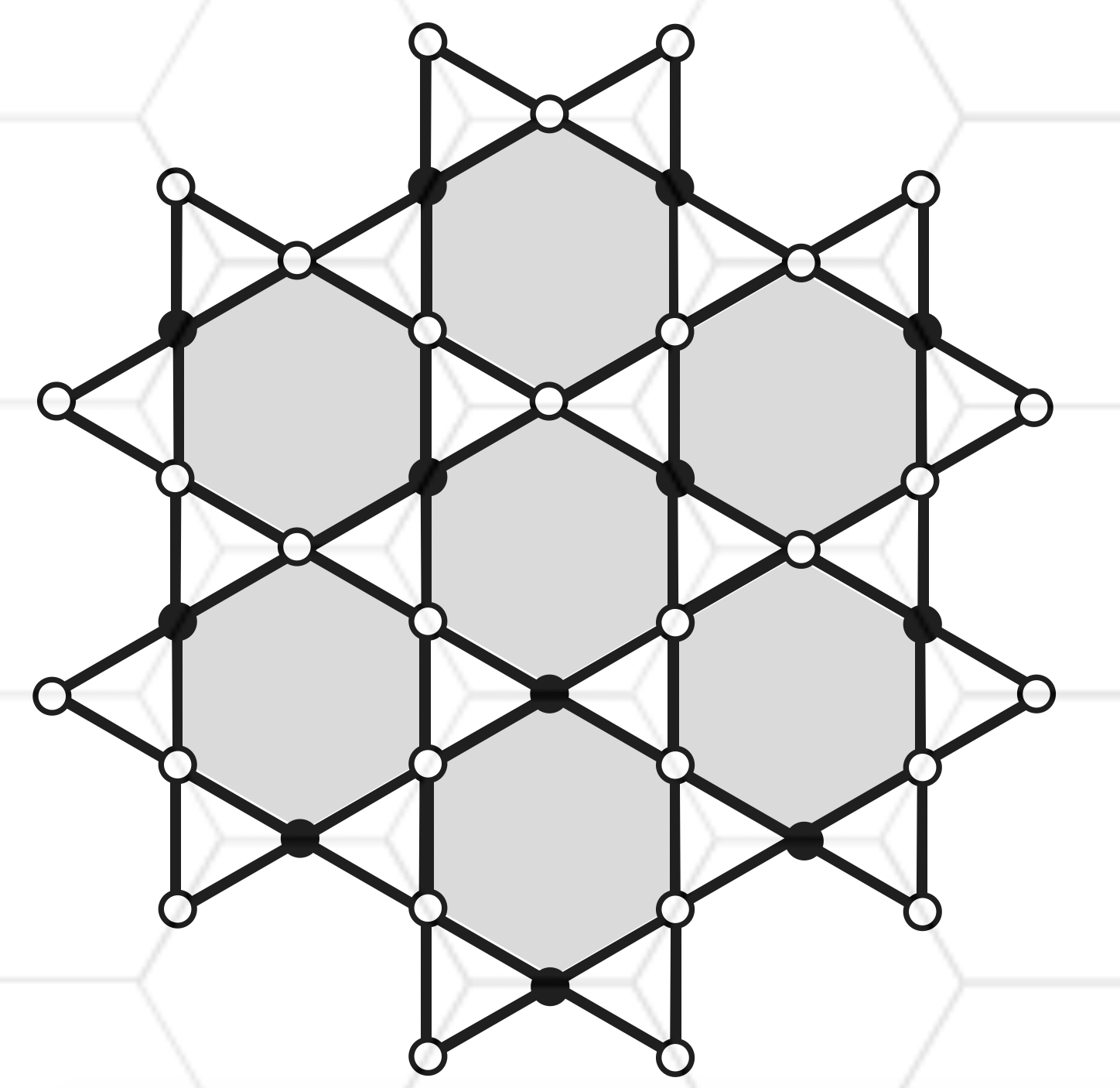}
		\caption{Empty partition in the Kagome lattice.}
		\label{fig:kagome-empty}
	\end{subfigure}
	\caption{Plane partitions in a dual Kagome lattice.}
\end{figure}

In~\cite{Araujo:2020opn}, we have shown that each dimer in the hexagonal lattice can be mapped to a dual point particle (and each empty corner is mapped to a hole), as in figure~\ref{fig:dimerparticle}. Using this map, we could finally prove that the plane partitions growth is dual to an occupation problem in a Kagome lattice, see figure~\ref{fig:kagome-empty}. The big advantage of this dual description is that now we can study the 3D quantum crystal melting problem from a 2D lattice perspective, and this is the viewpoint we want to push forward in the current work.

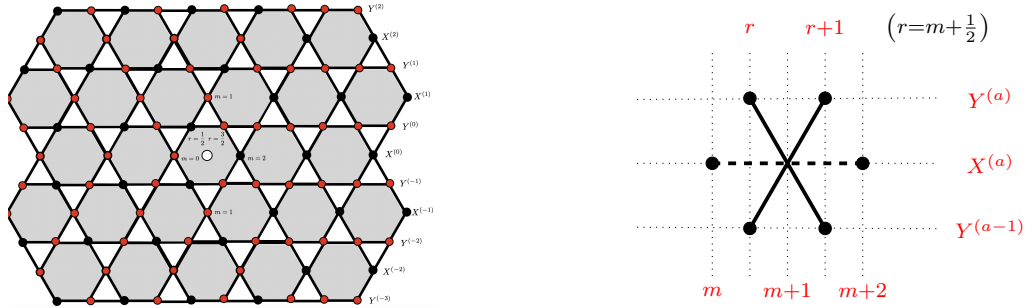
\begin{figure}[h!]	
\centering
\begin{subfigure}[h!]{0.56\textwidth}
	\centering
	\includegraphics[width=5.7cm]{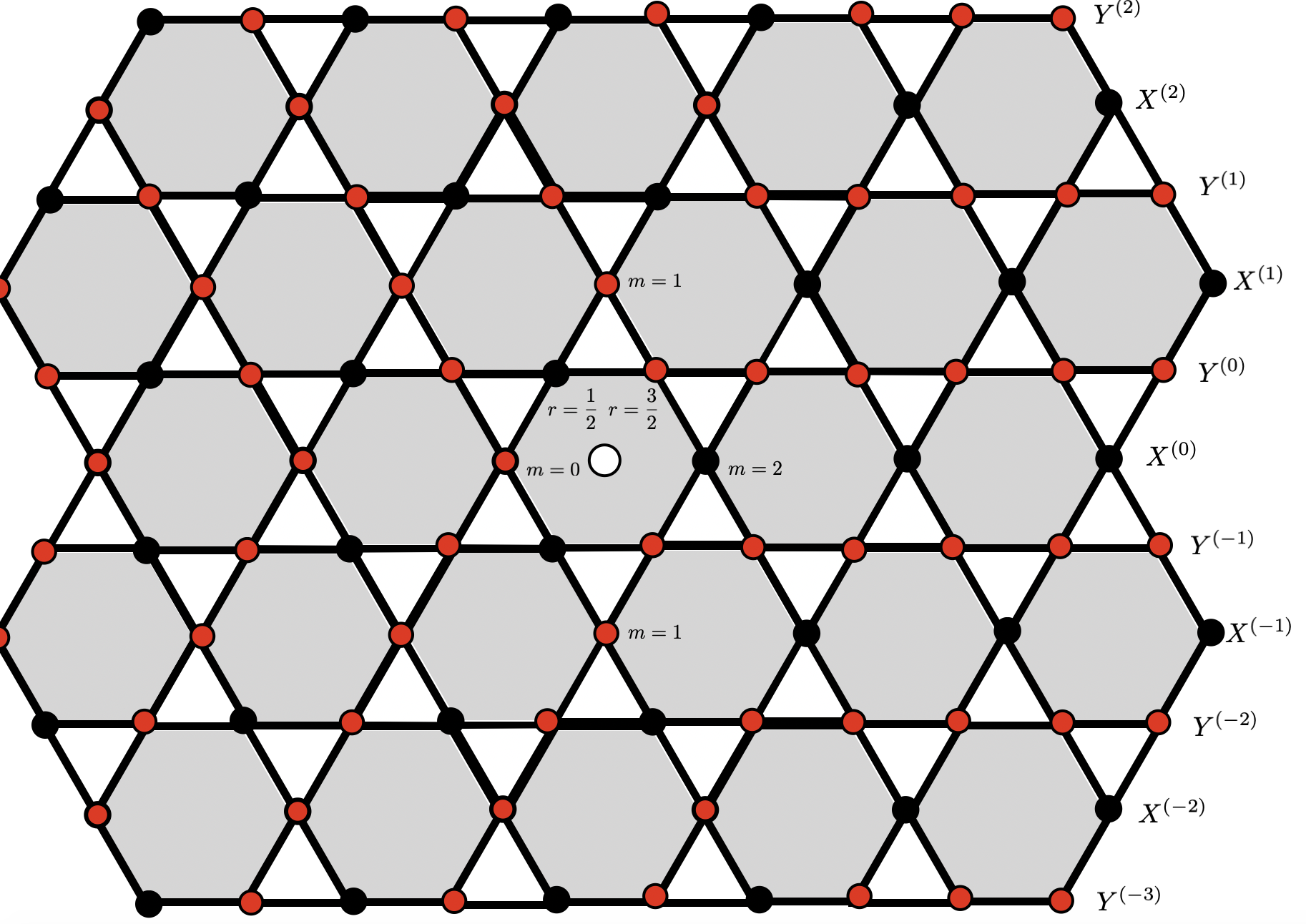}
	\caption{Slice of the empty configuration in the rotated lattice.}
	\label{fig:kagome}
\end{subfigure}
~
\begin{subfigure}[h!]{0.4\textwidth}
	\centering
	\begin{tikzpicture}
	\node at (2,1.8) {\({\scriptstyle \left(r=m+\frac{1}{2}\right) }\)}; 
	\draw[dotted] (-2,-0.85) -- (2,-0.87); \node at (2.7,-0.87) {\({\scriptstyle \red{Y^{(a-1)}}}\)};
	\draw[dotted] (-2,0) -- (-1,0); \draw[dotted] (1,0) -- (2,0); \node at (2.7,0) {\({\scriptstyle \red{X^{(a)}}}\)};
	\draw[dotted] (-2,0.85) -- (2,0.87); \node at (2.7,0.87) {\({\scriptstyle \red{Y^{(a)}}}\)};
	\draw[dotted] (-1,-1.5) -- (-1,1.5); \node at (-1,-1.7) {\({\scriptstyle \red{m}}\)};
	\draw[dotted] (0,-1.5) -- (0,1.5); \node at (0,-1.7) {\({\scriptstyle \red{m+1}}\)};
	\draw[dotted] (1,-1.5) -- (1,1.5); \node at (1,-1.7) {\({\scriptstyle \red{m+2}}\)};
	\draw[dotted] (-0.5,-1.5) -- (-0.5,1.5); \node at (-0.5,1.8) {\({\scriptstyle \red{r}}\)};
	\draw[dotted] (0.5,-1.5) -- (0.5,1.5); \node at (0.5,1.8) {\({\scriptstyle \red{r+1}}\)}; 
	\draw[dashed, line width=0.5mm] (-1,0) -- (1,0);
	\draw[black,fill=black] (1,0) circle (.5ex) ;
	\draw[black,fill=black] (-1,0) circle (.5ex) ;
	\draw[rotate=60, line width=0.5mm] (0,0) -- (1,0);
	\draw[black,fill=black][rotate=60] (1,0) circle (.5ex) ;
	\draw[rotate=-60, line width=0.5mm] (0,0) -- (1,0); 
	\draw[black,fill=black][rotate=-60] (1,0) circle (.5ex) ;
	\draw[rotate=120, line width=0.5mm] (0,0) -- (1,0);
	\draw[black,fill=black][rotate=120] (1,0) circle (.5ex) ; 
	\draw[rotate=-120, line width=0.5mm] (0,0) -- (1,0);
	\draw[black,fill=black][rotate=-120] (1,0) circle (.5ex) ; 
	\end{tikzpicture}
	\caption{Parametrization of a hexagon cell.}
	\label{fig:k-param}
\end{subfigure}
\caption{Empty partition and hexagon cell parametrization.}
\end{figure}

We start in section \(2\) with a rotation of \(-\pi/6\) in the Kagome lattice defined in~\cite{Araujo:2020opn}. We also show that the system can be written in terms of two independent families of free fermions with generating functions \(\psi^{(a)}(z)\) and \(\theta^{(a)}(z)\). In this free fermion formulation, it is straightforward to see that the lattice model is equivalent to two sets of spin chains -- we call them \(X^{(a)}\) and \(Y^{(b)}\)-type spin chains. The vacuum configuration is represented graphically in figure~\ref{fig:kagome}, where the white circle denotes the plaquette where the flip (box creation) can be performed.

In section \(3\), using the parametrization of figure~\ref{fig:k-param}, we rewrite the quantum Hamiltonian~(\ref{eq:qtcrystal}) as a series of coupled spin chains; and we also perform a Jordan-Wigner transformation to study the 3D problem using the fundamental representation of the \(su(2)\) algebra. This perspective is much more involved, and  should be contrasted with the 2D case of~\cite{Dijkgraaf:2008ua}, where the Jordan-Wigner transformation relates the 2D crystal melting problem to the XXZ model with kink boundary conditions. In the analysis of section \(4\), the growth operators for the crystal melting Hamiltonian are represented in terms of free fermions components, and we also find explicit expressions for the Kagome lattice states corresponding to the 0-, 1- and 2-boxes plane partition configurations. 

In spite of all these simplifications, we could not find an explicit expression for the quantum plane partition Hamiltonian. In section \(5\) we write all allowed local lattice hexagons, and then we assign an unspecified Boltzmann weight to each local configurations. Given that the quantum Hamiltonian eigenstates can be written as sums of products of local hexagon configurations, we ultimately want to determine the minute contributions to the quantum partition function given by the Boltzmann weights functional expressions. Furthermore, we need to remark that these Boltzmann weights associate a classical statistical model to the original quantum system. 

In section 5 and 6 we also start the analysis of this classical problem. The Yang-Baxter integrability of the crystal melting Hamiltonian can be settled if the Boltzmann weights can be arranged as a matrix satisfying the Yang-Baxter equation. Specifically, in section 5 the Lax operator for the classical statistical system is defined, but the initial attempts to define a consistent transfer matrix have been unsuccessful, and it naturally represents a major drawback for this approach. Surprisingly, we show in section 6 that with minor modifications on the rules that connect the local hexagon configurations, it is possible to define a subsystem analogous to the integrable 6-vertex model. This fact,and further evidences collected in~\cite{Dijkgraaf:2008ua, Araujo:2020opn}, might indicate the Yang-Baxter integrable structure underlying the original plane partition problem, but a definite mathematical proof is still elusive.

In section \(7\), we conclude this work with a discussion on the results we obtained and the limitations of our techniques. We also discuss some interesting unanswered questions, and a short overview on future research directions.

%%%%%%%%%%%%%%%%%%%%%%%%%%%%%%%%%%%%%
%%%%%%%%%%%%%%%%%%%%%%%%%%%%%%%%%%%%%
%%%%%%%%%%%%%%%%%%%%%%%%%%%%%%%%%%%%%
%%%%%%%%%%%%%%%%%%%%%%%%%%%%%%%%%%%%%
%%%%%%%%%%%%%%%%%%%%%%%%%%%%%%%%%%%%%
%%%%%%%%%%%%%%%%%%%%%%%%%%%%%%%%%%%%%
%%%%%%%%%%%%%%%%%%%%%%%%%%%%%%%%%%%%%

\section{Free fermions formulation}
\label{sec:Fermions}

This section defines a new parametrization for the study of plane partition growth in the Kagome lattice. A thorough description of the relation between plane partitions and their Kagome lattice formulation can be found in~\cite{Araujo:2020opn}.

\subsection{X-type spin chains}

In order to define the \(X^{(a)}\)-type spin chains, let us start with some well known facts~\cite{Hitchin1999, Miwa2000, Dijkgraaf:2008ua}. The components of the free fermions \(\psi^{(a)}(z)=\sum_m \psi^{(a)}_m z^m\), with \(a\in \mathbb{Z}\), satisfy the canonical anticommutation relations
\be 
\{\psi_m^{(a)}, \psi_n^{\ast(b)}\} = \delta^{ab}\delta_{mn}\qquad 
\{\psi_m^{(a)}, \psi_n^{(b)}\} = \{\psi_m^{(a)}, \psi_n^{(b)}\} = 0\qquad 
\forall \  m,n,a,b \in \mathbb{Z}\; .
\ee
The operators \(\eta_m^{(a)} = \psi_m^{(a)} \psi_m^{\ast (a)} \) and \(\bar{\eta}_m^{(a)} = \psi_m^{\ast (a)} \psi_m^{(a)} \) count, respectively, the number of holes and particles in the chain. The \emph{vacuum} \(|0\rangle\) is the state satisfying
\be 
\begin{split}
	\psi^{(a)}_n |0\rangle = 0, \ n\leq 0\; , \ \ \psi^{\ast (a)}_n |0\rangle = 0, \ n> 0\; , \  \forall a \in \mathbb{Z}\; .
\end{split}
\ee
This corresponds to a configuration where all positive (-labeled) sites, \(n> 0\), are occupied by particles, see figure~\ref{fig:vacuum}. 

\begin{figure}[h!]
	\centering
	\begin{tikzpicture}
	\node at (-3,0) {\(\cdots\)};
	\draw (-2,0) -- (2,0);
	\node at (-2,-0.5) {\({\scriptstyle -2}\)}; \draw[black,fill=white] (-2,0) circle (.5ex) ;
	\node at (-1,-0.5) {\({\scriptstyle -1}\)}; \draw[black,fill=white] (-1,0) circle (.5ex) ;
	\node at (0,-0.5) {\({\scriptstyle 0}\)}; \draw[black,fill=white] (0,0) circle (.5ex) ;
	\node at (1,-0.5) {\({\scriptstyle 1}\)}; \draw[black,fill=black] (1,0) circle (.5ex) ;
	\node at (2,-0.5) {\({\scriptstyle 2}\)}; \draw[black,fill=black] (2,0) circle (.5ex) ;
	\node at (3,0) {\(\cdots\)};
	\end{tikzpicture}
	\caption{Vacuum.}
	\label{fig:vacuum}
\end{figure}	

Additionally, the shifted vacuum is defined by the relations
\be 
|m\rangle^{(a)} = 
\left\{ 
\begin{array}{ll}
	\psi^{(a)}_m	\cdots \psi^{(a)}_2 \psi^{(a)}_1|0\rangle & m>0 \\
	\psi^{\ast (a)}_{m+1}	\cdots \psi^{\ast (a)}_{-1} \psi^{\ast (a)}_0|0\rangle & m<0 
\end{array}
\right. \; .
\ee
From figure~\ref{fig:kagome}, it is easy to see that there is a \emph{zigzag} between the even (-labeled) rows \(X^{(2a)}\) and the odd (-labeled) rows \(X^{(2a+1)}\). More specifically, if particles in the even rows are localized at the positions \(m {\rm d}_X\), where \(m\in \mathbb{Z}\) and \({\rm d}_X\) is the lattice distance, the odd chains particles are positioned at \((m+1/2) {\rm d}_X\). Additionally, direct inspection of figure~\ref{fig:kagome} also shows that the lattice distance \({\rm d}_X\) in the X-rows is twice the \(Y\)-chains lattice distance \({\rm d}_Y\). Without any loss of generality, one can  consider the normalization \({\rm d}_Y=1\). The \(X^{(a)}\)-spin chains are defined by the relations above and two additional properties that we just described.

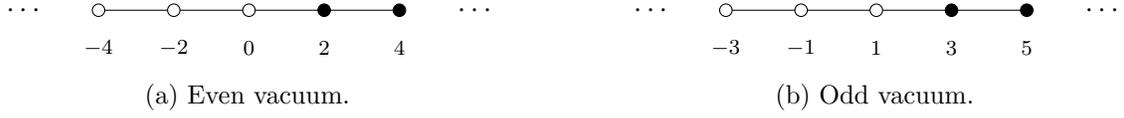
\begin{figure}[h!]
	\begin{subfigure}[t]{0.5\textwidth}
		\centering
		\begin{tikzpicture}
		\node at (-3,0) {\(\cdots\)};
		\draw (-2,0) -- (2,0);
		\node at (-2,-0.5) {\({\scriptstyle -4}\)}; \draw[black,fill=white] (-2,0) circle (.5ex) ;
		\node at (-1,-0.5) {\({\scriptstyle -2}\)}; \draw[black,fill=white] (-1,0) circle (.5ex) ;
		\node at (0,-0.5) {\({\scriptstyle 0}\)}; \draw[black,fill=white] (0,0) circle (.5ex) ;
		\node at (1,-0.5) {\({\scriptstyle 2}\)}; \draw[black,fill=black] (1,0) circle (.5ex) ;
		\node at (2,-0.5) {\({\scriptstyle 4}\)}; \draw[black,fill=black] (2,0) circle (.5ex) ;
		\node at (3,0) {\(\cdots\)};
		\end{tikzpicture}
		\caption{Even vacuum.}
		\label{fig:evenvacuum}	
	\end{subfigure}
	~
	\begin{subfigure}[t]{0.5\textwidth}
		\centering
		\begin{tikzpicture}
		\node at (-3,0) {\(\cdots\)};
		\draw (-2,0) -- (2,0);
		\node at (-2,-0.5) {\({\scriptstyle -3}\)}; \draw[black,fill=white] (-2,0) circle (.5ex) ;
		\node at (-1,-0.5) {\({\scriptstyle -1}\)}; \draw[black,fill=white] (-1,0) circle (.5ex) ;
		\node at (0,-0.5) {\({\scriptstyle 1}\)}; \draw[black,fill=white] (0,0) circle (.5ex) ;
		\node at (1,-0.5) {\({\scriptstyle 3}\)}; \draw[black,fill=black] (1,0) circle (.5ex) ;
		\node at (2,-0.5) {\({\scriptstyle 5}\)}; \draw[black,fill=black] (2,0) circle (.5ex) ;
		\node at (3,0) {\(\cdots\)};
		\end{tikzpicture}
		\caption{Odd vacuum.}
		\label{fig:oddvacuum}
	\end{subfigure}
	\caption{Even and odd vacua.}
\end{figure}

In order to address these two points at once, we parametrize even (odd) rows with even (odd) indices. Moreover, when we assumed \({\rm d}_Y=1\), we have automatically rescaled the lattice distance in the lines \(X^{(a)}\) by a factor of two. Consequently, if \([0], [1]\in \mathbb{Z}/2\mathbb{Z}\) are the equivalence classes of even and odd integers, respectively, we define the vacuum in the even \(X\) rows, 
the \emph{even vacuum} \(|{\bm 0}\rangle\), as
\be 
	\psi^{(a)}_n |{\bm 0}\rangle^{(a)} = 0\ \ \  n\in [0]\leq 0\; , \quad 
	\psi^{\ast (a)}_n |{\bm 0}\rangle^{(a)} = 0 \ \ \ n\in [0]>0  \; , \quad  \forall a \in [0]
\ee
represented as~\ref{fig:evenvacuum}; and the vacuum \(|{\bm 1}\rangle\) in the odd \(X\) rows, the \emph{odd vacuum}, as
\be 
\begin{array}{ll}
	\psi^{(a)}_n |{\bm 1}\rangle^{(a)} = 0 & n\in [1]\leq 0\; , \\ 
	\psi^{\ast (a)}_n |{\bm 1}\rangle^{(a)} = 0 & n\in [1]>0 
\end{array}\; , \ \forall a \in [1]
\ee
represented as~\ref{fig:oddvacuum}.

Finally, the shifted even and odd vacua are respectively
\bse
\be 
|{\bf m}\rangle^{(a)} = 
\left\{ 
\begin{array}{ll}
	\psi^{(a)}_m	\cdots \psi^{(a)}_4 \psi^{(a)}_2|{\bm 0}\rangle & [0]\ni m >0 \\
	\psi^{\ast (a)}_{m+2}	\cdots \psi^{\ast (a)}_{-2} \psi^{\ast (a)}_0|{\bm 0}\rangle & [0]\ni m <0 \\
\end{array}
\right. \; , \ \forall \ a\in [0] 
\ee
and 
\be 
|{\bf m}\rangle^{(a)} = 
\left\{ 
\begin{array}{ll}
	\psi^{(a)}_m	\cdots \psi^{(a)}_5 \psi^{(a)}_3|{\bm 1}\rangle & [1]\ni m>1 \\
	\psi^{\ast (a)}_{m+2}	\cdots \psi^{\ast (a)}_{-1} \psi^{\ast (a)}_1|{\bm 1}\rangle & [1]\ni m<1
\end{array}
\right. \; , \  \forall \ a\in [1] \; .
\ee
\ese
As we see in a later section, the definitions above fully characterize the \(X\)-spin chains in the Kagome lattice. Before addressing this point, let us now move to the \(Y\)-spin chains. 

\subsection{Y-type spin chains}

This case is more straightforward, and we simply consider fermionic operators defined by their anticommutation relations
\be 
\{\theta_r^{(a)}, \theta_s^{\ast(b)}\} = \delta^{ab}\delta_{rs}\qquad
\{\theta_r^{(a)}, \theta_s^{(b)}\} = \{\theta_r^{(a)}, \theta_s^{(b)}\} = 0\qquad 
\forall \ r,s \in \mathbb{Z}+\frac{1}{2}\; .
\ee
The differences with relation to the X-rows start with the number operators \(\zeta_r^{(a)} = \theta_r^{\ast(a)} \theta_r^{(a)} \) and \(\bar{\zeta}_r^{(a)} = \theta_r^{(a)} \theta_r^{\ast(a)} \). These operators count, respectively, the number of particles and holes in the \(Y^{(a)}\) chain. Observe that their structures are different from the number operators \(\eta\) and \(\bar{\eta}\). These definitions have been used in \cite{Dijkgraaf:2008ua} to rewrite the integer partition problem in terms of the XXZ-Hamiltonian with kink boundary conditions. 

\begin{figure}[h!]
	\centering
	\begin{tikzpicture}
	\node at (-2.3,0) {\(\cdots\)};
	\draw (-2,0) -- (2,0);
	\node at (-1.5,-0.5) {\({\scriptstyle -\frac{3}{2}}\)}; \draw[black,fill=black] (-1.5,0) circle (.5ex) ;
	\node at (-0.5,-0.5) {\({\scriptstyle -\frac{1}{2}}\)}; \draw[black,fill=black] (-0.5,0) circle (.5ex) ;
	\node at (0,0) {\({\scriptstyle |}\)}; \node at (0,0.3) {\({ \scriptstyle 0}\)};
	\node at (0.5,-0.5) {\({\scriptstyle \frac{1}{2}}\)}; \draw[black,fill=white] (0.5,0) circle (.5ex) ;
	\node at (1.5,-0.5) {\({\scriptstyle \frac{3}{2}}\)}; \draw[black,fill=white] (1.5,0) circle (.5ex) ;
	\node at (2.3,0) {\(\cdots\)};
	\end{tikzpicture}
	\caption{Vacuum in the \(Y\)-rows.}
	\label{fig:Yvacuum}
\end{figure}
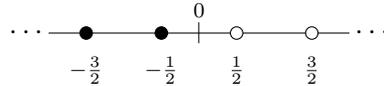

Fortunately, there is no zigzag between even and odd Y-rows as in the X-rows case, and we do not need to modify the conventions above. The vacuum \(|\bm{\tilde{0}}\rangle\) is given by
\be 
\theta^{(a)}_r |\bm{\tilde{0}}\rangle^{(a)} = 0, \ r>0\; , \quad \theta^{\ast (a)}_r 
|\bm{\tilde{0}}\rangle^{(a)} = 0, \ r< 0\; , \quad \forall a\in \mathbb{Z}\; ,
\ee
and it corresponds to the configuration in figure~\ref{fig:Yvacuum}, where all negative half-integers (\(r< 0\)) 
positions are occupied by particles.

Finally, the shifted vacua of the \(Y\)-rows are 
\be 
|\bm{\tilde{\ell}}\rangle^{(a)} = 
\left\{ 
\begin{array}{ll}
	\theta^{(a)}_{\ell+1/2}	\cdots \theta^{(a)}_{-3/2} \theta^{(a)}_{-1/2}|\bm{\tilde{0}}\rangle & \ell<0 \\
	\theta^{\ast (a)}_{\ell-1/2} \theta^{\ast (a)}_{\ell - 3/2}	\cdots \theta^{\ast (a)}_{1/2}|\bm{\tilde{0}}\rangle & \ell >0 
\end{array}
\right. \quad \; \ell \in \mathbb{Z}\; .
\ee

%%%%%%%%%%%%%%%%%%%%%%%%%%%%%%%%%%%%%
%%%%%%%%%%%%%%%%%%%%%%%%%%%%%%%%%%%%%
%%%%%%%%%%%%%%%%%%%%%%%%%%%%%%%%%%%%%
%%%%%%%%%%%%%%%%%%%%%%%%%%%%%%%%%%%%%
%%%%%%%%%%%%%%%%%%%%%%%%%%%%%%%%%%%%%

\section{Classifying states with plane partitions}
\label{sec:Part.states}

Now we want to use the definitions above to write the states classified by plane partitions. We first define the empty partition state, which is the subtler state in the Hilbert space. The construction of explicit expressions for states with more boxes is trivial from the formal viewpoint, but it is also an extremely laborious endeavor.

\begin{roundbox}
 \ \ \ \(0\)-\textbf{box}
\end{roundbox}

Let us start with the empty partition \(|{\bm \emptyset}\rangle\). Using figure~\ref{fig:kagome}, it is easy to see that the X-rows contribute as
\be 
\begin{split}
	|{\bm \emptyset}_X\rangle & = \bigotimes_{a\in \mathbb{Z}} ||\bm{a}|\rangle^{(\bm{a})}\\
	& = \cdots \otimes |{\bm 2}\rangle^{(-2)} \otimes |{\bm 1}\rangle^{(-1)}\otimes  |{\bm 0}\rangle^{(0)}\otimes |{\bm 1}\rangle^{(1)}
	\otimes |{\bm 2}\rangle^{(2)}\otimes\cdots
\end{split}
\ee

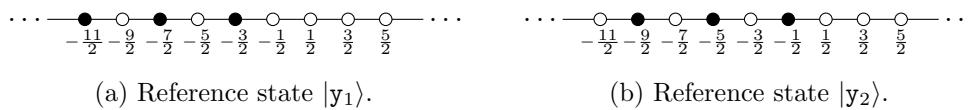
\begin{figure}[h!]
	\centering
\begin{subfigure}[h!]{0.4\textwidth}
	\centering
	\begin{tikzpicture}
	\draw (-3,0) -- (2,0); 
	\node at (-2.5,-0.3) {\({\scriptscriptstyle -\frac{11}{2}}\)}; \node at (-2.0,-0.3) {\({\scriptscriptstyle -\frac{9}{2}}\)};
	\node at (-1.5,-0.3) {\({\scriptscriptstyle -\frac{7}{2}}\)}; \node at (-1.0,-0.3) {\({\scriptscriptstyle -\frac{5}{2}}\)};
	\node at (-0.5,-0.3) {\({\scriptscriptstyle -\frac{3}{2}}\)}; \node at (-0,-0.3) {\({\scriptscriptstyle -\frac{1}{2}}\)};
	\node at (0.5,-0.3) {\({\scriptscriptstyle \frac{1}{2}}\)}; \node at (1,-0.3) {\({\scriptscriptstyle \frac{3}{2}}\)};
	\node at (1.5,-0.3) {\({\scriptscriptstyle \frac{5}{2}}\)}; 
	\node at (-3.3,0) {\(\dots\)}; \node at (2.3,0) {\(\dots\)}; 
	\draw[black,fill=black] (-2.5,0) circle (.5ex) ; \draw[black,fill=white] (-2,0) circle (.5ex) ;
	\draw[black,fill=black] (-1.5,0) circle (.5ex) ; \draw[black,fill=white] (-1,0) circle (.5ex) ;
	\draw[black,fill=black] (-0.5,0) circle (.5ex) ; \draw[black,fill=white] (0,0) circle (.5ex) ;
	\draw[black,fill=white] (0.5,0) circle (.5ex) ; \draw[black,fill=white] (1,0) circle (.5ex) ;
	\draw[black,fill=white] (1.5,0) circle (.5ex) ;
	\end{tikzpicture}
	\caption{Reference state \(|\mathtt{y}_1 \rangle\). }
	\label{fig:vacuumy1}	
\end{subfigure}
~
\begin{subfigure}[h!]{0.4\textwidth}
	\centering
	\begin{tikzpicture}
	\draw (-3,0) -- (2,0); 
	\node at (-2.5,-0.3) {\({\scriptscriptstyle -\frac{11}{2}}\)}; \node at (-2.0,-0.3) {\({\scriptscriptstyle -\frac{9}{2}}\)};
	\node at (-1.5,-0.3) {\({\scriptscriptstyle -\frac{7}{2}}\)}; \node at (-1.0,-0.3) {\({\scriptscriptstyle -\frac{5}{2}}\)};
	\node at (-0.5,-0.3) {\({\scriptscriptstyle -\frac{3}{2}}\)}; \node at (-0,-0.3) {\({\scriptscriptstyle -\frac{1}{2}}\)};
	\node at (0.5,-0.3) {\({\scriptscriptstyle \frac{1}{2}}\)}; \node at (1,-0.3) {\({\scriptscriptstyle \frac{3}{2}}\)};
	\node at (1.5,-0.3) {\({\scriptscriptstyle \frac{5}{2}}\)}; 
	\node at (-3.3,0) {\(\dots\)}; \node at (2.3,0) {\(\dots\)}; 
	\draw[black,fill=white] (-2.5,0) circle (.5ex) ; \draw[black,fill=black] (-2,0) circle (.5ex) ;
	\draw[black,fill=white] (-1.5,0) circle (.5ex) ; \draw[black,fill=black] (-1,0) circle (.5ex) ;
	\draw[black,fill=white] (-0.5,0) circle (.5ex) ; \draw[black,fill=black] (0,0) circle (.5ex) ;
	\draw[black,fill=white] (0.5,0) circle (.5ex) ; \draw[black,fill=white] (1,0) circle (.5ex) ;
	\draw[black,fill=white] (1.5,0) circle (.5ex) ;
	\end{tikzpicture}
	\caption{Reference state \(|\mathtt{y}_2 \rangle\).}
	\label{fig:vacuumy2}
\end{subfigure}
	\caption{Fiducial states \(|\mathtt{y}_1 \rangle\) and \(|\mathtt{y}_2 \rangle\). }
\end{figure}

The Y-rows contributions, on the other hand, are more involved. We start with the definition of two fiducial states, namely
\be 
|{\tt y}_1\rangle^{(a)} := \prod_{i\in \mathbb{Z}^{\geq 0}} \theta_{-(4i+1)/2}|\bm{\tilde{0}}\rangle\; ,
\ee
that corresponds to the configuration in figure~\ref{fig:vacuumy1} and 
\be 
|{\tt y}_2\rangle^{(a)} := \prod_{i\in \mathbb{Z}^{\geq 0}} \theta_{-(4i+3)/2}|\bm{\tilde{0}}\rangle\; ,
\ee
that corresponds to the configuration in figure~\ref{fig:vacuumy2}.

We also need to define the states
\bse
\be 
|\bm{(4\ell + 1)/2}\rangle^{(a)} = \prod^{\ell}_{i=0} \theta_{(4i+1)/2}|{\tt y}_1\rangle^{(a)} 
\ee
and 
\be 
|\bm{(4\ell + 3)/2}\rangle^{(a)} = \prod^{\ell}_{i=0} \theta_{(4i+3)/2}|{\tt y}_2\rangle^{(a)}\; ,
\ee
for \(\ell \in \mathbb{Z}^{\geq 0}\).
\ese
The Y-rows contribute to the empty configuration state as 
\be 
\begin{split}
	|{\bm \emptyset}_Y\rangle & = \bigotimes_{a\in \mathbb{Z}^{\geq 0} }\left(|\bm{(2a+1)/2}\rangle^{(a)} \otimes |\bm{(2a+1)/2}\rangle^{(-a-1)}\right)\\
	& = |\bm{1/2}\rangle^{(0)}\otimes |\bm{1/2}\rangle^{(-1)}\otimes |\bm{3/2}\rangle^{(1)}\otimes 
	|\bm{3/2}\rangle^{(-2)}\otimes \cdots
\end{split}
\ee

All in all, the lattice configuration corresponding to the 0-boxes plane partition is written as
\begin{shaded*}
	{\bf 0-box configuration}
	\be 
	|{\bm \emptyset}\rangle=|{\bm \emptyset}_X\rangle \otimes |{\bm \emptyset}_Y\rangle\; .
	\ee
\end{shaded*}

Before studying states with more boxes, let us write the growth operators in terms of the free fermions components. We first write the number operator
\bse 
\be 
\label{eq:hop01}
\sum| \square \rangle \langle \square | = 
\sum_{\substack{a\in \mathbb{Z}; m\in [a]\\ r\in \mathbb{Z}+\frac{1}{2}}}
\zeta_{r}^{(a)} \bar{\zeta}_{r+1}^{(a)} \zeta_{r}^{(a-1)} \bar{\zeta}_{r+1}^{(a-1)}  \eta_{m}^{(a)} \bar{\eta}_{m+2}^{(a)} 
\ee
that counts the number of available places in a given configuration, and
\be 
\label{eq:hop02}
\sum| \blacksquare \rangle \langle \blacksquare | = 
\sum_{\substack{a\in \mathbb{Z}; m\in [a]\\ r\in \mathbb{Z}+\frac{1}{2}}}
\zeta_{r+1}^{(a)} \bar{\zeta}_{r}^{(a)} \zeta_{r+1}^{(a-1)} \bar{\zeta}_{r}^{(a-1)}  \eta_{m+2}^{(a)} \bar{\eta}_{m}^{(a)}
\ee
that gives the number of boxes that can be consistently removed from a given plane partition.
\ese

The acting of these operators on the empty partition state gives
\bse 
\be 
\begin{split} 
	\sum |\square\rangle \langle \square | {\bm \emptyset}\rangle
	= \zeta^{(0)}_{1/2} \bar{\zeta}^{(0)}_{3/2} \zeta^{(-1)}_{1/2} \bar{\zeta}^{(-1)}_{3/2} 
	\eta^{(0)}_{0} \bar{\eta}^{(0)}_{2}|{\bm \emptyset}\rangle = |{\bm \emptyset}\rangle\; ,
\end{split}
\ee	
and 
\be 
\begin{split} 
	\sum |\blacksquare\rangle \langle \blacksquare | {\bm \emptyset}\rangle
	= \zeta^{(0)}_{3/2} \bar{\zeta}^{(0)}_{1/2} \zeta^{(-1)}_{3/2} \bar{\zeta}^{(-1)}_{1/2} 
	\eta^{(0)}_{2} \bar{\eta}^{(0)}_{0}|{\bm \emptyset}\rangle = 0\; .
\end{split}
\ee	
It naturally means that this state does not have any box to be removed and there is one available place to add a box. Therefore, it indeed corresponds to a 0-box configuration, and our definitions are consistent.
\ese

Finally, the box--annihilation and -creation operators are defined as follows
\bse
\be 
\label{eq:hop03}
\sum| \square \rangle \langle \blacksquare | = 
\sum_{\substack{a\in \mathbb{Z}; m\in [a]\\ r\in \mathbb{Z}+\frac{1}{2}}}
\theta_{r}^{\ast (a)}  \theta_{r+1}^{(a)} \theta_{r}^{\ast(a-1)}  \theta_{r+1}^{(a-1)} \psi_{m+2}^{\ast (a)} \psi_{m}^{(a)}
\ee
and 
\be 
\label{eq:hop04}
\sum| \blacksquare \rangle \langle \square | = 
\sum_{\substack{a\in \mathbb{Z}; m\in [a]\\ r\in \mathbb{Z}+\frac{1}{2}}}
\theta_{r+1}^{\ast (a)}  \theta_{r}^{(a)} \theta_{r+1}^{\ast(a-1)}  \theta_{r}^{(a-1)} \psi_{m}^{\ast (a)} \psi_{m+2}^{(a)}\; .
\ee
\ese

Let us now act with \(\sum |\square\rangle \langle \blacksquare |\) and 
\(\sum |\blacksquare\rangle \langle \square |\) on \(|{\bm \emptyset}\rangle\). Initially we have 
\be 
\sum |\square\rangle \langle \blacksquare | {\bm \emptyset}\rangle = 0\; ,
\ee
that logically annihilates the empty configuration. In other words, it is the highest weight state in this representation, 
and states with more boxes are obtained from the action of the box-creation operator. 

\begin{roundbox}
	\ \ \ \(1\)-\textbf{box}
\end{roundbox}

\begin{figure}[h!]
	\centering
	\includegraphics[width=4.0cm]{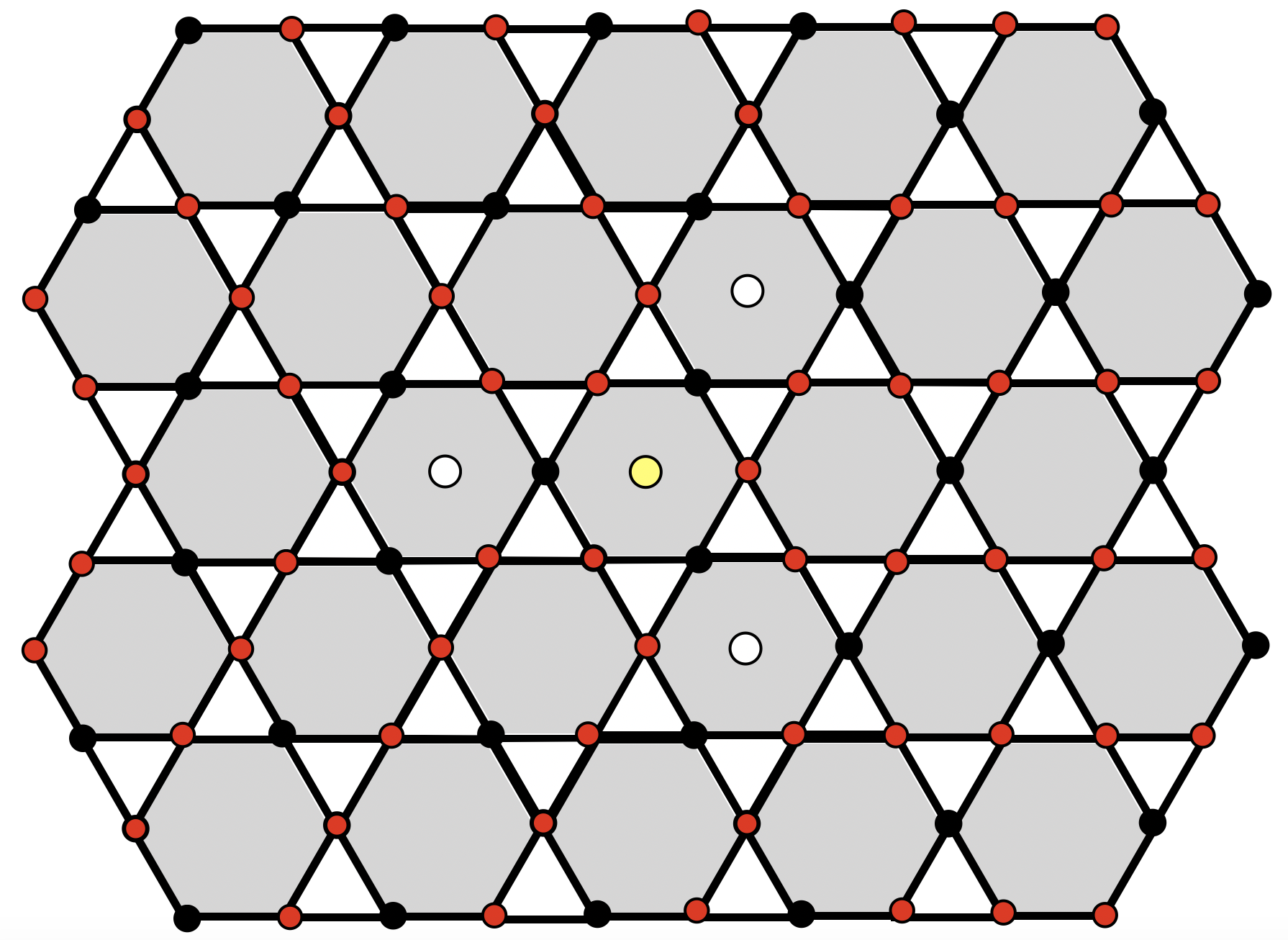}
	\caption{Lattice configuration \(|\one \ \rangle\).}
	\label{fig:1-box}
\end{figure}

From the considerations above, the 1-box configuration is
\begin{shaded*}{\bf \(1\)-box configuration}
	\be 
	\begin{split}
		\ytableausetup{centertableaux, smalltableaux}
		\left| \one\ \right\rangle \equiv   & 
		\sum |\blacksquare\rangle \langle \square | {\bm \emptyset}\rangle\\
		= & \theta^{\ast(0)}_{3/2} \theta^{(0)}_{1/2} \theta^{\ast(-1)}_{3/2} \theta^{(-1)}_{1/2} \psi^{\ast(0)}_0 \psi^{(0)}_2 |{\bm \emptyset}\rangle
	\end{split}
	\ee
\end{shaded*}
We say that it is the 1-box plane partition state with lattice configuration given by figure~\ref{fig:1-box}. As we have mentioned before, the white circles in~\ref{fig:1-box} denote the plaquettes where boxes can be created, and the big yellow circle denotes the plaquette where the box-annihilation operator acts in a nontrivial manner. 

The number operators now give
\bse
\be 
\left(\sum |\blacksquare\rangle \langle \blacksquare |\right) |  \one \ \rangle
= \zeta^{(0)}_{3/2} \bar{\zeta}^{(0)}_{1/2} \zeta^{(-1)}_{3/2} \bar{\zeta}^{(-1)}_{1/2} 
\eta^{(0)}_{2} \bar{\eta}^{(0)}_{0}|  \one \ \rangle = |  \one \ \rangle\; ,
\ee		
which says that we have one box to remove from this configuration;  and
\be 
\begin{split} 
	\left(\sum |\square\rangle \langle \square |\right) & |  \one \ \rangle
	= \zeta^{(-1)}_{3/2} \bar{\zeta}^{(-1)}_{5/2} \zeta^{(-2)}_{3/2} \bar{\zeta}^{(-2)}_{5/2} 
	\eta^{(-1)}_1 \bar{\eta}^{(-1)}_3|  \one \ \rangle\\
	& +  \zeta^{(0)}_{-3/2} \bar{\zeta}^{(0)}_{-1/2} \zeta^{(-1)}_{-3/2} \bar{\zeta}^{(-1)}_{-1/2} 
	\eta^{(0)}_{-2} \bar{\eta}^{(0)}_0|  \one \ \rangle\\
	& + \zeta^{(1)}_{3/2} \bar{\zeta}^{(1)}_{5/2} \zeta^{(0)}_{3/2} \bar{\zeta}^{(0)}_{5/2} 
	\eta^{(1)}_1 \bar{\eta}^{(1)}_3|  \one \ \rangle\\
	& = 3 |  \one \ \rangle\; ,
\end{split}
\ee	
\ese	
that indicates the three available places where one can add a second box.

We now apply the box-annihilation operators to the lattice state corresponding to the 1-box state. Therefore
\be 
\ytableausetup{centertableaux, smalltableaux}
\left(\sum |\square \rangle \langle \blacksquare| \right)
\left| \one\ \right\rangle = |{\bm \emptyset}\rangle \; ,
\ee
since there is one box to be removed consistently.

\begin{roundbox}
	\ \ \(2\)-\textbf{boxes}
\end{roundbox}

\begin{figure}[h!]
	\centering
\begin{subfigure}[h!]{0.3\textwidth}
	\centering
	\includegraphics[width=3.5cm]{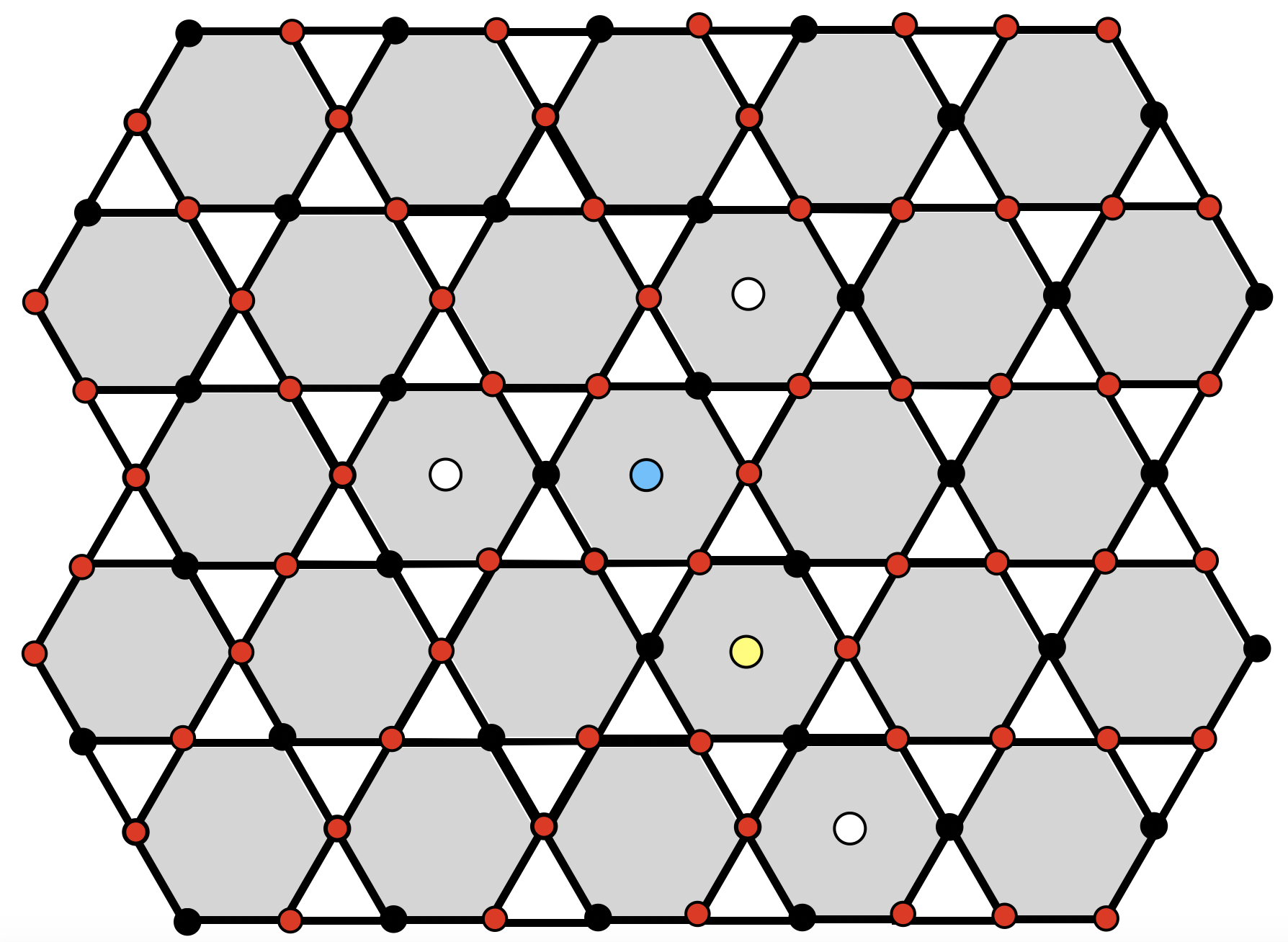}
	\caption{Lattice configuration \(|\twoa \ \rangle\).}
	\label{fig:2a-boxes}
\end{subfigure}
~
\begin{subfigure}[h!]{0.3\textwidth}
	\centering
	\includegraphics[width=3.5cm]{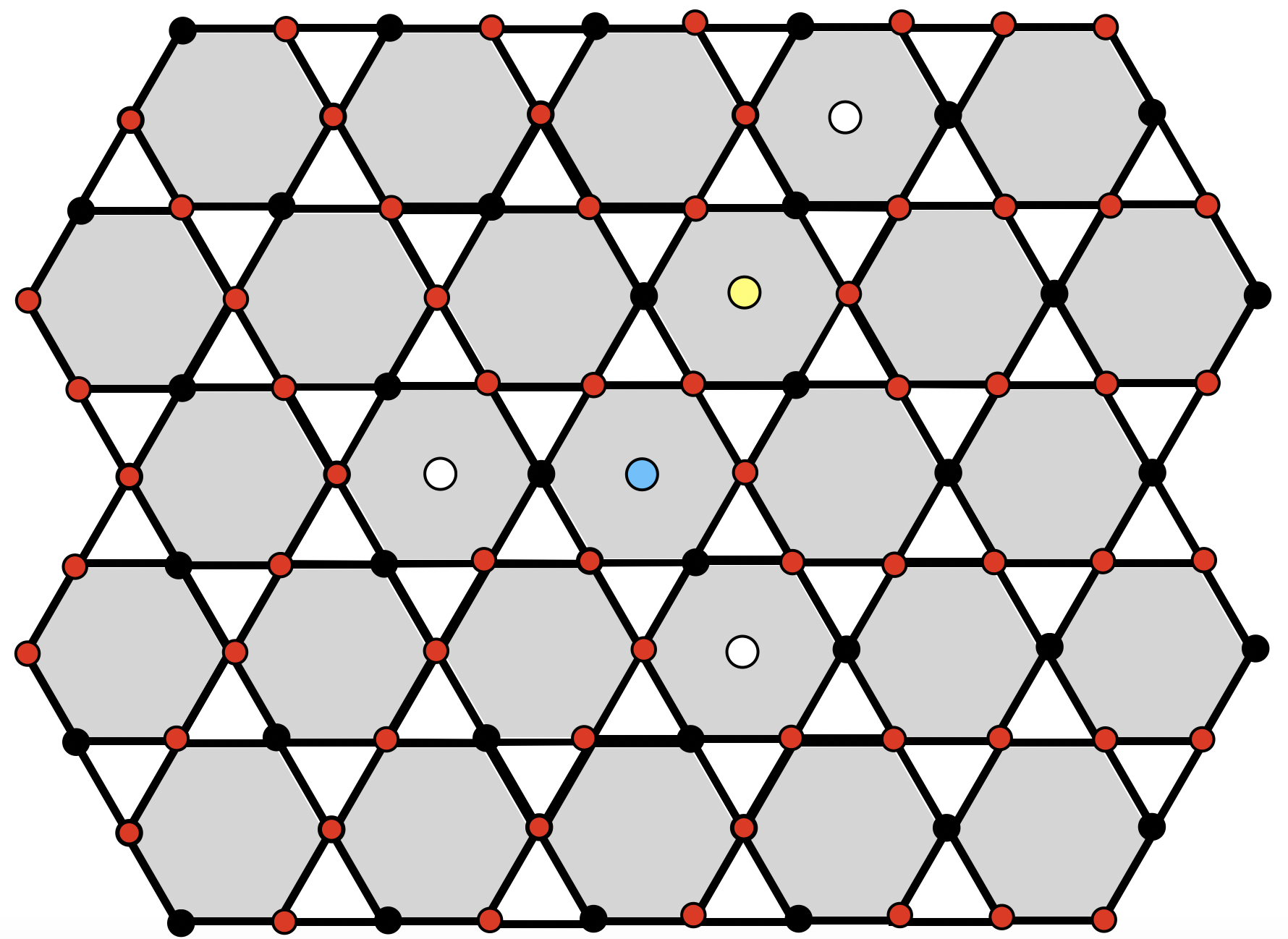}
	\caption{Lattice configuration \(|\twob \ \rangle\).}
	\label{fig:2b-boxes}
\end{subfigure}
~
\begin{subfigure}[h!]{0.3\textwidth}
	\centering
	\includegraphics[width=3.5cm]{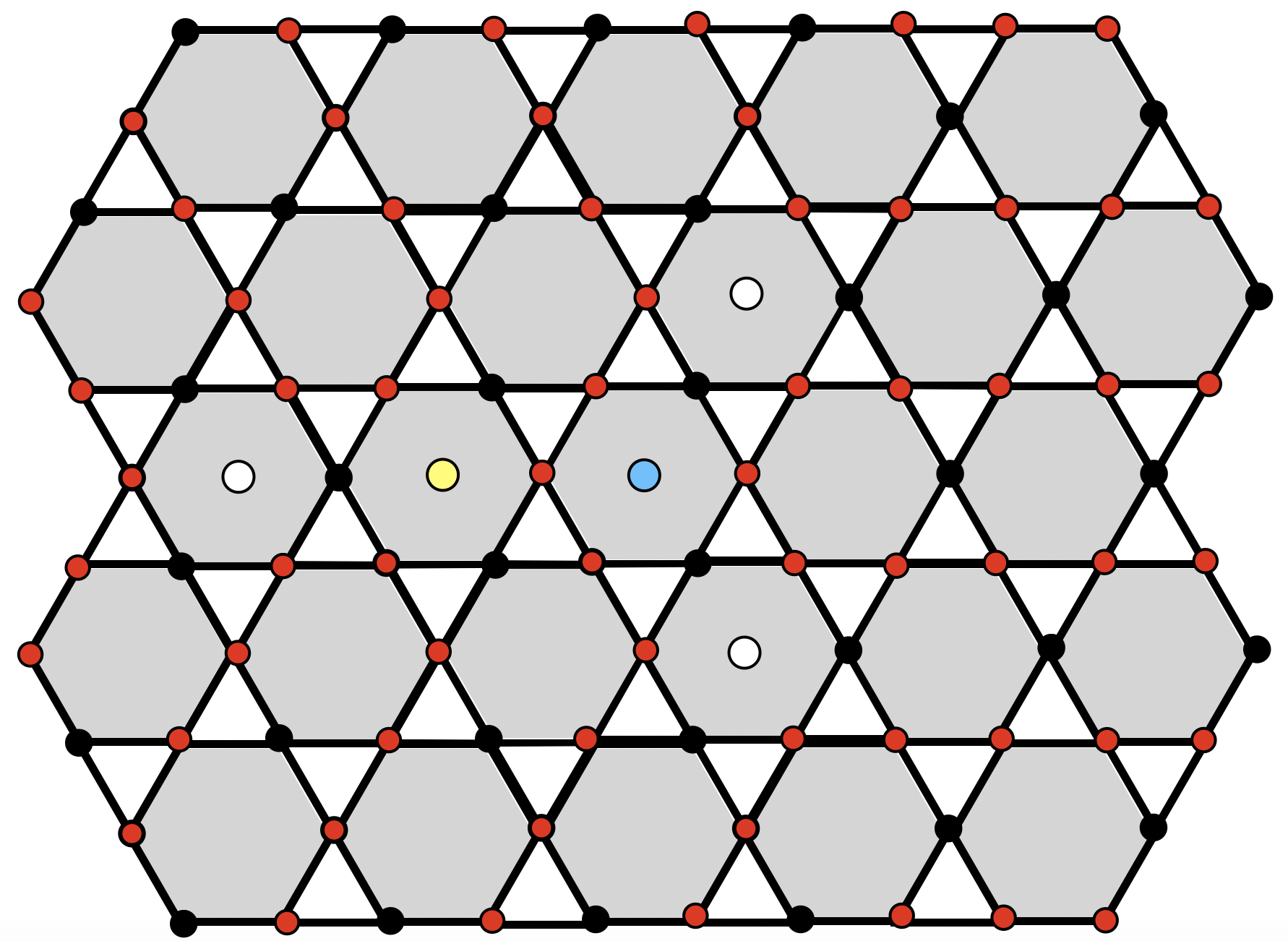}
	\caption{Lattice configuration \(|\twoc \ \rangle\).}
	\label{fig:2c-boxes}
\end{subfigure}
	\label{fig:2-boxes-conf}
	\caption{States corresponding to 2-boxes plane partitions.}
\end{figure}

Finally, acting with the creation operator we have
\bse
	\be 
	\begin{split}
		\ytableausetup{centertableaux, smalltableaux}
		\left(\sum |\blacksquare \rangle \langle \square| \right)
		\left| \ydiagram[*(lightgray)]{1}\right\rangle & =  
		\sum_{m\in [1]}\theta^{\ast(-1)}_{m+3/2} \theta^{(-1)}_{m+1/2} \theta^{\ast(-2)}_{m+3/2} \theta^{(-2)}_{m+1/2} \psi^{\ast(-1)}_m \psi^{(-1)}_{m+2} \left| \ydiagram[*(lightgray)]{1}\right\rangle + \\
		& + \sum_{m\in [1]}\theta^{\ast(1)}_{m+3/2} \theta^{(1)}_{m+1/2} \theta^{\ast(0)}_{m+3/2} \theta^{(0)}_{m+1/2} \psi^{\ast(1)}_m \psi^{(1)}_{m+2} \left| \one\ \right\rangle\\
		& + 
		\sum_{m\in [0]}
		\theta^{\ast(0)}_{m+3/2} \theta^{(0)}_{m+1/2} \theta^{\ast(-1)}_{m+3/2} \theta^{(-1)}_{m+1/2} \psi^{\ast(0)}_m \psi^{(0)}_{m+2} 
		\left| \one\ \right\rangle
	\end{split}
	\ee	
where we represent these states in terms of \(2\)-boxes partitions as
\begin{shaded*}{\bf \(2\)-boxes configurations}
	\be 
	\left| \twoa\ \right\rangle = 
	\theta^{\ast(-1)}_{5/2} \theta^{(-1)}_{3/2} \theta^{\ast(-2)}_{5/2} \theta^{(-2)}_{3/2} \psi^{\ast(-1)}_1 \psi^{(-1)}_3 \left|\one \ \right\rangle 
	\ee	
	\be 
	\left| \twob \ \right\rangle = 
	\theta^{\ast(1)}_{5/2} \theta^{(1)}_{3/2} \theta^{\ast(0)}_{5/2} \theta^{(0)}_{3/2} \psi^{\ast(1)}_1 \psi^{(1)}_3 \left|\one \ \right\rangle 
	\ee	
	\be 
	\left| \twoc \ \right\rangle = 
	\theta^{\ast(0)}_{-1/2} \theta^{(0)}_{-3/2} \theta^{\ast(-1)}_{-1/2} \theta^{(-1)}_{-3/2} \psi^{\ast(1)}_{-2} \psi^{(1)}_0 \left|\one \ \right\rangle 
	\ee	
\end{shaded*}
These states are written diagrammatically in figures~\ref{fig:2a-boxes},~\ref{fig:2b-boxes}
and~\ref{fig:2c-boxes}, respectively. We also represent the locked plane partition boxes with a big blue circle. More precisely, the blue circles denote the boxes that cannot be removed in a way that respects the plane partition rules.
\ese

Evidently we can build states with more boxes in a similar way, but now we want to study other aspects of this model. 
Diagonalization of the Hamiltonian~(\(\ref{eq:Hamiltonian}\)) is a problem we could not address in this work, but from~\cite{Dijkgraaf:2008ua}, 
we know that its ground state is
\be 
|\Psi_0\rangle = \sum_{\Lambda} q^{\frac{\# Boxes(\Lambda)}{2}}|\Lambda\rangle \; ,
\ee
in such a way that its norm squared gives that famous MacMahon function
\be 
\langle \Psi_0 | \Psi_0 \rangle = \prod_{n}\frac{1}{(1-q^n)^n}\; .
\ee 

%%%%%%%%%%%%%%%%%%%%%%%%%%%%%%%%%%%%%
%%%%%%%%%%%%%%%%%%%%%%%%%%%%%%%%%%%%%
%%%%%%%%%%%%%%%%%%%%%%%%%%%%%%%%%%%%%
%%%%%%%%%%%%%%%%%%%%%%%%%%%%%%%%%%%%%

\section{Kagome lattice Hamiltonian}
\label{sec:Hamiltonian}

As we said before, in~\cite{Araujo:2020opn} we have rewritten the Hamiltonian~(\ref{eq:qtcrystal}) in terms of fermions in a Kagome lattice. Furthermore, the plane partition classification of states is a level grading; and for each plane partition configuration there is a corresponding state in the Kagome lattice. In this section we want to continue this analysis using the parametrization of section~(\ref{sec:Fermions}). 

\begin{figure}[h!]	
	\centering
	\begin{subfigure}[h!]{0.4\textwidth}
		\centering
		\begin{tikzpicture}
		\draw[dashed, line width=0.5mm] (-1,0) -- (1,0);
		\draw[black,fill=black] (1,0) circle (.8ex) ;
		\draw[black,fill=white] (-1,0) circle (.8ex) ;
		\draw[rotate=60, line width=0.5mm] (0,0) -- (1,0);
		\draw[black,fill=white][rotate=60] (1,0) circle (.8ex) ;
		\draw[rotate=-60, line width=0.5mm] (0,0) -- (1,0); 
		\draw[black,fill=white][rotate=-60] (1,0) circle (.8ex) ;
		\draw[rotate=120, line width=0.5mm] (0,0) -- (1,0);
		\draw[black,fill=black][rotate=120] (1,0) circle (.8ex) ; 
		\draw[rotate=-120, line width=0.5mm] (0,0) -- (1,0);
		\draw[black,fill=black][rotate=-120] (1,0) circle (.8ex) ; 
		\node at (1.7,0) {\(\Longrightarrow\)};
		\end{tikzpicture}
		\begin{tikzpicture}
		\draw[dashed, line width=0.5mm] (-1,0) -- (1,0);
		\draw[black,fill=white] (1,0) circle (.8ex) ;
		\draw[black,fill=black] (-1,0) circle (.8ex) ;
		\draw[rotate=60, line width=0.5mm] (0,0) -- (1,0);
		\draw[black,fill=black][rotate=60] (1,0) circle (.8ex) ;
		\draw[rotate=-60, line width=0.5mm] (0,0) -- (1,0); 
		\draw[black,fill=black][rotate=-60] (1,0) circle (.8ex) ;
		\draw[rotate=120, line width=0.5mm] (0,0) -- (1,0);
		\draw[black,fill=white][rotate=120] (1,0) circle (.8ex) ; 
		\draw[rotate=-120, line width=0.5mm] (0,0) -- (1,0);
		\draw[black,fill=white][rotate=-120] (1,0) circle (.8ex) ; 
		\end{tikzpicture}
		\caption{Box creation.}
		\label{fig:box-creat}
	\end{subfigure}
	~
	\begin{subfigure}[h!]{0.4\textwidth}
		\centering
		\begin{tikzpicture}
		\draw[dashed, line width=0.5mm] (-1,0) -- (1,0);
		\draw[black,fill=white] (1,0) circle (.8ex) ;
		\draw[black,fill=black] (-1,0) circle (.8ex) ;
		\draw[rotate=60, line width=0.5mm] (0,0) -- (1,0);
		\draw[black,fill=black][rotate=60] (1,0) circle (.8ex) ;
		\draw[rotate=-60, line width=0.5mm] (0,0) -- (1,0); 
		\draw[black,fill=black][rotate=-60] (1,0) circle (.8ex) ;
		\draw[rotate=120, line width=0.5mm] (0,0) -- (1,0);
		\draw[black,fill=white][rotate=120] (1,0) circle (.8ex) ; 
		\draw[rotate=-120, line width=0.5mm] (0,0) -- (1,0);
		\draw[black,fill=white][rotate=-120] (1,0) circle (.8ex) ; 
		\node at (1.7,0) {\(\Longrightarrow\)};
		\end{tikzpicture}
		\begin{tikzpicture}
		\draw[dashed, line width=0.5mm] (-1,0) -- (1,0);
		\draw[black,fill=black] (1,0) circle (.8ex) ;
		\draw[black,fill=white] (-1,0) circle (.8ex) ;
		\draw[rotate=60, line width=0.5mm] (0,0) -- (1,0);
		\draw[black,fill=white][rotate=60] (1,0) circle (.8ex) ;
		\draw[rotate=-60, line width=0.5mm] (0,0) -- (1,0); 
		\draw[black,fill=white][rotate=-60] (1,0) circle (.8ex) ;
		\draw[rotate=120, line width=0.5mm] (0,0) -- (1,0);
		\draw[black,fill=black][rotate=120] (1,0) circle (.8ex) ; 
		\draw[rotate=-120, line width=0.5mm] (0,0) -- (1,0);
		\draw[black,fill=black][rotate=-120] (1,0) circle (.8ex) ; 
		\end{tikzpicture}
		\caption{Box Annihilation.}
		\label{fig:box-annih}
	\end{subfigure}
	\caption{Box creation and annihilation from the hexagon cell perspective.}
\end{figure}
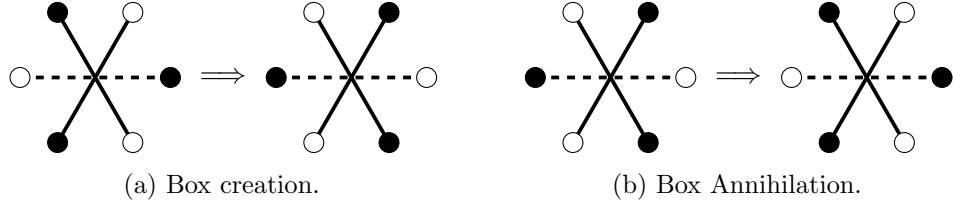

Remember that the box creation and annihilation are equivalent to simultaneous particles hopping in the Kagome lattice hexagon. In the \(X\) and \(Y\)-spin chains parametrization, such hexagons live in the triads \(Y^{(a)}\)-\(X^{(a)}\)-\(Y^{(a-1)}\), and box creation and annihilation are represented as in figures~\ref{fig:box-creat} and~\ref{fig:box-annih}, respectively. 

Using these simple observations, the parametrization defined in section~(\ref{sec:Fermions}) and the operators~(\ref{eq:hop01}),~(\ref{eq:hop02}),~(\ref{eq:hop03}) and~\ref{eq:hop04}), it is easy to convince ourselves that the Hamiltonian~(\ref{eq:qtcrystal}) is equivalent to
\bse
\be 
\label{eq:Hamiltonian}
\begin{split}
	H =  & - J \sum_{\substack{a\in \mathbb{Z}; m\in [a] \\ r= m +\frac{1}{2} }}  \left[ 
	\theta_{r}^{\ast (a)}  \theta_{r+1}^{(a)} \theta_{r}^{\ast (a-1)}  \theta_{r+1}^{(a-1)} \psi_{m+2}^{\ast (a)} \psi_{m}^{(a)} + 
	\theta_{r+1}^{\ast (a)}  \theta_{r}^{(a)} \theta_{r+1}^{\ast (a-1)}  \theta_{r}^{(a-1)} \psi_{m}^{\ast (a)} \psi_{m+2}^{(a)}
	\right. \\
	& \left. 
	- \frac{V \sqrt{q}}{J} \zeta_{r}^{(a)} \bar{\zeta}_{r+1}^{(a)} \zeta_{r}^{(a-1)} \bar{\zeta}_{r+1}^{(a-1)}  \eta_{m}^{(a)} \bar{\eta}_{m+2}^{(a)} - 
	\frac{V}{J \sqrt{q} } \zeta_{r+1}^{(a)} \bar{\zeta}_{r}^{(a)} \zeta_{r+1}^{(a-1)} \bar{\zeta}_{r}^{(a-1)}  \eta_{m+2}^{(a)} \bar{\eta}_{m}^{(a)}
	\right]\; ,
\end{split}	
\ee
where \([a]\) denotes the equivalence classes of \(\mathbb{Z}/2\mathbb{Z}\), which is a manifestation of the 
\(X\)-rows zigzag between even and odd rows. It is easy to see that the Hamiltonian has two parts: one of them acts on the triad \(Y^{(2a)}\)-\(X^{(2a)}\)-\(Y^{(2a-1)}\), and another on 
\(Y^{(2a+1)}\)-\(X^{(2a+1)}\)-\(Y^{(2a)}\), therefore
\be 
\boxed{
	H = \sum_{\substack{a,m\\ (r=2m+ 1/2)} } H_{2a, 2m} +
	\sum_{\substack{a,m\\ (r=2m+3/2)} } H_{2a+1, 2m+1} }
\ee
More explicitly
\be 
\label{eq:Hamiltonian02}
\begin{split}
	H& = -J \sum_{\substack{a\in \mathbb{Z}; m\in [a] \\ r= m+\frac{1}{2} }}  \left[
	\left( \theta^{\ast (2a)}_{r} \theta_{r+1}^{(2a)} \theta^{\ast (2a-1)}_{r} \theta_{r+1}^{(2a-1)} 
	\psi^{\ast (2a)}_{2m+2} \psi_{2m}^{(2a)} 
	+ \theta^{\ast (2a)}_{r+1} \theta_{r}^{(2a)} \theta^{\ast (2a-1)}_{r+1} \theta_{r}^{(2a-1)} 
	\psi^{\ast (2a)}_{2m} \psi_{2m+2}^{(2a)}  + 
	\right.\right.\\
	& + \left. 
	\theta^{\ast (2a+1)}_{r} \theta_{r+1}^{(2a+1)} \theta^{\ast (2a)}_{r} \theta_{r+1}^{(2a)} 
	\psi^{\ast (2a+1)}_{2m+3} \psi_{2m+1}^{(2a+1)} 
	+ \theta^{\ast (2a+1)}_{r+1} \theta_{r}^{(2a+1)} \theta^{\ast (2a)}_{r+1} \theta_{r}^{(2a)} 
	\psi^{\ast (2a+1)}_{2m+1} \psi_{2m+3}^{(2a+1)} 
	\right) \\
	& -\frac{V \sqrt{q} }{J} \left(
	\zeta_{r}^{(2a)} \bar{\zeta}_{r+1}^{(2a)} \zeta_{r}^{(2a-1)} \bar{\zeta}_{r+1}^{(2a-1)}   \eta_{2m}^{(2a)} \bar{\eta}_{2m+2}^{(2a)} + 
	\zeta_{r}^{(2a+1)} \bar{\zeta}_{r+1}^{(2a+1)} \zeta_{r}^{(2a)} \bar{\zeta}_{r+1}^{(2a)}   \eta_{2m+1}^{(2a+1)} \bar{\eta}_{2m+3}^{(2a+1)} 
	\right) \\
	& -\left.\frac{V}{J \sqrt{q}} \left(
	\zeta_{r+1}^{(2a)} \bar{\zeta}_{r}^{(2a)} \zeta_{r+1}^{(2a-1)} \bar{\zeta}_{r}^{(2a-1)}   \eta_{2m+2}^{(2a)} \bar{\eta}_{2m}^{(2a)} + 
	\zeta_{r+1}^{(2a+1)} \bar{\zeta}_{r}^{(2a+1)} \zeta_{r+1}^{(2a)} \bar{\zeta}_{r}^{(2a)}   \eta_{2m+3}^{(2a+1)} \bar{\eta}_{2m+1}^{(2a+1)} 
	\right)\right].
\end{split}
\ee
\ese

As we mentioned before, the 2D version of~(\ref{eq:qtcrystal}) has been mapped to the XXZ-Hamiltonian in~\cite{Dijkgraaf:2008ua} using a Jordan-Wigner transformation. With the parametrization that we have defined in section~(\ref{sec:Fermions}), we can apply the same idea to the 3D problem but, as we see below, the final result is not so appealing. 
Let us first denote the site configurations as \(|\bullet\rangle\equiv |+\rangle \equiv |\uparrow\rangle\) and \(|\circ\rangle\equiv |-\rangle \equiv |\downarrow\rangle \). The Jordan-Wigner transformation gives the following map between free fermions and the fundamental representation of the \(su(2)\) algebra
\be
\begin{array}{l}
	\psi^{\ast(a)}_{m} |\bullet\rangle_{m} = 0 \\
	\psi^{(a)}_{m} |\circ\rangle_{m} = 0
\end{array} 
\Leftrightarrow
\begin{array}{l}
	\chi^{+ (a)}_{m} |+\rangle_{m} = 0 \\
	\chi^{- (a)}_{m} |-\rangle_{m} = 0\\
	\chi^{z (a)}_{m} |\pm\rangle_{m} = \pm\frac{1}{2} |\pm\rangle_{m}
\end{array}
\qquad 
\begin{array}{l}
	\theta^{\ast(a)}_{r} |\bullet\rangle_{r} = 0\\	
	\theta^{(a)}_{r} |\circ\rangle_{r} = 0
\end{array}
\Leftrightarrow
\begin{array}{l}
	\tau^{+ (a)}_{r} |+\rangle_{r} = 0 \\
	\tau^{- (a)}_{r} |-\rangle_{r} = 0\\
	\tau^{z (a)}_{r} |\pm\rangle_{r} = \pm \frac{1}{2} |\pm\rangle_{r} 
\end{array}
\ee
where \(\vec{\chi}\) and \(\vec{\xi}\) satisfy the \(su(2)\) algebra
\be 
\begin{array}{lll}
	\left[\chi_+, \chi_-\right] = 2\chi^z && \left[\chi_z, \chi_{\pm}\right] = \pm \chi^{\pm}\\
	\left[\xi_+, \xi_-\right] = 2\xi^z && \left[\xi_z, \xi_{\pm}\right] = \pm \xi^{\pm}
\end{array}\; .
\ee
The number operators act in a very straightforward manner
\be 
\begin{array}{lll}
	\eta_m | \bullet\rangle_m = 0 && \eta_m | \circ\rangle_m = | \circ\rangle_m \\
	\bar{\eta}_m | \bullet\rangle_m = | \bullet\rangle_m && \bar{\eta}_m | \circ\rangle_m = 0
\end{array}\
\qquad
\begin{array}{lll}
	\zeta_r | \bullet\rangle_r = | \bullet\rangle_r && \zeta_r | \circ\rangle_r = 0 \\
	\bar{\zeta}_r | \bullet\rangle_r = 0 && \bar{\zeta}_r | \circ\rangle_r = | \circ\rangle_r
\end{array}\; .
\ee

In terms of the \(su(2)\) generators, the Hamiltonian~(\ref{eq:Hamiltonian}) can be inelegantly expressed as
\be 
\label{eq:Hamiltonian03}
\begin{split}
	H = &  - J \sum_{\substack{a\in \mathbb{Z}; m\in [a]\\ r\in \mathbb{Z}+\frac{1}{2}}} \left\{
	\tau_{r}^{+ (a)}  \tau_{r+1}^{-(a)} \tau_{r}^{+(a-1)}  \tau_{r+1}^{-(a-1)} \chi_{m+2}^{+ (a)} \chi_{m}^{-(a)} + 
	\tau_{r+1}^{+ (a)}  \tau_{r}^{-(a)} \tau_{r+1}^{+(a-1)}  \tau_{r}^{-(a-1)} \chi_{m}^{+ (a)} \chi_{m+2}^{-(a)}
	\right. \\
	-\frac{V\sqrt{q}}{J} & \left[\frac{1}{2}\left(\tau^{z (a)}_r - \tau^{z (a)}_{r+1}\right) - \tau^{z (a) }_r \tau^{z (a) }_{r+1} + \frac{1}{4}\right] 
	\left[\frac{1}{2}\left(\tau^{z (a-1)}_r - \tau^{z (a-1)}_{r+1}\right) - \tau^{z (a-1) }_r \tau^{z (a-1) }_{r+1} + \frac{1}{4} \right] \times \\
	& \left[\frac{1}{2}\left(\chi^{z (a)}_{m+2} - \chi^{z (a)}_m\right) - \chi^{z (a) }_m \chi^{z (a) }_{m+2} + \frac{1}{4} \right]\\
	-\frac{V}{J \sqrt{q}} & \left[\frac{1}{2}\left(\tau^{z (a)}_{r+1} - \tau^{z (a)}_{r}\right) - \tau^{z (a) }_{r+1} \tau^{z (a) }_{r} + \frac{1}{4}\right] 
	\left[\frac{1}{2}\left(\tau^{z (a-1)}_{r+1} - \tau^{z (a-1)}_{r}\right) - \tau^{z (a-1) }_{r+1} \tau^{z (a-1) }_{r} + \frac{1}{4} \right] \times \\
	& \left. \left[\frac{1}{2}\left(\chi^{z (a)}_{m} - \chi^{z (a)}_{m+2}\right) - \chi^{z (a) }_{m+2} \chi^{z (a) }_m + \frac{1}{4}\right]
	\right\}\; ,
\end{split}
\ee
and we see that contrary to the 2D case, the resulting plane partition Hamiltonian in terms of \(su(2)\) generators does not make the problem any easier. 

%%%%%%%%%%%%%%%%%%%%%%%%%%%%%%%%%%%%%
%%%%%%%%%%%%%%%%%%%%%%%%%%%%%%%%%%%%%
%%%%%%%%%%%%%%%%%%%%%%%%%%%%%%%%%%%%%
%%%%%%%%%%%%%%%%%%%%%%%%%%%%%%%%%%%%%

\section{Boltzmann weights \& Classical Problem}
\label{sec:boltzmann}

As we have seen in section~(\ref{sec:Hamiltonian}), the Kagome lattice Hamiltonian can be decomposed in even and odd parts
\be 
\begin{split}
H & =  \sum_{\substack{a, m \in \mathbb{Z}\\ r\in \mathbb{Z}+\frac{1}{2}}} {\cal H}_{a,m}  
= \sum_{\substack{a, m \in \mathbb{Z}\\ r\in \mathbb{Z}+\frac{1}{2}}} \left(H_{2a,2m } + H_{2a+1, 2m+1} \right)\\
&\equiv H_{(even)} + H_{(odd)}\; .
\end{split}
\ee
We consider now that the system lives in a finite lattice, say with \(N\) triads \(YXY\)-rows and \(M\) columns (or hexagons) in each \(YXY\)-triad. Moreover, periodic boundary conditions are imposed in both directions; 
in other words, the system lives in a torus \(\mathbb{Z}_{M} \times\mathbb{Z}_{N}\). Evidently we ultimately want to consider the limits \(N, M\to \infty\), but let us retain the periodic boundary conditions now. The quantum partition function is given by
\bse
\be 
\begin{split}
	Z & = \mathrm{Tr}\left( e^{-\beta H}\right) \\ 
	& = \mathrm{Tr}\left[ \exp\left(-\beta \sum_{a, m\in \mathbb{Z}} (H_{2a,2m} +H_{2a+1,2m+1}\right)\right]\; ,
\end{split}
\ee
where \(\beta\equiv1/\kappa_B T\). Since \([{\cal H}_{a,m}, {\cal H}_{a,m\pm 1}]\neq 0\),  \([{\cal H}_{a,m}, {\cal H}_{a\pm 1,m}]\neq 0\) and \([{\cal H}_{a,m}, {\cal H}_{a\pm 1,m\pm 1}]\neq 0\), we cannot split the partition function in terms of products of local operators. Using the Zassenhaus formula -- the dual Baker-Campbell-Hausdorff formula -- the partition function can be written as
\be 
\label{eq:p.func01}
Z = \mathrm{Tr}\left( e^{-\beta H_{(even)}}  e^{-\beta H_{(odd)}}
\prod_{n\geq 2}  e^{C_{n}(H_{(even)}, H_{(odd)})} \right)\; ,
\ee
\ese
where \(C_{n}(H_{(even)}, H_{(odd)})\) are homogeneous 
polynomials in \(H_{(even)}\) and \(H_{(odd)}\) of degree \(n\). 

As usual, in order to compute this partition function we need to diagonalize the Hamiltonian \(H\) to determine the eigenvalues \(E_{\mathcal{C}}\) of the eigenstates \(\left|\Psi(\mathcal{C})\right\rangle \). Therefore, the Boltzmann weight for each eigen-configuration \({\cal C}\) is
\be 
\label{eq:Boltzmann}
{\cal W}_B(\mathcal{C}) = \exp \left( -\beta E_{\mathcal{C}}\right)\; .
\ee
As we said before, diagonalization of \(H\) is a Herculean task, and the very existence of the current work can be traced to our attempt to dodge these difficulties as much as possible. Given the impossibility of solving the quantum system above, let us step back and scrutinize a classical counterpart of it.

Let us first observe that each eigen-configuration is composed by a sum of states classified by plane partitions, that is
\be 
|\Psi_{\cal C}\rangle = 
\sum_{\Lambda} \psi_{\cal C}(\Lambda, J, V, q)|\Lambda\rangle\; ,
\ee
where each plane partition state corresponds to a Kagome lattice configuration. In other words, each plane partition state \(|\Lambda\rangle\) can be written as a formal product of local hexagon states of the type represented in figure~\ref{fig:hexagon}. Moreover, since we assume that the system lives in a torus, we obviously have a finite number \(\ell\) of local configurations -- we will see below that \(\ell=18\) and that it is independent of \(N\) and \(M\).

\begin{hypo}
At this point we hypothesize that each hexagon configuration \(\hexagon_i\), with \(i=1, \dots, \ell\), defines a well defined and unique statistical weight \({\cal W}(\hexagon_i)\). In other words, we suppose that these weights do not depend on the specific position of the hexagon \(\hexagon_i\) inside the lattice, the \((a,m)\) parametrization, but only on the local structure defined in figure~\ref{fig:hexagon}.
\end{hypo}

Consequently, each hexagon cell parametrized as in figure~\ref{fig:hexagon} gives a minute contribution to the partition function. With these observations, one could try a reverse engineering process: build eigenstates from sums of cell configurations products.
\begin{figure*}[t!]
	\centering
	\begin{tikzpicture}
	\draw[dashed, line width=0.5mm] (-1,0) -- (1,0);
	\node at (-1,-0.3) {\({\scriptstyle \alpha_m}\)}; \node at (1,-0.3) {\({\scriptstyle \alpha_{m+2}}\)};
	\draw[black,fill=black] (1,0) circle (.5ex) ;
	\draw[black,fill=black] (-1,0) circle (.5ex) ;
	\draw[rotate=60, line width=0.5mm] (0,0) -- (1,0); \node at (0.5,1.2) {\({\scriptstyle \mu^{(a)}_{r+1}}\)};
	\draw[black,fill=black][rotate=60] (1,0) circle (.5ex) ;
	\draw[rotate=-60, line width=0.5mm] (0,0) -- (1,0); \node at (0.5,-1.2) {\({\scriptstyle \mu^{(a-1)}_{r+1}}\)};
	\draw[black,fill=black][rotate=-60] (1,0) circle (.5ex) ;
	\draw[rotate=120, line width=0.5mm] (0,0) -- (1,0);
	\draw[black,fill=black][rotate=120] (1,0) circle (.5ex) ; \node at (-0.5,1.2) {\({\scriptstyle \mu^{(a)}_{r}}\)};
	\draw[rotate=-120, line width=0.5mm] (0,0) -- (1,0);
	\draw[black,fill=black][rotate=-120] (1,0) circle (.5ex) ; \node at (-0.5,-1.2) {\({\scriptstyle \mu^{(a-1)}_{r}}\)};
	\end{tikzpicture}
	\caption{Hexagon.}
	\label{fig:hexagon}
\end{figure*}

The most important question to be addressed is how these weights can be determined. Given a local term in the Hamiltonian \({\cal H}_{a,m}\), its local eigenvectors are, obviously, linear combinations of the local configurations~\ref{fig:hexagon}; and the latter are the building blocks for the eigen-configurations \({\cal C}\). In practical terms, we would like to use the local eigenvectors of \({\cal H}_{a,m}\) to define the statistical weights  \({\cal W}(\hexagon_i)\). This idea, on the other hand, defines a rather tautological program, since it obviously demands the solution of the problem we originally want to solve. Therefore, the functional forms \({\cal W}(\hexagon_i)\) cannot be determined now, so we can alternatively try to impose constraints on these objects and see what their solutions can teach us about the quantum problem. This is what we want to do in the remainder of the current work.

It is worth repeating the points above. We want to understand the dynamics of these local configurations parametrized by the triads \(Y^{(2a)} X^{(2a)} Y^{(2a-1)}\) or \(Y^{(2a+1)} X^{(2a+1)} Y^{(2a)}\). Ultimately, we would like to assign Boltzmann weights to the hexagons~\ref{fig:hexagon} and insert them back into the partition function expression; unfortunately, there is no (known) natural way to assign statistical weights to each hexagon configuration. Throughout the remainder of this work we try to find functional relations for the weights imposed by integrability.

Evidently, the associated classical problem is defined by a partition function of the form
\be 
\label{eq:partfunc}
\boxed{
	Z_{cl} = \sum_{n_1, n_2, \dots}\prod_{\hexagon} {\cal W}(\hexagon_i)^{n_i}}\; ,
\ee
where \({\cal W}(\hexagon)\) are local weights for the quantum problem, and that are interpreted as Boltzmann weights for each local classical  configurations. We discuss this problem carefully next section. Hopefully, a better understanding of the classical system will unveil details of the quantum Hamiltonian~(\ref{eq:Hamiltonian}).

\subsection{Local configurations}
\label{sec:LocalConf}

In this section we want to find the local configurations that appear in the quantum problem, and using these allowed states we define the associated classical statistical physics problem we mentioned before. Let us start with the local hexagon configuration represented by figure~\ref{fig:hexagon} with corresponding local state
\be 
|\hexagon_{a, m}\rangle 
= \left|\begin{smallmatrix} \mu^{(a)}_{r} & \mu^{(a)}_{r+1} \\ \alpha_m & \alpha_{m+2} \\ \mu^{(a-1)}_{r} & \mu^{(a-1)}_{r+1} \end{smallmatrix}\right\rangle
\quad \mu, \alpha\equiv \circ, \bullet \; .
\ee
Na\"{i}vely, there are \(2^6=64\) such states, but since we are ultimately interested in the plane partition growth, there are constraints which send most of these configurations to zero. Let us start seeing how this reduction works.

Remember that there are two characteristic lattice distances: \({\rm d}_Y=1\), the lattice distance in the \(Y\)-spin chains, and \({\rm d}_X=2\) that is the lattice distance in the \(X\)-spin chain. 
Additionally, observe that the distance between a given site in the \(X\)-chain and the two nearest sites 
in the \(Y\)-spin chain is also \({\rm d}_Y=1\). 

In the 0-box configuration \(|{\bm{\emptyset}}\rangle\), there is no nearest-neighbor pair of sites 
(those with \({\rm d}_Y=1\)) occupied simultaneously by particles. In other words, given two 
sites with distance \(=1\), the configurations \( \left| \begin{smallmatrix} \bullet & \bullet \end{smallmatrix} \right\rangle \) and \( \left| \begin{smallmatrix} \bullet \\ \bullet \end{smallmatrix} \right\rangle \) do not 
appear anywhere in the 0-box state. Using this \emph{distance embargo}, we can get rid of many local hexagon 
states in the lattice. For example, local configurations in the two sets \(\left\{\left|\begin{smallmatrix} \bullet & \bullet \\ \ast & \ast \\ \ast & \ast \end{smallmatrix}\right\rangle\right\} \) and  \(\left\{\left|\begin{smallmatrix} \ast & \ast \\ \ast & \ast \\ \bullet & \bullet\end{smallmatrix}\right\rangle\right\} \) are absent in \(|{\bm{\emptyset}}\rangle\). Similarly, states of the type \(\left\{\left|\begin{smallmatrix} \bullet & \ast \\ \bullet & \ast \\ \ast & \ast \end{smallmatrix}\right\rangle\right\} \), \(\left\{\left|\begin{smallmatrix} \ast & \bullet \\ \ast & \bullet \\ \ast & \ast \end{smallmatrix}\right\rangle\right\} \), \(\left\{\left|\begin{smallmatrix} \ast & \ast\\ \bullet & \ast \\ \bullet & \ast \end{smallmatrix}\right\rangle\right\} \) and \(\left\{\left|\begin{smallmatrix} \ast & \ast\\  \ast & \bullet \\ \ast & \bullet \end{smallmatrix}\right\rangle\right\} \) are also forbidden. Finally, in the set \(\left\{\left|\begin{smallmatrix} \ast & \ast \\ \bullet & \bullet \\  \ast & \ast \end{smallmatrix}\right\rangle\right\} \), only the state \(\left|\begin{smallmatrix} \circ & \circ \\ \bullet & \bullet \\  \circ & \circ \end{smallmatrix}\right\rangle \) is allowed. 

\begin{figure}[t!]
	\centering
	\begin{tikzpicture}
	%below
	\draw[dashed] (-4,1.5) -- (-2,1.5); \draw[black,fill=black] (-4,1.5) circle (.5ex); \draw[black,fill=black] (-2,1.5) circle (.5ex);
	\draw[ line width=0.5mm] (-4.5,0) -- (-3.5,1); \draw[black,fill=black] (-4.5,0) circle (.5ex) ; \draw[black,fill=black] (-3.5,1) circle (.5ex) ;
	\draw[ line width=0.5mm] (-4.5,1) -- (-3.5,0); \draw[black,fill=black] (-4.5,1) circle (.5ex) ; \draw[black,fill=black] (-3.5,0) circle (.5ex) ;
	%above
	\draw[dashed] (-5,0.5) -- (-3,0.5); \draw[black,fill=black] (-5,0.5) circle (.5ex); \draw[black,fill=black] (-3,0.5) circle (.5ex);
	\draw[ line width=0.5mm] (-3.5,1) -- (-2.5,2); \draw[black,fill=black] (-3.5,1) circle (.5ex) ; \draw[black,fill=black] (-2.5,2) circle (.5ex) ;
	\draw[ line width=0.5mm] (-3.5,2) -- (-2.5,1); \draw[black,fill=black] (-3.5,2) circle (.5ex) ; \draw[black,fill=black] (-2.5,1) circle (.5ex) ;
	%boundary terms
	\end{tikzpicture}
	\caption{Diagrammatic representation of the tensor product state.}
	\label{fig:tensorstate}
\end{figure}
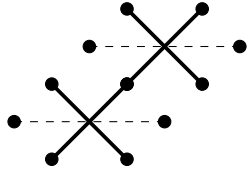

Observe that this restriction also forbids certain tensor product states. Making a list of all forbidden configurations is not a simple task anymore (and it is not necessary either), so let us understand the general idea with an example. Consider the product of two hexagons
\be 
|\hexagon_{a, m}\rangle \otimes |\hexagon_{a+1, m+2}\rangle
= \left|\begin{smallmatrix} \mu^{(a)}_{r} & \mu^{(a)}_{r+1} \\ \alpha_m & \alpha_{m+2} \\ \mu^{(a-1)}_{r} & \mu^{(a-1)}_{r+1} \end{smallmatrix}\right\rangle
\otimes 
\left|\begin{smallmatrix} \mu^{(a+1)}_{r+1} & \mu^{(a+1)}_{r+2} \\ \alpha_{m+1} & \alpha_{m+3} \\ \mu^{(a)}_{r+1} 
	& \mu^{(a)}_{r+2}
\end{smallmatrix}\right\rangle \; ,
\ee
that we represent diagrammatically as in figure~\ref{fig:tensorstate}. It is easy to see that states in the set \(\left\{\left|\begin{smallmatrix} \bullet & \ast \\ \ast & \ast \\ \ast & \ast \end{smallmatrix}\right\rangle \otimes \left|\begin{smallmatrix} \ast & \ast \\ \bullet & \ast \\ \ast & \ast \end{smallmatrix}\right\rangle\right\} \) are not allowed since \(distance (\mu^{(a)}_r, \alpha_{m+1}) = {\rm d}_Y=1\). 

But now we need to ask ourselves if the absence of these states is a general feature or an artifact of the ground state. In order to answer this point, we need to understand some properties of the system we are interested in. First, observe that the Hamiltonian~(\ref{eq:Hamiltonian}) is written as a sum of local operators \(H_{a,m}\) acting independently in each hexagon cell. There are just two states that give nontrivial results under its action, namely 
\(\left\{\left|\begin{smallmatrix} \bullet & \circ\\  \circ & \bullet \\ \bullet & \circ \end{smallmatrix}\right\rangle\right\}\) and \(\left\{\left|\begin{smallmatrix} \circ & \bullet \\ \bullet & \circ\\ \circ & \bullet \end{smallmatrix}\right\rangle\right\}\). As a matter of fact, the kinetic terms of \(H_{a,m}\) exchange these two states whilst the potential term acts diagonally with eigenvalues \(0\) and \(1\). Therefore, if we consider the 
zero box state \(|{\bm \emptyset}\rangle\) as the ground state, the local states with nearest-neighbors are not generated in configurations corresponding to a generic plane partition. As an important remark, observe that the embargo above is a feature of the expansion around the zero box ground state. 

From the restrictions above, the classification of the nontrivial hexagon states is just a simple combinatorial problem. There is, naturally, just \(1\) empty hexagon state
\bse
\be 
\mathfrak{R}_0 = \left\{ \left|[0]\right\rangle\equiv 
\left|\begin{smallmatrix}
	\circ & \circ \\
	\circ & \circ \\
	\circ & \circ
\end{smallmatrix}\right\rangle
\right\}\; ,
\ee
and we define the notation \(| [n]_{abc}\rangle\), where \(n\) denotes the number of particles, and \(a, b,c\) their positions with relation to the defining configuration \( \left|\begin{smallmatrix} 5 & 6 \\ 3 & 4 \\ 1 & 2 \end{smallmatrix}\right\rangle\). This notation will be clearer in the next example. 

There are \(6=\left( \begin{smallmatrix} 6 \\ 1 \end{smallmatrix} \right)\) states with \(1\) particle, and all of them are nontrivial, namely
\be 
\mathfrak{R}_1 = \left\{
\begin{matrix}
	\left|[1]_1\right\rangle\equiv \left|\begin{smallmatrix} \circ & \circ \\ \circ & \circ \\ \bullet & \circ
	\end{smallmatrix}\right\rangle\; ; &
	\left|[1]_2\right\rangle\equiv \left| \begin{smallmatrix} \circ & \circ \\ \circ & \circ\\ \circ & \bullet 
	\end{smallmatrix}\right\rangle \; ;  &
	\left|[1]_3\right\rangle\equiv \left| \begin{smallmatrix} \circ & \circ \\ \bullet & \circ\\ \circ & \circ 
	\end{smallmatrix}\right\rangle\\[0.2cm]
	\left|[1]_4\right\rangle\equiv \left| \begin{smallmatrix} \circ & \circ \\ \circ & \bullet \\ \circ & \circ
	\end{smallmatrix}\right\rangle\; ; &
	\left|[1]_5\right\rangle\equiv \left| \begin{smallmatrix} \bullet & \circ \\ \circ & \circ \\ \circ & \circ
	\end{smallmatrix}\right\rangle\; ; &
	\left|[1]_6\right\rangle\equiv \left| \begin{smallmatrix} \circ & \bullet \\ \circ & \circ \\ \circ & \circ
	\end{smallmatrix}\right\rangle
\end{matrix}
\right\}\; .
\ee

Furthermore, we have \(15=\left( \begin{smallmatrix} 6 \\ 2 \end{smallmatrix} \right)\) states with \(2\) particles, 
and \(9\) of them are nontrivial, namely
\be 
\mathfrak{R}_2 =\left\{
\begin{matrix}
	\left|[2]_{14}\right\rangle\equiv \left| \begin{smallmatrix} \circ & \circ \\ \circ & \bullet \\ \bullet & \circ
	\end{smallmatrix}\right\rangle\; ;&
	\left|[2]_{15}\right\rangle\equiv \left| \begin{smallmatrix} \bullet & \circ \\ \circ & \circ \\ \bullet & \circ
	\end{smallmatrix}\right\rangle\; ;& 
	\left|[2]_{16}\right\rangle\equiv \left| \begin{smallmatrix} \circ & \bullet \\ \circ & \circ \\ \bullet & \circ 
	\end{smallmatrix}\right\rangle\\[0.2cm]
	\left|[2]_{23}\right\rangle\equiv \left| \begin{smallmatrix} \circ & \circ \\ \bullet & \circ\\ \circ & \bullet 
	\end{smallmatrix}\right\rangle\; ;& 
	\left|[2]_{25}\right\rangle\equiv \left| \begin{smallmatrix} \bullet & \circ \\ \circ & \circ \\ \circ & \bullet 
	\end{smallmatrix}\right\rangle\; ;& 
	\left|[2]_{26}\right\rangle\equiv \left| \begin{smallmatrix} \circ & \bullet \\ \circ & \circ \\ \circ & \bullet
	\end{smallmatrix}\right\rangle \\[0.2cm]	
	\left|[2]_{34}\right\rangle\equiv \left| \begin{smallmatrix} \circ & \circ \\ \bullet & \bullet \\ \circ & \circ
	\end{smallmatrix}\right\rangle\; ; &
	\left|[2]_{36}\right\rangle\equiv \left| \begin{smallmatrix} \circ & \bullet \\ \bullet & \circ\\ \circ & \circ 
	\end{smallmatrix}\right\rangle\; ; & \\[0.2cm]
	\left|[2]_{45}\right\rangle\equiv \left|\begin{smallmatrix} \bullet & \circ \\ \circ & \bullet \\ \circ & \circ 
	\end{smallmatrix}\right\rangle \; \phantom{;} && 
\end{matrix}\right\}
\ee
There are \(20=\left( \begin{smallmatrix} 6 \\ 3 \end{smallmatrix} \right)\) with \(3\) particles, and \(2\) of them are relevant to our discussion, namely
\be 
\mathfrak{R}_3 =\left\{ 
\left|[3]_{236}\right\rangle\equiv 
\left|\begin{smallmatrix}
	\circ & \bullet \\
	\bullet & \circ \\
	\circ & \bullet
\end{smallmatrix}\right\rangle\; ; \
\left|[3]_{145}\right\rangle\equiv \left| 
\begin{smallmatrix}
	\bullet & \circ \\
	\circ & \bullet\\
	\bullet & \circ 
\end{smallmatrix}\right\rangle
\right\}\; ,
\ee
and all other states are null. 
\ese

Finally, there is just \(1\) state with \(6\) particles, \(6=\left( \begin{smallmatrix} 6 \\ 5 \end{smallmatrix} \right)\) 
states with \(5\) particles, and \(15=\left( \begin{smallmatrix} 6 \\ 4 \end{smallmatrix} \right)\) states with \(4\) particles. All these states are null since there are necessarily two particles occupying nearest-neighbor sites. 

Putting all these facts together, we immediately see that there are \(18\) nontrivial states. We should remark that, in principle, there is absolutely nothing wrong with the null states above, they simply do not appear in the system we are interested in. The set of physical hexagon states is
\be 
\label{eq:States}
\mathfrak{R}=\mathfrak{R}_0\cup \mathfrak{R}_1\cup \mathfrak{R}_2\cup \mathfrak{R}_3\; .
\ee

\subsection{Macroscopic configurations \& Classical statistical system}
\label{sec:classicalsystem}

In this section we study the rules to consistently combine local states to create macroscopic configurations \({\cal C}\). In fact, we have already considered the \(X\) and \(Y\)-rows parametrization, thus we know how to do connect the hexagons consistently; but now we want to put in words the rules we have been using. From the parametrization of figure~\ref{fig:hexagon}, the consistent configurations are obtained with the following rules:
\begin{shaded}{\bf Rules}
\begin{itemize}
	\item Consider two hexagons \(A\) and \(B\) in the triad \(Y^{(a-1)}X^{(a)}Y^{(a)}\) and their generic parameterizations
	\((a,m)\) and \((a,m')\) respectively. They are connected horizontally if one of the following conditions is satisfied:
	\bse
	\begin{align}
	\texttt{Rule \#1}:&  \quad \alpha^A_{m+2}= \alpha^B_{m'} \qquad \text{for} \ m+2=m' \label{eq:rule1}\\
	\texttt{Rule \#2}:&  \quad \alpha^A_{m}= \alpha^B_{m'+2} \qquad \text{for} \ m=m'+2 \label{eq:rule2} \; .
	\end{align}
	\item Consider a hexagon \(A\) in the triad \(Y^{(a-1)}X^{(a)}Y^{(a)}\) parametrized by \((a,m)\), 
	and another hexagon \(B\) in the triad \(Y^{(a)}X^{(a+1)}Y^{(a+1)}\) parametrized by \((a+1,m')\). 
	They are connected diagonally if one of the following conditions is satisfied:
	\begin{align}
	\texttt{Rule \#3}:&  \quad \mu^{A (a)}_{m+3/2}= \mu^{B (a)}_{m'+1/2} \qquad \text{for} \ m+1=m' \label{eq:rule3} \\
	\texttt{Rule \#4}:& \quad \mu^{A (a)}_{m+1/2}= \mu^{B (a)}_{m'+3/2} \qquad \text{for} \ m=m'+1 \label{eq:rule4}\; .
	\end{align}
	\ese
\end{itemize}	 
\end{shaded}

Using the rules above we can build the configurations \({\cal C}\), and counting all possible macroscopic states \({\cal C}\) is the essence of the classical statistical problem we want to study. It should also be clear now that the vast majority of the consistent configurations \({\cal C}\) do not define plane partitions states. In particular, we do not impose any restriction on particles occupying sites with distance \(=1\). Evidently, at some point we need to impose this extra layer of complexity, but now we are interested in this simplified classical problem. 

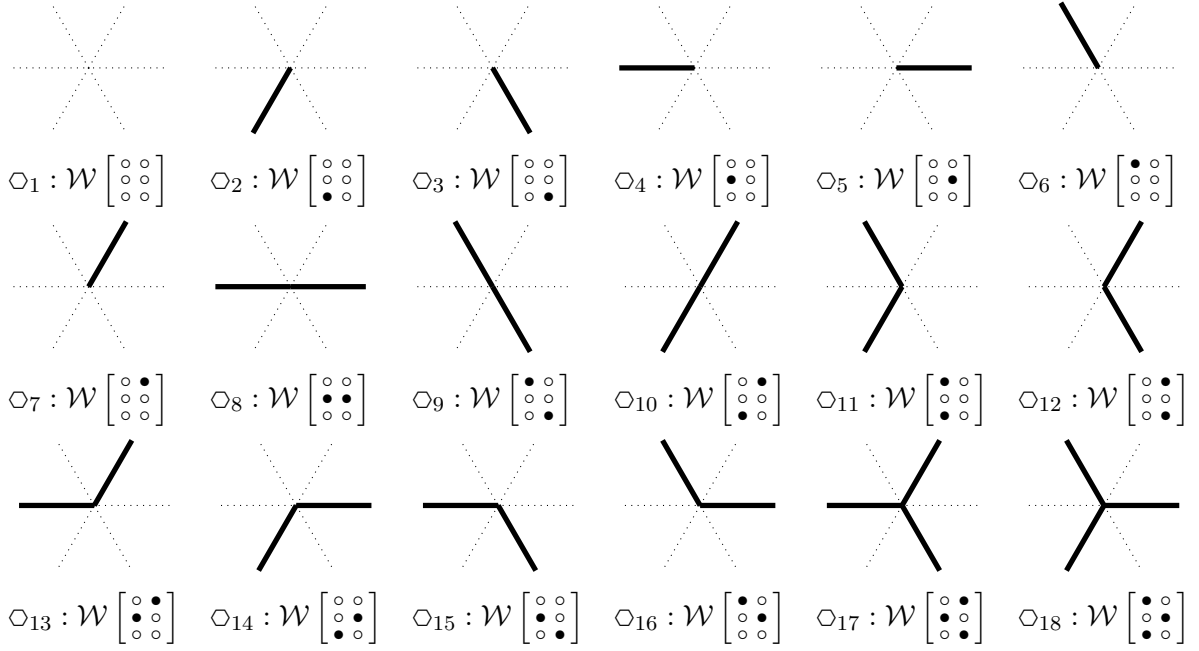
\begin{figure}[h!]
	\centering
	\begin{subfigure}[t]{2.3cm}
		\begin{tikzpicture}
		\draw[rotate=60, dotted] (0,0) -- (1,0); \draw[rotate=-60, dotted] (0,0) -- (1,0);
		\draw[rotate=120, dotted] (0,0) -- (1,0); \draw[rotate=-120, dotted] (0,0) -- (1,0);
		\draw[dotted] (1,0) -- (0,0); \draw[dotted] (-1,0) -- (0,0);
		\node at (0,-1.5) {\(\hexagon_1: {\cal W}\left[ \begin{smallmatrix} \circ & \circ \\  \circ & \circ 
			\\ \circ & \circ \end{smallmatrix} \right]\)};
		\end{tikzpicture}
	\end{subfigure}
	~
	\begin{subfigure}[t]{2.3cm}
		\begin{tikzpicture}
		\draw[rotate=60, dotted] (0,0) -- (1,0); \draw[rotate=-60, dotted] (0,0) -- (1,0);
		\draw[rotate=120, dotted] (0,0) -- (1,0); \draw[rotate=-120, line width=0.7mm] (0,0) -- (1,0);
		\draw[dotted] (1,0) -- (0,0); \draw[dotted] (-1,0) -- (0,0);
		\node at (0,-1.5) {\(\hexagon_2: {\cal W}\left[ \begin{smallmatrix} \circ & \circ \\  \circ & \circ
			\\ \bullet & \circ \end{smallmatrix} \right]\)};
		\end{tikzpicture}
	\end{subfigure}
	~
	\begin{subfigure}[t]{2.3cm}
		\begin{tikzpicture}
		\draw[rotate=60, dotted] (0,0) -- (1,0); \draw[rotate=-60, line width=0.7mm] (0,0) -- (1,0);
		\draw[rotate=120, dotted] (0,0) -- (1,0); \draw[rotate=-120, dotted] (0,0) -- (1,0);
		\draw[dotted] (1,0) -- (0,0); \draw[dotted] (-1,0) -- (0,0);
		\node at (0,-1.5) {\(\hexagon_3:{\cal W}\left[ \begin{smallmatrix} \circ & \circ \\  \circ & \circ
			\\ \circ & \bullet \end{smallmatrix} \right]\)};
		\end{tikzpicture}
	\end{subfigure}
	~	
	\begin{subfigure}[t]{2.3cm}
		\begin{tikzpicture}
		\draw[rotate=60, dotted] (0,0) -- (1,0); \draw[rotate=-60, dotted] (0,0) -- (1,0);
		\draw[rotate=120, dotted] (0,0) -- (1,0); \draw[rotate=-120, dotted] (0,0) -- (1,0);
		\draw[dotted] (1,0) -- (0,0); \draw[line width=0.7mm] (-1,0) -- (0,0);
		\node at (0,-1.5) {\(\hexagon_4: {\cal W}\left[ \begin{smallmatrix} \circ & \circ \\  \bullet & \circ
			\\ \circ & \circ \end{smallmatrix} \right]\)};
		\end{tikzpicture}
	\end{subfigure}
	~	
	\begin{subfigure}[t]{2.3cm}
		\begin{tikzpicture}
		\draw[rotate=60, dotted] (0,0) -- (1,0); \draw[rotate=-60, dotted] (0,0) -- (1,0);
		\draw[rotate=120, dotted] (0,0) -- (1,0); \draw[rotate=-120, dotted] (0,0) -- (1,0);
		\draw[line width=0.7mm] (1,0) -- (0,0); \draw[dotted] (-1,0) -- (0,0);
		\node at (0,-1.5) {\(\hexagon_5: {\cal W}\left[ \begin{smallmatrix} \circ & \circ \\  \circ & \bullet
			\\ \circ & \circ \end{smallmatrix} \right]\)};
		\end{tikzpicture}
	\end{subfigure}
	~	
	\begin{subfigure}[t]{2.3cm}
		\begin{tikzpicture}
		\draw[rotate=60, dotted] (0,0) -- (1,0); \draw[rotate=-60, dotted] (0,0) -- (1,0);
		\draw[rotate=120, line width=0.7mm] (0,0) -- (1,0); \draw[rotate=-120, dotted] (0,0) -- (1,0);
		\draw[dotted] (1,0) -- (0,0); \draw[dotted] (-1,0) -- (0,0);
		\node at (0,-1.5) {\(\hexagon_6 : {\cal W}\left[ \begin{smallmatrix} \bullet & \circ \\  \circ & \circ
			\\ \circ & \circ \end{smallmatrix} \right]\)};
		\end{tikzpicture}
	\end{subfigure}
	~	
	\begin{subfigure}[t]{2.3cm}
		\begin{tikzpicture}
		\draw[rotate=60, line width=0.7mm] (0,0) -- (1,0); \draw[rotate=-60, dotted] (0,0) -- (1,0);
		\draw[rotate=120, dotted] (0,0) -- (1,0); \draw[rotate=-120, dotted] (0,0) -- (1,0);
		\draw[dotted] (1,0) -- (0,0); \draw[dotted] (-1,0) -- (0,0);
		\node at (0,-1.5) {\(\hexagon_7:{\cal W}\left[ \begin{smallmatrix} \circ & \bullet \\  \circ & \circ
			\\ \circ & \circ \end{smallmatrix} \right]\)};
		\end{tikzpicture}
	\end{subfigure}
	~	
	\begin{subfigure}[t]{2.3cm}
		\begin{tikzpicture}
		\draw[rotate=60, dotted] (0,0) -- (1,0); \draw[rotate=-60, dotted] (0,0) -- (1,0);
		\draw[rotate=120, dotted] (0,0) -- (1,0); \draw[rotate=-120, dotted] (0,0) -- (1,0);
		\draw[line width=0.7mm] (1,0) -- (0,0); \draw[line width=0.7mm] (-1,0) -- (0,0);
		\node at (0,-1.5) {\(\hexagon_8:{\cal W}\left[ \begin{smallmatrix} \circ & \circ \\  \bullet & \bullet 
			\\ \circ & \circ \end{smallmatrix} \right]\)};
		\end{tikzpicture}
	\end{subfigure}
	~
	\begin{subfigure}[t]{2.3cm}
		\begin{tikzpicture}
		\draw[rotate=60, dotted] (0,0) -- (1,0); \draw[rotate=-60, line width=0.7mm] (0,0) -- (1,0);
		\draw[rotate=120, line width=0.7mm] (0,0) -- (1,0); \draw[rotate=-120, dotted] (0,0) -- (1,0);
		\draw[dotted] (1,0) -- (0,0); \draw[dotted] (-1,0) -- (0,0);
		\node at (0,-1.5) {\(\hexagon_9:{\cal W}\left[ \begin{smallmatrix} \bullet & \circ \\  \circ & \circ 
			\\ \circ & \bullet \end{smallmatrix} \right]\)};
		\end{tikzpicture}
	\end{subfigure}
	~
	\begin{subfigure}[t]{2.3cm}
		\begin{tikzpicture}
		\draw[rotate=60, line width=0.7mm] (0,0) -- (1,0); \draw[rotate=-60, dotted] (0,0) -- (1,0);
		\draw[rotate=120, dotted] (0,0) -- (1,0); \draw[rotate=-120, line width=0.7mm] (0,0) -- (1,0);
		\draw[dotted] (1,0) -- (0,0); \draw[dotted] (-1,0) -- (0,0);
		\node at (0,-1.5) {\(\hexagon_{10}:{\cal W}\left[ \begin{smallmatrix} \circ & \bullet \\  \circ & \circ 
			\\ \bullet & \circ \end{smallmatrix} \right]\)};
		\end{tikzpicture}
	\end{subfigure}
	~
	\begin{subfigure}[t]{2.3cm}
		\begin{tikzpicture}
		\draw[rotate=60, dotted] (0,0) -- (1,0); \draw[rotate=-60, dotted] (0,0) -- (1,0);
		\draw[rotate=120, line width=0.7mm] (0,0) -- (1,0); \draw[rotate=-120, line width=0.7mm] (0,0) -- (1,0);
		\draw[dotted] (1,0) -- (0,0); \draw[dotted] (-1,0) -- (0,0);
		\node at (0,-1.5) {\(\hexagon_{11}:{\cal W}\left[ \begin{smallmatrix} \bullet & \circ \\  \circ & \circ 
			\\ \bullet & \circ \end{smallmatrix} \right]\)};
		\end{tikzpicture}
	\end{subfigure}
	~
	\begin{subfigure}[t]{2.3cm}
		\begin{tikzpicture}
		\draw[rotate=60, line width=0.7mm] (0,0) -- (1,0); \draw[rotate=-60, line width=0.7mm] (0,0) -- (1,0);
		\draw[rotate=120, dotted] (0,0) -- (1,0); \draw[rotate=-120, dotted] (0,0) -- (1,0);
		\draw[dotted] (1,0) -- (0,0); \draw[dotted] (-1,0) -- (0,0);
		\node at (0,-1.5) {\(\hexagon_{12}:{\cal W}\left[ \begin{smallmatrix} \circ & \bullet \\  \circ & \circ 
			\\ \circ & \bullet \end{smallmatrix} \right]\)};
		\end{tikzpicture}
	\end{subfigure}
	~
	\begin{subfigure}[t]{2.3cm}
		\begin{tikzpicture}
		\draw[rotate=60, line width=0.7mm] (0,0) -- (1,0); \draw[rotate=-60, dotted] (0,0) -- (1,0);
		\draw[rotate=120, dotted] (0,0) -- (1,0); \draw[rotate=-120, dotted] (0,0) -- (1,0);
		\draw[dotted] (1,0) -- (0,0); \draw[line width=0.7mm] (-1,0) -- (0,0);
		\node at (0,-1.5) {\(\hexagon_{13}:{\cal W}\left[ \begin{smallmatrix} \circ & \bullet \\  \bullet & \circ \\ \circ & \circ \end{smallmatrix} \right]\)};
		\end{tikzpicture}
	\end{subfigure}
	~
	\begin{subfigure}[t]{2.3cm}
		\begin{tikzpicture}
		\draw[rotate=60, dotted] (0,0) -- (1,0); \draw[rotate=-60, dotted] (0,0) -- (1,0);
		\draw[rotate=120, dotted] (0,0) -- (1,0); \draw[rotate=-120, line width=0.7mm] (0,0) -- (1,0);
		\draw[line width=0.7mm] (1,0) -- (0,0); \draw[dotted] (-1,0) -- (0,0);
		\node at (0,-1.5) {\(\hexagon_{14}:{\cal W}\left[ \begin{smallmatrix} \circ & \circ \\  \circ & \bullet \\ \bullet & \circ \end{smallmatrix} \right]\)};
		\end{tikzpicture}
	\end{subfigure}
	~
	\begin{subfigure}[t]{2.3cm}
		\begin{tikzpicture}
		\draw[rotate=60, dotted] (0,0) -- (1,0); \draw[rotate=-60, line width=0.7mm] (0,0) -- (1,0);
		\draw[rotate=120, dotted] (0,0) -- (1,0); \draw[rotate=-120, dotted] (0,0) -- (1,0);
		\draw[dotted] (1,0) -- (0,0); \draw[line width=0.7mm] (-1,0) -- (0,0);
		\node at (0,-1.5) {\(\hexagon_{15}:{\cal W}\left[ \begin{smallmatrix} \circ & \circ \\  \bullet & \circ \\ \circ & \bullet \end{smallmatrix} \right]\)};
		\end{tikzpicture}
	\end{subfigure}
	~
	\begin{subfigure}[t]{2.3cm}
		\begin{tikzpicture}
		\draw[rotate=60, dotted] (0,0) -- (1,0); \draw[rotate=-60,dotted] (0,0) -- (1,0);
		\draw[rotate=120, line width=0.7mm] (0,0) -- (1,0); \draw[rotate=-120, dotted] (0,0) -- (1,0);
		\draw[line width=0.7mm] (1,0) -- (0,0); \draw[dotted] (-1,0) -- (0,0);
		\node at (0,-1.5) {\(\hexagon_{16}:{\cal W}\left[ \begin{smallmatrix} \bullet & \circ \\  \circ & \bullet \\ \circ & \circ \end{smallmatrix} \right]\)};
		\end{tikzpicture}
	\end{subfigure}
	~
	\begin{subfigure}[t]{2.3cm}
		\begin{tikzpicture}
		\draw[rotate=60, line width=0.7mm] (0,0) -- (1,0); \draw[rotate=-60, line width=0.7mm] (0,0) -- (1,0);
		\draw[rotate=120, dotted] (0,0) -- (1,0); \draw[rotate=-120, dotted] (0,0) -- (1,0);
		\draw[dotted] (1,0) -- (0,0); \draw[line width=0.7mm] (-1,0) -- (0,0);
		\node at (0,-1.5) {\(\hexagon_{17}:{\cal W}\left[ \begin{smallmatrix} \circ & \bullet \\  \bullet & \circ \\ \circ & \bullet \end{smallmatrix} \right]\)};
		\end{tikzpicture}
	\end{subfigure}
	~
	\begin{subfigure}[t]{2.3cm}
		\begin{tikzpicture}
		\draw[rotate=60, dotted] (0,0) -- (1,0); \draw[rotate=-60,dotted] (0,0) -- (1,0);
		\draw[rotate=120, line width=0.7mm] (0,0) -- (1,0); \draw[rotate=-120, line width=0.7mm] (0,0) -- (1,0);
		\draw[line width=0.7mm] (1,0) -- (0,0); \draw[dotted] (-1,0) -- (0,0);
		\node at (0,-1.5) {\(\hexagon_{18}:{\cal W}\left[ \begin{smallmatrix} \bullet & \circ \\  \circ & \bullet \\ \bullet & \circ \end{smallmatrix} \right]\)};
		\end{tikzpicture}
	\end{subfigure}
	\caption{18 hexagon vertex configurations with their respective weights.}
	\label{fig:18hvertex}
\end{figure}

Given a macroscopic configuration \({\cal C}\), its Boltzmann weight is defined by \({\cal W}({\cal C})\), and since it is a combination of the \(18\) hexagon states we defined before, we write 
\be 
{\cal W}({\cal C})= \prod_{i=1}^{18} {\cal W}({\cal C},\hexagon_i)
\ee
where \({\cal C}\) is defined by a string of integers \(\vec{n}=\{n_1, n_2, \dots, n_{18}\}\) which determines how many times the hexagon \(\hexagon_i\) appears in \({\cal C}\). Moreover, if we write 
\({\cal W}(1,\hexagon_i)\equiv {\cal W}(\hexagon_i) = \exp(-\beta\varepsilon_i)\), the classical partition function~(\ref{eq:partfunc}) reads
\be 
\label{eq:partfunc2}
Z_{cl} = \sum_{\vec{n}} \exp\left(-\beta \sum_{i=1}^{18}n_i\varepsilon_i \right) .
\ee
We can denote diagrammatically the hexagon configurations as in figure~\ref{fig:18hvertex}. In section~\ref{sec:transfer} we show that integrability of this system cannot be settled using a na\"ive row-to-row formalism, but with small modification of the rules~(\ref{eq:rule1} -- \ref{eq:rule4}), we can make contact between this model and a descendant of the 6-vertex model. We need to remark that albeit 
the similarities, the statistical system above does not define an ordinary vertex model since the thick lines can finish within the lattice configuration \({\cal C}\), and not just at the boundaries.

\subsection{Lax operator}

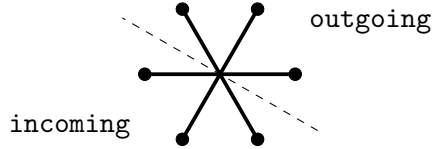
\begin{figure*}[h!]
	\centering
	\begin{tikzpicture}
	\draw[line width=0.5mm] (-1,0) -- (1,0);
	\draw[black,fill=black] (1,0) circle (.5ex) ;
	\draw[black,fill=black] (-1,0) circle (.5ex) ;
	\draw[rotate=60, line width=0.5mm] (0,0) -- (1,0);
	\draw[black,fill=black][rotate=60] (1,0) circle (.5ex) ;
	\draw[rotate=-60, line width=0.5mm] (0,0) -- (1,0); 
	\draw[black,fill=black][rotate=-60] (1,0) circle (.5ex) ;
	\draw[rotate=120, line width=0.5mm] (0,0) -- (1,0);
	\draw[black,fill=black][rotate=120] (1,0) circle (.5ex) ;
	\draw[rotate=-120, line width=0.5mm] (0,0) -- (1,0);
	\draw[black,fill=black][rotate=-120] (1,0) circle (.5ex);	
	\draw[dashed][rotate=150]  (-1.5,0) -- (1.5,0);
	\node at (-2,-0.7) {\texttt{incoming}}; \node at (2,0.7) {\texttt{outgoing}};
	\end{tikzpicture}
	\caption{Scattering process}
	\label{fig:scattering}
\end{figure*}

We now decompose each hexagon cell into three incoming and three outgoing particles, as in figure~\ref{fig:scattering}. This decomposition also defines an S-matrix -- that we recklessly call Lax operator. From the S-matrix expressions we build the Boltzmann weights to be the scattering amplitudes of these processes. See~\cite{Gomez:1996az} for an example of this idea in the 6- and 8-vertex models. 

\begin{figure}[b!]
	\centering
	\includegraphics[height=1.2in]{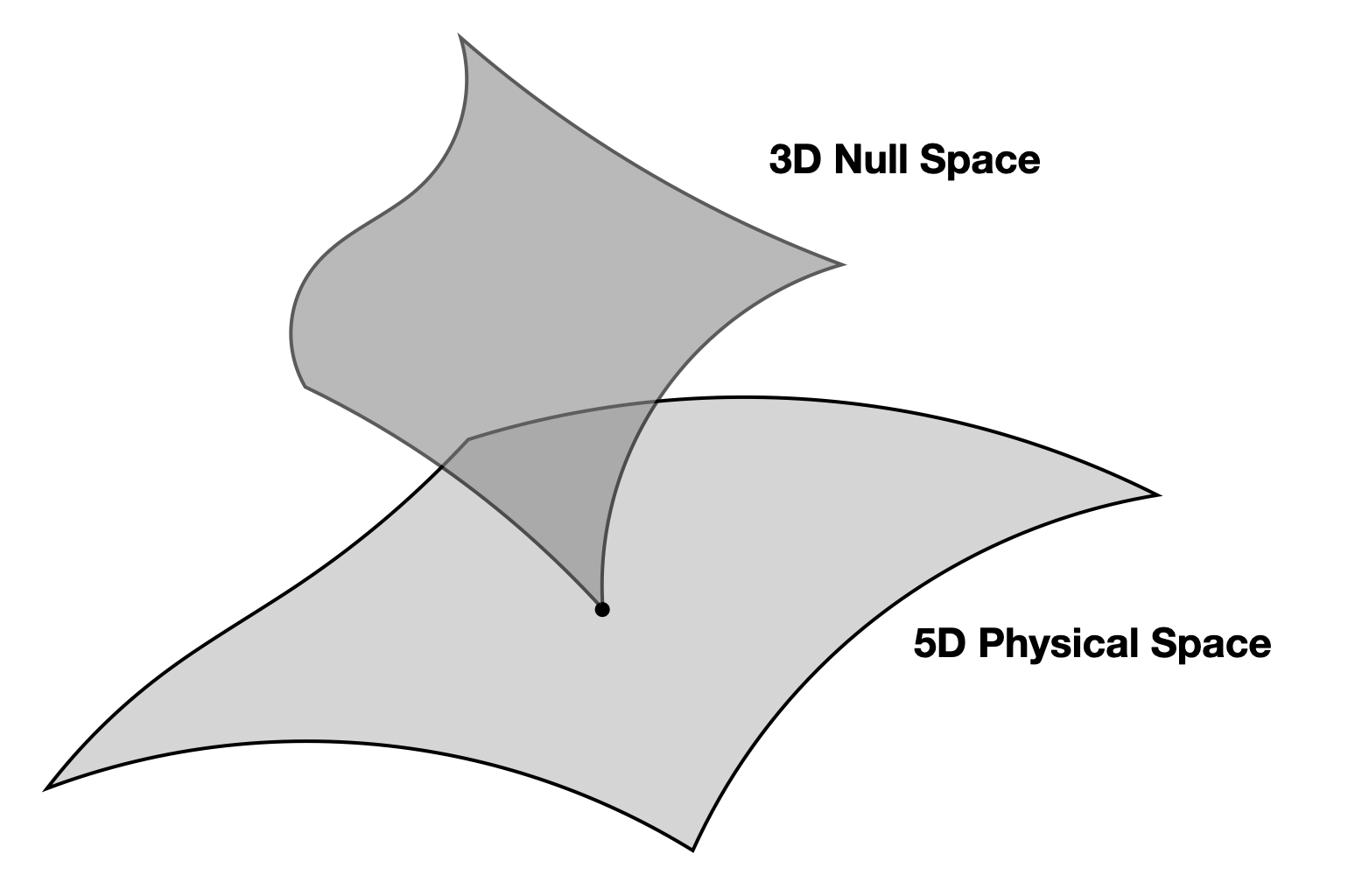}
	\caption{Physical and null spaces.}
	\label{fig:physspace}
\end{figure}

In principle, there \(64\) different processes of the type represented by figure~\ref{fig:scattering} with 8 incoming and 8 outgoing different configurations each. But as we said before, there are several forbidden configurations. 
From the diagrams~\ref{fig:18hvertex}, it is easy to see that there are just \(5\) incoming and \(5\) outgoing 
nontrivial configurations
\be 
\label{eq:phystates}
\begin{split}
	{\tt incoming} = & \{ \left| \begin{smallmatrix} \circ \\ \circ & \circ  \end{smallmatrix}\right\rangle\; , 
	\left|\begin{smallmatrix} \bullet \\ \circ & \circ  \end{smallmatrix}\right\rangle\; , 
	\left|\begin{smallmatrix} \circ \\ \bullet  & \circ  \end{smallmatrix}\right\rangle \}\; , 
	\left|\begin{smallmatrix} \circ \\ \circ & \bullet  \end{smallmatrix}\right\rangle\; ,
	\left|\begin{smallmatrix} \bullet \\ \circ & \bullet  \end{smallmatrix}\right\rangle\}\\
	{\tt outgoing} = & \{ \left| \begin{smallmatrix} \circ & \circ \\ & \circ  \end{smallmatrix}\right\rangle\; , 
	\left|\begin{smallmatrix} \bullet & \circ \\ & \circ  \end{smallmatrix}\right\rangle\; , 
	\left|\begin{smallmatrix} \circ & \bullet \\ & \circ  \end{smallmatrix}\right\rangle \}\; , 
	\left|\begin{smallmatrix} \circ & \circ \\ & \bullet  \end{smallmatrix}\right\rangle\; ,
	\left|\begin{smallmatrix} \bullet & \circ \\ & \bullet  \end{smallmatrix}\right\rangle\}
\end{split}\; .
\ee
In other words, the initial
\(\left|\begin{smallmatrix} \bullet \\ \bullet & \circ  \end{smallmatrix}\right\rangle, 
\left|\begin{smallmatrix} \circ \\ \bullet & \bullet  \end{smallmatrix}\right\rangle
\left|\begin{smallmatrix} \bullet \\ \bullet & \bullet  \end{smallmatrix}\right\rangle\)
and final
\(\left|\begin{smallmatrix} \bullet & \bullet \\ \circ  \end{smallmatrix}\right\rangle, 
\left|\begin{smallmatrix} \bullet & \circ \\ \bullet  \end{smallmatrix}\right\rangle
\left|\begin{smallmatrix} \bullet & \bullet \\ \bullet  \end{smallmatrix}\right\rangle\)
states decouple from the space of physical states, see figure~\ref{fig:physspace}. 

Let us denote the Hilbert space associated to the \(X^{(a)}\)-chains by 
\(\tilde{\mathscr{H}}^{(a)} = \bigotimes_m \tilde{V}_m^{(a)}\) and the one associated to the \(Y^{(a)}\)-chains by \(\mathscr{H}^{(a)} = \bigotimes_r  V_r^{(a)}\), where \(\tilde{V}_m\simeq V_r \simeq \mathbb{C}^2\). The space
\(\tilde{V}\simeq \mathbb{C}^2\) is called the \emph{auxiliary} or \emph{horizontal} space, and \(V\times V \simeq \mathbb{C}^4\) the \emph{quantum} or \emph{vertical} space.

The Lax operator \({\cal L}_{a,m} \) is then defined as
\be 
{\cal L}_{a,m}: \tilde{V}_m^{(a)}  \otimes V^{(a-1)}_{m+1/2}\otimes V^{(a-1)}_{m+3/2} \to  
\tilde{V}_{m+2}^{(a)} \otimes V^{(a)}_{m+1/2}\otimes V^{(a)}_{m+3/2}\; ,
\ee
with action
\be 
\label{eq:laxoperator}
{\cal L}_{a,m} \left|\begin{smallmatrix} \alpha \\ \mu_1 & \mu_2  \end{smallmatrix}\right\rangle = 
\sum_{\beta, \nu_1, \nu_2} 
{\cal W}\left[ 
\begin{smallmatrix} \nu_1 & \nu_2\\  \alpha & \beta \\ \mu_1 & \mu_2 \end{smallmatrix}
\right]
\left|\begin{smallmatrix} \nu_1 & \nu_2\\ & \beta \end{smallmatrix}\right\rangle\; ,
\ee
where \({\cal W}\left[ \begin{smallmatrix} \nu_1 & \nu_2\\ \alpha & \beta \\ \mu_1 & \mu_2 \end{smallmatrix} \right]\) is the 
Boltzmann weight associated to a state parametrized accordingly. Therefore
\begin{flalign} 
	\nn 
	{\cal L}_{a,m} \left|\begin{smallmatrix} \circ \\ \circ & \circ  \end{smallmatrix}\right\rangle =  
	{\cal W}\left[ \begin{smallmatrix} \circ & \circ\\  \circ & \circ \\ \circ & \circ \end{smallmatrix} \right] 
	\left|\begin{smallmatrix} \circ & \circ \\ & \circ \end{smallmatrix}\right\rangle +
	{\cal W}\left[ \begin{smallmatrix} \bullet & \circ\\  \circ & \circ \\ \circ & \circ \end{smallmatrix} \right] 
	\left|\begin{smallmatrix} \bullet & \circ \\ & \circ \end{smallmatrix}\right\rangle 
	+{\cal W}\left[ \begin{smallmatrix} \circ & \bullet\\  \circ & \circ \\ \circ & \circ \end{smallmatrix} \right] 
	\left|\begin{smallmatrix} \circ & \bullet \\ & \circ \end{smallmatrix}\right\rangle +
	{\cal W}\left[ \begin{smallmatrix} \circ & \circ\\  \circ & \bullet \\ \circ & \circ \end{smallmatrix} \right] 
	\left|\begin{smallmatrix} \circ & \circ \\ & \bullet \end{smallmatrix}\right\rangle + 
	{\cal W}\left[ \begin{smallmatrix} \bullet & \circ\\  \circ & \bullet \\ \circ & \circ \end{smallmatrix} \right] 
	\left|\begin{smallmatrix} \bullet & \circ \\ & \bullet \end{smallmatrix}\right\rangle &&
\end{flalign} 
\begin{flalign} 
	\nn
	{\cal L}_{a,m} \left|\begin{smallmatrix} \circ \\ \bullet & \circ  \end{smallmatrix}\right\rangle = 
	{\cal W}\left[ \begin{smallmatrix} \circ & \circ\\  \circ & \circ \\ \bullet & \circ \end{smallmatrix} \right] 
	\left|\begin{smallmatrix} \circ & \circ \\ & \circ \end{smallmatrix}\right\rangle +
	{\cal W}\left[ \begin{smallmatrix} \bullet & \circ\\  \circ & \circ \\ \bullet & \circ \end{smallmatrix} \right] 
	\left|\begin{smallmatrix} \bullet & \circ \\ & \circ \end{smallmatrix}\right\rangle
	+ {\cal W}\left[ \begin{smallmatrix} \circ & \bullet\\  \circ & \circ \\ \bullet & \circ \end{smallmatrix} \right] 
	\left|\begin{smallmatrix} \circ & \bullet \\ & \circ \end{smallmatrix}\right\rangle + 
	{\cal W}\left[ \begin{smallmatrix} \circ & \circ\\  \circ & \bullet \\ \bullet & \circ \end{smallmatrix} \right] 
	\left|\begin{smallmatrix} \circ & \circ \\ & \bullet \end{smallmatrix}\right\rangle + 
	{\cal W}\left[ \begin{smallmatrix} \bullet & \circ\\  \circ & \bullet \\ \bullet & \circ \end{smallmatrix} \right] 
	\left|\begin{smallmatrix} \bullet & \circ \\ & \bullet \end{smallmatrix}\right\rangle &&
\end{flalign} 
\begin{flalign} 
	\label{eq:L-expression}
	{\cal L}_{a,m} \left|\begin{smallmatrix} \circ \\ \circ & \bullet  \end{smallmatrix}\right\rangle =
	{\cal W}\left[ \begin{smallmatrix} \circ & \circ\\  \circ & \circ \\ \circ & \bullet \end{smallmatrix} \right] 
	\left|\begin{smallmatrix} \circ & \circ \\ & \circ \end{smallmatrix}\right\rangle +
	{\cal W}\left[ \begin{smallmatrix} \bullet & \circ\\  \circ & \circ \\ \circ & \bullet \end{smallmatrix} \right] 
	\left|\begin{smallmatrix} \bullet & \circ \\ & \circ \end{smallmatrix}\right\rangle
	+ {\cal W}\left[ \begin{smallmatrix} \circ & \bullet\\  \circ & \circ \\ \circ & \bullet \end{smallmatrix} \right] 
	\left|\begin{smallmatrix} \circ & \bullet \\ & \circ \end{smallmatrix}\right\rangle &&
\end{flalign} 
\begin{flalign} 
\nn
{\cal L}_{a,m} \left|\begin{smallmatrix} \bullet \\ \circ & \circ  \end{smallmatrix}\right\rangle = 
{\cal W}\left[ \begin{smallmatrix} \circ & \circ\\  \bullet & \circ \\ \circ & \circ \end{smallmatrix} \right] 
\left|\begin{smallmatrix} \circ & \circ \\ & \circ \end{smallmatrix}\right\rangle + 
{\cal W}\left[ \begin{smallmatrix} \circ & \bullet\\  \bullet & \circ \\ \circ & \circ \end{smallmatrix} \right] 
\left|\begin{smallmatrix} \circ & \bullet \\ & \circ \end{smallmatrix}\right\rangle
+ {\cal W}\left[ \begin{smallmatrix} \circ & \circ\\  \bullet & \bullet \\ \circ & \circ \end{smallmatrix} \right] 
\left|\begin{smallmatrix} \circ & \circ \\ & \bullet \end{smallmatrix}\right\rangle &&
\end{flalign} 
\begin{flalign} 
	\nn
	{\cal L}_{a,m} \left|\begin{smallmatrix} \bullet \\ \circ & \bullet  \end{smallmatrix}\right\rangle = 
	{\cal W}\left[ \begin{smallmatrix} \circ & \circ\\  \bullet & \circ \\ \circ & \bullet \end{smallmatrix} \right] 
	\left|\begin{smallmatrix} \circ & \circ \\ & \circ \end{smallmatrix}\right\rangle + 
	{\cal W}\left[ \begin{smallmatrix} \circ & \bullet\\  \bullet & \circ \\ \circ & \bullet \end{smallmatrix} \right] 
	\left|\begin{smallmatrix} \circ & \bullet \\ & \circ \end{smallmatrix}\right\rangle &&
\end{flalign} 

In other words, the most general Lax operator is given by
\bse 
\be 
{\cal L}_{a,m } =
\begin{pmatrix}
	%line 01
		\left\langle \begin{smallmatrix} \circ & \circ \\ & \circ  \end{smallmatrix}\right| 
		{\cal L}_{a,m} \left|\begin{smallmatrix} \circ \\ \circ & \circ  \end{smallmatrix}\right\rangle  &
		\left\langle \begin{smallmatrix} \circ & \circ \\ & \circ  \end{smallmatrix}\right| 
		{\cal L}_{a,m} \left|\begin{smallmatrix} \circ \\ \circ & \bullet \end{smallmatrix}\right\rangle &
		\left\langle \begin{smallmatrix} \circ & \circ \\ & \circ  \end{smallmatrix}\right| 
		{\cal L}_{a,m} \left|\begin{smallmatrix} \circ \\ \bullet & \circ \end{smallmatrix}\right\rangle &
		\left\langle \begin{smallmatrix} \circ & \circ \\ & \circ  \end{smallmatrix}\right| 
		{\cal L}_{a,m} \left|\begin{smallmatrix} \bullet \\ \circ & \circ \end{smallmatrix}\right\rangle &
		\left\langle \begin{smallmatrix} \circ & \circ \\ & \circ  \end{smallmatrix}\right| 
		{\cal L}_{a,m} \left|\begin{smallmatrix} \bullet \\ \circ & \bullet \end{smallmatrix}\right\rangle \\
	%line 02	
		\left\langle \begin{smallmatrix} \bullet & \circ \\ & \circ  \end{smallmatrix}\right| 
		{\cal L}_{a,m} \left|\begin{smallmatrix} \circ \\ \circ & \circ  \end{smallmatrix}\right\rangle &
		\left\langle \begin{smallmatrix} \bullet & \circ \\ & \circ  \end{smallmatrix}\right| 
		{\cal L}_{a,m} \left|\begin{smallmatrix} \circ \\ \circ & \bullet  \end{smallmatrix}\right\rangle & 
		\left\langle \begin{smallmatrix} \bullet & \circ \\ & \circ  \end{smallmatrix}\right| 
		{\cal L}_{a,m} \left|\begin{smallmatrix} \circ \\ \bullet & \circ  \end{smallmatrix}\right\rangle & 
		\left\langle \begin{smallmatrix} \bullet & \circ \\ & \circ  \end{smallmatrix}\right| 
		{\cal L}_{a,m} \left|\begin{smallmatrix} \bullet \\ \circ & \circ  \end{smallmatrix}\right\rangle &
		\left\langle \begin{smallmatrix} \bullet & \circ \\ & \circ  \end{smallmatrix}\right| 
		{\cal L}_{a,m} \left|\begin{smallmatrix} \bullet \\ \circ & \bullet \end{smallmatrix}\right\rangle \\ 
	%line 03
		\left\langle \begin{smallmatrix} \circ & \bullet \\ & \circ  \end{smallmatrix}\right| 
		{\cal L}_{a,m} \left|\begin{smallmatrix} \circ \\ \circ & \circ  \end{smallmatrix}\right\rangle &  
		\left\langle \begin{smallmatrix} \circ & \bullet \\ & \circ  \end{smallmatrix}\right| 
		{\cal L}_{a,m} \left|\begin{smallmatrix} \circ \\ \circ & \bullet  \end{smallmatrix}\right\rangle & 
		\left\langle \begin{smallmatrix} \circ & \bullet \\ & \circ  \end{smallmatrix}\right| 
		{\cal L}_{a,m} \left|\begin{smallmatrix} \circ \\ \bullet & \circ  \end{smallmatrix}\right\rangle & 
		\left\langle \begin{smallmatrix} \circ & \bullet \\ & \circ  \end{smallmatrix}\right| 
		{\cal L}_{a,m} \left|\begin{smallmatrix} \bullet \\ \circ & \circ  \end{smallmatrix}\right\rangle &
		\left\langle \begin{smallmatrix} \circ & \bullet \\ & \circ  \end{smallmatrix}\right| 
		{\cal L}_{a,m} \left|\begin{smallmatrix} \bullet \\ \circ & \bullet \end{smallmatrix}\right\rangle \\ 
	%line 04
		\left\langle \begin{smallmatrix} \circ & \circ \\ & \bullet  \end{smallmatrix}\right| 
		{\cal L}_{a,m} \left|\begin{smallmatrix} \circ \\ \circ & \circ  \end{smallmatrix}\right\rangle &
		\left\langle \begin{smallmatrix} \circ & \circ \\ & \bullet  \end{smallmatrix}\right| 
		{\cal L}_{a,m} \left|\begin{smallmatrix} \circ \\ \circ & \bullet  \end{smallmatrix}\right\rangle & 
		\left\langle \begin{smallmatrix} \circ & \circ \\ & \bullet  \end{smallmatrix}\right| 
		{\cal L}_{a,m} \left|\begin{smallmatrix} \circ \\ \bullet & \circ  \end{smallmatrix}\right\rangle & 
		\left\langle \begin{smallmatrix} \circ & \circ \\ & \bullet  \end{smallmatrix}\right| 
		{\cal L}_{a,m} \left|\begin{smallmatrix} \bullet \\ \circ & \circ  \end{smallmatrix}\right\rangle & 
		\left\langle \begin{smallmatrix} \circ & \circ \\ & \bullet  \end{smallmatrix}\right| 
		{\cal L}_{a,m} \left|\begin{smallmatrix} \bullet \\ \circ & \bullet \end{smallmatrix}\right\rangle \\
	%line 05
		\left\langle \begin{smallmatrix} \bullet & \circ \\ & \bullet  \end{smallmatrix}\right| 
		{\cal L}_{a,m} \left|\begin{smallmatrix} \circ \\ \circ & \circ  \end{smallmatrix}\right\rangle &
		\left\langle \begin{smallmatrix} \circ & \circ \\ & \bullet  \end{smallmatrix}\right| 
		{\cal L}_{a,m} \left|\begin{smallmatrix} \bullet \\ \circ & \bullet  \end{smallmatrix}\right\rangle & 
		\left\langle \begin{smallmatrix} \bullet & \circ \\ & \bullet  \end{smallmatrix}\right| 
		{\cal L}_{a,m} \left|\begin{smallmatrix} \circ \\ \bullet & \circ  \end{smallmatrix}\right\rangle & 
		\left\langle \begin{smallmatrix} \bullet & \circ \\ & \bullet  \end{smallmatrix}\right| 
		{\cal L}_{a,m} \left|\begin{smallmatrix} \bullet \\ \circ & \circ  \end{smallmatrix}\right\rangle &
		\left\langle \begin{smallmatrix} \bullet & \circ \\ & \bullet  \end{smallmatrix}\right| 
		{\cal L}_{a,m} \left|\begin{smallmatrix} \bullet \\ \circ & \bullet \end{smallmatrix}\right\rangle  \\
		&&&&&  {\bm 0}_{3\times 3}
	\end{pmatrix}
	\ee
where the nonsingular \(5\times 5\) matrix reads
	\be 
	\label{eq:Lax}
	{\cal L}^{(5\times 5)}_{a,m }
	= \begin{pmatrix}
		{\cal W}([0]) & {\cal W}([1]_2) & {\cal W}([1]_1) & {\cal W}([1]_3) & {\cal W}([2]_{23}) \\
		{\cal W}([1]_5) & {\cal W}([2]_{25}) & {\cal W}([2]_{15}) & 0 & 0 \\
		{\cal W}([1]_6) & {\cal W}([2]_{26}) & {\cal W}([2]_{16}) & {\cal W}([2]_{36}) & {\cal W}([3]_{236}) \\
		{\cal W}([1]_4) & 0 & {\cal W}([2]_{14}) & {\cal W}([2]_{34}) & 0 \\
		{\cal W}([2]_{45}) & 0 & {\cal W}([3]_{145}) & 0 & 0 \\
	\end{pmatrix}\; .
	\ee
\ese
As we explained before, we would like to see this operator as an \(S\)-matrix. Remember that the physical space (defined by the incoming or outgoing vectors~(\ref{eq:phystates})) is \(5\) dimensional, and the other \(3\) directions decouple from the scattering process. Therefore, the fact that \({\cal L}_{a,m }\) is a singular matrix is just a spurious result of our insistence in writing it with its legs along the \(3\)D null space, 
see~\ref{fig:physspace}. As we see next section, there are some advantages in considering these terms. 

One important aspect we have ignored so far is the dependence of all Boltzmann weights on the parameters \(J, V\) and \(q\). Since we want a Hermitian Hamiltonian, these are real parameters. Moreover, in the 1D and 2D problems the mass gap relates the parameters \(J\) and \(q\), so that a complex parameter \(u\in \mathbb{C}\) can be introduced so that \(V=\Re(u)\) and \(q=\Im(u)\). We can now suppose that the conjecture in~\cite{Dijkgraaf:2008ua} is correct, therefore the same reasoning applies to the 3D case, and we can simply write the coupling constants in terms of a spectral parameter \(u\), that is \( J= J(u), V=V(u)\) and \(q=q(u)\). Therefore, the Boltzmann weights can also be denoted as
\be 
{\cal W}\left[ \begin{smallmatrix} \nu_1 & \nu_2\\ \alpha & \beta \\ \mu_1 & \mu_2 \end{smallmatrix} \right] \equiv 
{\cal W}\left[ u \Big|\begin{smallmatrix} \nu_1 & \nu_2\\ \alpha & \beta \\ \mu_1 & \mu_2 \end{smallmatrix}\right]\; .
\ee

In order to simplify the notation, we will keep the dependence on the spectral parameter implicit, and we reintroduce its dependence conveniently, for example, in the expression for \emph{row-to-row transfer matrix} that we define now
\be 
\label{eq:transfermatrix}
\begin{split}
	{\bm t}_a(u) =\mathrm{Tr}_{\tilde{\mathscr{H}}^{(a)}}\left(\prod_{m} {\cal L}_{a,m}(u) \right)\; .
\end{split}
\ee
As usual, we can think of it as an operator that transfers the state \(|{\bm \mu}\rangle \in \mathscr{H}^{(a-1)}\) 
to a linear combination of \(|{\bm \nu}\rangle \in \mathscr{H}^{(a)}\). More specifically
\bse
\be 
{\bm t}_a |{\bm \mu}\rangle 
=\sum_{\bm \nu } ({\bm T}_a)^{\bm \nu}_{\bm \mu} 
|{\bm \nu}\rangle 
\ee
where 
\be 
\begin{split}
	|{\bm \mu}\rangle &=|\mu_1\; , \tilde{\mu}_1\; , \mu_2\; , \tilde{\mu}_2\; , \dots \; , \mu_{N}\; , \tilde{\mu}_{N}\rangle \\
	|{\bm \nu}\rangle &= |\nu_1\; , \tilde{\nu}_1\; , \mu_2\; , \tilde{\nu}_2\; , \dots \; , \nu_{N}\; , \tilde{\nu}_{N}\rangle \; ,
\end{split}
\ee
and 
\be 
	({\bm t}_a)^{\bm \nu}_{\bm \mu}  = \langle {\bm \nu} | {\bm T}_a | {\bm \mu }\rangle  = \sum_{\bm \alpha}
	{\cal W}
	\Big[\begin{smallmatrix}
		\nu_1 & \tilde{\nu}_1 \\ \alpha_1 & \alpha_2 \\ \mu_1 & \tilde{\mu}_1
	\end{smallmatrix}\Big]
	{\cal W}\Big[\begin{smallmatrix}
		\nu_2 & \tilde{\nu}_2 \\ \alpha_2 & \alpha_3 \\ \mu_2 & \tilde{\mu}_3
	\end{smallmatrix}\Big]\cdots
	\cdots  {\cal W}\Big[\begin{smallmatrix}
		\nu_{N-1} & \tilde{\nu}_{N-1} \\ \alpha_{N-1} & \alpha_{N} \\ \mu_{N-1} & \tilde{\mu}_{N-1} \end{smallmatrix}\Big]
	{\cal W}\Big[\begin{smallmatrix}
		\nu_N & \tilde{\nu}_N \\ \alpha_N & \alpha_1 \\ \mu_N & \tilde{\mu}_N
	\end{smallmatrix}\Big]\; .
\ee
\ese
We have more to say about the transfer matrix next section.

%%%%%%%%%%%%%%%%%%%%%%%%%%%%%%%%%%%%%
%%%%%%%%%%%%%%%%%%%%%%%%%%%%%%%%%%%%%
%%%%%%%%%%%%%%%%%%%%%%%%%%%%%%%%%%%%%
%%%%%%%%%%%%%%%%%%%%%%%%%%%%%%%%%%%%%

\section{Integrable (sub-) systems \& Vertex models}
\label{sec:transfer}

We have stressed in the hypothesis in section~\ref{sec:boltzmann} that the hexagons weights do not depend on their lattice parametrization -- the position \((a,m)\) -- but depend on the position and number of particles inside the hexagon. This hypothesis allowed us to define the weights of figure~\ref{fig:18hvertex} regardless of their particular location in the lattice. On the other hand, we need to be precise on the parametrization since the relative hexagon positions change from even to odd rows. This also implies that we actually have two types of Lax operators~(\ref{eq:laxoperator}), namely
\bse
\be 
	{\cal L}_{2a,2m} : \tilde{V}_{2m}^{(2a)} \otimes V^{(2a-1)}_{2m+1/2}\otimes V^{(2a-1)}_{2m+3/2} 
	\to \tilde{V}_{2m+2}^{(2a)} \otimes V^{(2a)}_{2m+1/2}\otimes V^{(2a)}_{2m+3/2}
\ee
and 
\be 
	{\cal L}_{2a+1,2m+1} : \tilde{V}_{2m+1}^{(2a+1)}  \otimes V^{(2a)}_{2m+3/2}\otimes V^{(2a)}_{2m+5/2}
	\to \tilde{V}_{2m+3}^{(2a+1)} \otimes V^{(2a+1)}_{2m+3/2}\otimes V^{(2a+1)}_{2m+5/2}\; .
\ee
\ese
Observe that our parametrization does not allow operators of the form \({\cal L}_{2a,2m+1}\) or \({\cal L}_{2a+1,2m} \). Evidently, since all these vector spaces are isomorphic the flamboyant notation above is unnecessary and we can simply write \({\cal L}_{2a,2m}, {\cal L}_{2a+1,2m+1} \in\textrm{End}(\tilde{V}\otimes V^2)\). Consequently, there are two types of row-to-row transfer matrices
\be 
\begin{split}
	& {\bm t}(u) =\mathrm{Tr}_{\tilde{\mathscr{H}}^{(2a)}}\left(\prod_{m} {\cal L}_{2a,2m}(u)\right)\\
	& \tilde{\bm t}(u) =\mathrm{Tr}_{\tilde{\mathscr{H}}^{(2a+1)}}\left(\prod_{m} {\cal L}_{2a+1,2m+1}(u)\right)\; .
\end{split}
\ee

As we said before, the system has \(M\)-columns (where we count by the number of hexagons), which 
means that there are \(2M\) slots in the X- and Y-rows. Remember that we have also imposed periodic boundary 
conditions, \(2M\sim 0\). Therefore, the transfer matrices act as
\bse 
\be
	{\bm t} |\mu_{1/2} \mu_{3/2}\cdots \mu_{2M-3/2} \mu_{2M-1/2}\rangle = 
	\sum_{\bm \nu} 
	{\bm t}^{\nu_{1/2} \nu_{3/2}\cdots \nu_{2M-3/2} \nu_{2M-1/2}}_{\mu_{1/2} \mu_{3/2}\cdots \mu_{2M-3/2} \mu_{2M-1/2}}
	| \nu_{1/2} \nu_{3/2}\cdots \nu_{2M-3/2} \nu_{2M-1/2} \rangle \; ,
\ee
and
\be
\begin{split}
	\tilde{\bm t} |\nu_{1/2} \nu_{3/2}\cdots \nu_{2M-3/2} \nu_{2M-1/2}\rangle & \equiv
	\tilde{\bm t} |\nu_{3/2} \nu_{5/2} \cdots \nu_{2M-1/2} \nu_{2M+1/2} \rangle \\
	& =\sum_{{\bm \mu}'} 
	\tilde{\bm t}_{\nu_{3/2} \nu_{5/2}\cdots \nu_{2M-1/2} \nu_{2M+1/2}}^{\mu'_{3/2} \mu'_{5/2}\cdots \mu'_{2M-1/2} \mu'_{2M+1/2}}
	|\mu'_{3/2}\mu'_{5/2}\cdots \mu'_{2M-1/2} \mu'_{2M+1/2}\rangle \; ,
\end{split}
\ee
where we have used that \(2M+1/2\sim 1/2\) and  \(2M+1\sim 1\).
\ese
Finally, the transfer matrices components can be easily calculated
\bse
\be 
	{\bm t}^{\nu_{1/2} \nu_{3/2}\cdots \nu_{2M-3/2} \nu_{2M-1/2}}_{\mu_{1/2} \mu_{3/2} \cdots \mu_{2M-3/2} \mu_{2M-1/2}}
	= \sum_{\bm \alpha}{\cal W} \left[\begin{smallmatrix}
	\nu_{1/2} & \nu_{3/2} \\ \alpha_0 & \alpha_2 \\ \mu_{1/2} & \mu_{3/2} \end{smallmatrix}\right]
	{\cal W} \left[\begin{smallmatrix} \nu_{5/2} & \nu_{7/2} \\ \alpha_2 & \alpha_4 \\ \mu_{5/2} & \mu_{7/2} \end{smallmatrix}\right]
	\cdots {\cal W} \left[\begin{smallmatrix} \nu_{2M-3/2} & \nu_{2M-1/2} \\ \alpha_{2M} & \alpha_0 \\ \mu_{2M-3/2} & \mu_{2M-1/2}
	\end{smallmatrix}\right] 
\ee
and 
\be 
	\tilde{\bm t}_{\nu_{3/2} \nu_{5/2}\cdots \nu_{2M-1/2} \nu_{2M+1/2}}^{\mu'_{3/2} \mu'_{5/2}\cdots \mu'_{2M-1/2} \mu'_{2M+1/2}}
	= \sum_{\bm \alpha} {\cal W} \left[\begin{smallmatrix} \mu'_{3/2} & \mu'_{5/2} \\ \alpha_1 & \alpha_3 \\ \nu_{3/2} & \nu_{5/2} 
	\end{smallmatrix}\right]
	{\cal W} \left[\begin{smallmatrix} \mu'_{7/2} & \mu'_{9/2} \\ \alpha_3 & \alpha_5 \\ \nu_{7/2} & \nu_{9/2}
	\end{smallmatrix}\right]\cdots {\cal W}
	\left[\begin{smallmatrix} \mu'_{2M-1/2} & \mu'_{1/2} \\ \alpha_{2M-1} & \alpha_1 \\ \nu_{2M-1/2} & \nu_{1/2}
	\end{smallmatrix}\right]\; .
\ee
\ese

\subsection{Transfer matrices commutation relations}

With this knowledge, we would like to study commutation relations of the transfer matrices. It is well known 
that for a generic integrable system, the conserved quantities \(I_n\) are obtained from an expansion of the form
\(\ln t(u) = \sum_n I_n u^n\) where \(t(u)\) is the transfer matrix. Finally, the commutativity of \(I_n\) is a 
consequence of the generalized commutation relation \([t(u), t(v)]=0\). 

Using the transfer matrices we have defined above, one can ask if their existence implies the integrability of the classical statistical problem we have defined. The answer turns out to be no, but in an interesting and promising way. Let us start with the products
\be 
\begin{split}
	{\bm t}\; {\bm t}' & =  \mathrm{Tr}_{ \tilde{\mathscr{H}}^{(2a)}\times \tilde{\mathscr{H}}^{(2b)}}\left( 
	{\cal L}_{2a,0} {\cal L}'_{2b,0} {\cal L}_{2a,1} {\cal L}'_{2b,1}\cdots {\cal L}_{2a,2M-2} {\cal L}'_{2b,2M-2} \right)
\end{split}
\ee
and 
\be 
\begin{split}
	{\bm t}'\; {\bm t} & =  \mathrm{Tr}_{ \tilde{\mathscr{H}}^{(2b)}\times \tilde{\mathscr{H}}^{(2a)}}\left( 
	{\cal L}'_{2b,0} {\cal L}_{2a,0} {\cal L}'_{2b,1} {\cal L}_{2a,1} \cdots {\cal L}'_{2b,2M-2} {\cal L}_{2a,2M-2} \right)
\end{split}
\ee
where \({\bm t} \equiv {\bm t}(u)\) and \({\bm t}'\equiv {\bm t}(v)\). It is easy to see that these two expressions are equal if there exists an operator \(\mathcal{R}\) such that
\be 
\label{eq:even-ybe}
\boxed{
\mathcal{R}_{2a,2b}{\cal L}_{2a,2m} {\cal L}'_{2b,2m} \stackrel{!}{=} {\cal L}'_{2b,2m} {\cal L}_{2a,2m}  \mathcal{R}_{2a,2b}
}\; .
\ee
Similarly, one can show that \([\tilde{\bm t}, \tilde{\bm t}']=0\) iff
\be 
\label{eq:odd-ybe}
\boxed{
\mathcal{R}_{2a+1,2b+1}{\cal L}_{2a+1,2m+1} {\cal L}'_{2b+1,2m+1} \stackrel{!}{=} {\cal L}'_{2b+1,2m+1} {\cal L}_{2a+1,2m+1}  \mathcal{R}_{2a+1,2b+1}
}\; .
\ee

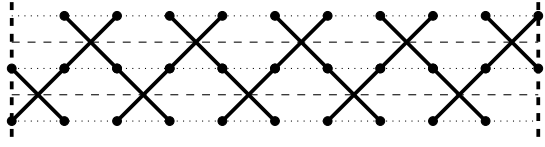
\begin{figure}[t!]
	\centering
	\begin{tikzpicture}[scale=0.7]
	\draw[line width=0.5mm, dashed] (-4.5,-0.3) -- (-4.5,2.3);
	\draw[line width=0.5mm, dashed] (5.5,-0.3) -- (5.5,2.3);	
	\draw[dashed] (-4.5,1.5) -- (5.5,1.5);
	\draw[dashed] (-4.5,0.5) -- (5.5,0.5);
	\draw[dotted] (-4.5,0) -- (5.5,0);
	\draw[dotted] (-4.5,1) -- (5.5,1);
	\draw[dotted] (-4.5,2) -- (5.5,2);
	%below
	\draw[ line width=0.5mm] (-4.5,0) -- (-3.5,1); \draw[black,fill=black] (-4.5,0) circle (.5ex) ; \draw[black,fill=black] (-3.5,1) circle (.5ex) ;
	\draw[ line width=0.5mm] (-4.5,1) -- (-3.5,0); \draw[black,fill=black] (-4.5,1) circle (.5ex) ; \draw[black,fill=black] (-3.5,0) circle (.5ex) ;
	\draw[ line width=0.5mm] (-2.5,0) -- (-1.5,1); \draw[black,fill=black] (-2.5,0) circle (.5ex) ; \draw[black,fill=black] (-1.5,1) circle (.5ex) ;
	\draw[ line width=0.5mm] (-2.5,1) -- (-1.5,0); \draw[black,fill=black] (-2.5,1) circle (.5ex) ; \draw[black,fill=black] (-1.5,0) circle (.5ex) ;
	\draw[ line width=0.5mm] (-0.5,0) -- (0.5,1); \draw[black,fill=black] (-0.5,0) circle (.5ex) ; \draw[black,fill=black] (0.5,1) circle (.5ex) ;
	\draw[ line width=0.5mm] (-0.5,1) -- (0.5,0); \draw[black,fill=black] (-0.5,1) circle (.5ex) ; \draw[black,fill=black] (0.5,0) circle (.5ex) ;
	\draw[ line width=0.5mm] (1.5,0) -- (2.5,1); \draw[black,fill=black] (1.5,0) circle (.5ex) ; \draw[black,fill=black] (2.5,1) circle (.5ex) ;
	\draw[ line width=0.5mm] (1.5,1) -- (2.5,0); \draw[black,fill=black] (1.5,1) circle (.5ex) ; \draw[black,fill=black] (2.5,0) circle (.5ex) ;
	\draw[ line width=0.5mm] (4.5,0) -- (3.5,1); \draw[black,fill=black] (4.5,0) circle (.5ex) ; \draw[black,fill=black] (3.5,1) circle (.5ex) ;
	\draw[ line width=0.5mm] (4.5,1) -- (3.5,0); \draw[black,fill=black] (4.5,1) circle (.5ex) ; \draw[black,fill=black] (3.5,0) circle (.5ex) ;
	%above
	\draw[ line width=0.5mm] (-3.5,1) -- (-2.5,2); \draw[black,fill=black] (-3.5,1) circle (.5ex) ; \draw[black,fill=black] (-2.5,2) circle (.5ex) ;
	\draw[ line width=0.5mm] (-3.5,2) -- (-2.5,1); \draw[black,fill=black] (-3.5,2) circle (.5ex) ; \draw[black,fill=black] (-2.5,1) circle (.5ex) ;
	\draw[ line width=0.5mm] (-1.5,1) -- (-0.5,2); \draw[black,fill=black] (-1.5,1) circle (.5ex) ; \draw[black,fill=black] (-0.5,2) circle (.5ex) ;
	\draw[ line width=0.5mm] (-1.5,2) -- (-0.5,1); \draw[black,fill=black] (-1.5,2) circle (.5ex) ; \draw[black,fill=black] (-0.5,1) circle (.5ex) ;
	\draw[ line width=0.5mm] (0.5,1) -- (1.5,2); \draw[black,fill=black] (0.5,1) circle (.5ex) ; \draw[black,fill=black] (1.5,2) circle (.5ex) ;
	\draw[ line width=0.5mm] (0.5,2) -- (1.5,1); \draw[black,fill=black] (0.5,2) circle (.5ex) ; \draw[black,fill=black] (1.5,1) circle (.5ex) ;
	\draw[ line width=0.5mm] (2.5,1) -- (3.5,2); \draw[black,fill=black] (2.5,1) circle (.5ex) ; \draw[black,fill=black] (3.5,2) circle (.5ex) ;
	\draw[ line width=0.5mm] (2.5,2) -- (3.5,1); \draw[black,fill=black] (2.5,2) circle (.5ex) ; \draw[black,fill=black] (3.5,1) circle (.5ex) ;
	\draw[ line width=0.5mm] (4.5,1) -- (5.5,2); \draw[black,fill=black] (4.5,1) circle (.5ex) ; \draw[black,fill=black] (5.5,2) circle (.5ex) ;
	\draw[ line width=0.5mm] (4.5,2) -- (5.5,1); \draw[black,fill=black] (4.5,2) circle (.5ex) ; \draw[black,fill=black] (5.5,1) circle (.5ex) ;
	\end{tikzpicture}
	\caption{Transfer matrices product \({\bm t} \tilde{\bm t}\) for \(M=5\).}
	\label{fig:transfermat}
\end{figure}

The case \([{\bm t}(u), \tilde{\bm t}(v)]\) is more interesting. The product can be diagrammatically represented in figure~\ref{fig:transfermat}, where the vertical lines denote periodic boundary condition. Observe that
\be 
\tilde{\bm t} \; {\bm t}= \mathrm{Tr}_{ \tilde{\mathscr{H}}^{(2b+1)} \times \tilde{\mathscr{H}}^{(2a)} } \left( 
{\cal L}_{2b+1,1} {\cal L}_{2a,0} {\cal L}_{2b+1,3} {\cal L}_{2a,2}  {\cal L}_{2b+1, 5}  {\cal L}_{2a, 4}\cdots 
{\cal L}_{2b+1,2M-1} {\cal L}_{2a,2M-2}
\right)
\ee
and
\be 
{\bm t}\; \tilde{\bm t} = \mathrm{Tr}_{ \tilde{\mathscr{H}}^{(2a)}\times \tilde{\mathscr{H}}^{(2b+1)}} \left( 
{\cal L}_{2a,0} {\cal L}_{2a,2} {\cal L}_{2b+1,1} {\cal L}_{2a, 4} {\cal L}_{2b+1,3}\cdots 
{\cal L}_{2a,2M-2} {\cal L}_{2a,2M-3} {\cal L}_{2b+1,2M-1}
\right)\; .
\ee 
The expressions above are not equal since the commutators 
\([{\cal L}_{a,m}, {\cal L}_{b,n}]\) do not necessarily vanish for \(m\neq n\). In fact, it is easy to see that 
for states with corresponding diagrams~\ref{fig:tensorstate}, we have 
\([{\cal L}_{a,m}, {\cal L}_{a+1,m+1}]\neq 0\) and \([{\cal L}_{a,m}, {\cal L}_{a+1,m'}]= 0\)
for \(m'>m+1\). Therefore, we have
\be 
[{\bm t}(u), \tilde{\bm t}(v)] \neq 0\; ,
\ee
which spoils the na\"ive integrability of the classical statistical system we are considering. It is worth
stressing that this non-commutativity does not disprove integrable, it just says that the row-to-row transfer matrix above is not appropriate. In summary, it is possible that the system is, indeed, non-integrable, but we have not yet ruled out its integrability \footnote{A simplicial homology approach for exact solvability has been studied in~\cite{Ogura2020}. Although the Hamiltonian considered in the current work does not satisfy the generic form of~\cite{Ogura2020}, one might try generalize some of these ideas to the present problem, using in particular the generalization of the Jordan-Wigner transformation in~\cite{Kazuhiko2016, *Minami:2017lld, *Minami:2019bry}.}.

\subsection{Integrability for even and odd subsystems}

Observe that the commutation relations for the even and odd transfer matrices~(\ref{eq:even-ybe}) and~(\ref{eq:odd-ybe}) satisfy the properties we are looking for. We want to finish this work explaining how subsystems defined by even or odd rows are connected to vertex models. 

We first modify the rules~(\ref{eq:rule1}),~(\ref{eq:rule2}),~(\ref{eq:rule3}) and~(\ref{eq:rule4}) defined in section~\ref{sec:classicalsystem}. We still assume the same set of local configurations of figure~\ref{fig:18hvertex}, but now we connect them using the following rules
\begin{shaded}
	
	{\bf Rules}
	
	\begin{itemize}
		\item Consider two hexagons \(A\) and \(B\) in the triad \(Y^{(a-1)}X^{(a)}Y^{(a)}\), with generic parametrization
		\((a,m)\) and \((a,m')\) respectively. They are connected horizontally if one of the following conditions is satisfied:
		\bse
		\begin{align}
		\texttt{Rule \#1}:&  \quad \alpha^A_{m+2}= \alpha^B_{m'} \qquad \text{for} \ m+2=m' \label{eq:newrule1}\\
		\texttt{Rule \#2}:&  \quad \alpha^A_{m}= \alpha^B_{m'+2} \qquad \text{for} \ m=m'+2 \label{eq:newrule2} \; .
		\end{align}
		\item Consider a hexagon \(A\) in the triad \(Y^{(a-1)}X^{(a)}Y^{(a)}\) parametrized by \((a,m)\), and another hexagon \(B\) in the triad \(Y^{(a)}X^{(a+1)}Y^{(a+1)}\) parametrized by \((a+1,m')\). They are connected vertically if the following conditions are satisfied:
		\begin{align}
		\texttt{Rule \#5}:&  \quad \mu^{A (a)}_{m+1/2}= \mu^{B (a)}_{m'+1/2} \qquad \text{for} \ m=m' \label{eq:newrule3} \\
		\texttt{Rule \#6}:& \quad \mu^{A (a)}_{m+3/2}= \mu^{B (a)}_{m'+3/2} \qquad \text{for} \ m=m' \label{eq:newrule4}\; .
		\end{align}
		\ese
	\end{itemize}	 
\end{shaded}

In other words, we have not changed the rules~(\ref{eq:rule1}),~(\ref{eq:rule2}), but now we link the rows vertically and not diagonally. It should be clear that in this case we do not have a Kagome lattice anymore, instead, we consider a triangular tiling of the plane. These rules imply that even and odd rows are completely equivalent, so we consider them simultaneously. Therefore, one can write~(\ref{eq:even-ybe}) and~(\ref{eq:odd-ybe}) as 
\be 
\boxed{
	\label{eq:fcr}
	\mathcal{R}_{a,b}(u,v) {\cal L}_{a, m}(u) {\cal L}_{b,m}(v) = {\cal L}_{b, m}(v) {\cal L}_{a,m}(u) \mathcal{R}_{a,b} (u,v) } \; ,
\ee
that is known as the \emph{fundamental commutation relation} (fcc).

Since integrability is a global property, we need to show the implications of these relations on 
the monodromy matrices. Let us consider a right multiplication of~(\ref{eq:fcr}) by \({\cal L}_{a, m+1} {\cal L}'_{b,m+1} \), 
then
\bse
\be 
\begin{split}
	\mathcal{R}_{a,b} {\cal L}_{a, m} {\cal L}'_{b,m} {\cal L}_{a, m+1} {\cal L}'_{b,m+1} 
	& = {\cal L}'_{b, m} {\cal L}_{a,m} \mathcal{R}_{a,b} {\cal L}_{a, m+1} {\cal L}'_{b,m+1}\\
	& = {\cal L}'_{b, m} {\cal L}_{a,m} {\cal L}'_{b,m+1} {\cal L}_{a, m+1}\mathcal{R}_{a,b}
\end{split}
\ee
where we use~(\ref{eq:even-ybe}) or~(\ref{eq:odd-ybe}) in the second line. 
Additionally, ultralocality implies
\be 
	\mathcal{R}_{a,b} \left({\cal L}_{a, m} {\cal L}_{a, m+1}\right) \left({\cal L}'_{b,m}  {\cal L}'_{b,m+1}\right) =
	\left({\cal L}'_{b, m} {\cal L}'_{b,m+1}\right)  \left( {\cal L}_{a,m} {\cal L}_{a, m+1} \right) \mathcal{R}_{a,b}\; ,
\ee
\ese
that has the form~(\ref{eq:fcr}). Repeating this idea \(M\)-times, one can find the RTT-relation
\be 
\label{eq:rtt}
\boxed{
	\mathcal{R}_{a,b}(u,v) {\bm T}_{a}(u) {\bm T}_{b}(v) = {\bm T}_{b}(v) {\bm T}_{a}(u) \mathcal{R}_{a,b}(u,v)}\; ,
\ee
where \({\bm T}(u)\) represents the monodromy operators
\be 
\begin{split}
	{\bm T}_{2a}(u) & = {\cal L}_{2a, 0}(u) {\cal L}_{2a, 2}(u)\cdots {\cal L}_{2a, 2M-2}(u)\\
	{\bm T}_{2a+1}(u) & = {\cal L}_{2a+1, 1}(u) {\cal L}_{2a+1, 3}(u) \cdots {\cal L}_{2a, 2M-1}(u)\; .
\end{split}
\ee

\begin{figure}[h]
	\centering
	\begin{tikzpicture}
	\draw (-.5,-.5) -- (.5,.5); \draw (.5,-.5) -- (-.5,.5);
	\draw (.5,-.5) -- (2.5,-.5); \draw (.5,.5) -- (2.5,.5);
	\draw (1.5,-.5) -- (1,-1); \draw (1.5,-.5) -- (2,-1);
	\draw (1.5,.5) -- (1,1);  \draw (1.5,.5) -- (2,1);
	\draw plot [smooth cycle] coordinates {(1.5,-.5) (1.3,0) (1.5,.5) (1.7,0)};
	\node at (-0.7,.55) {\(\tilde{\beta}\)}; \node at (-0.7,-.55) {\(\beta\)};
	\node at (2.7,.55) {\(\alpha\)}; \node at (2.7,-.55) {\(\tilde{\alpha}\)};
	\node at (1,-1.3) {\(\mu_1\)}; \node at (2,-1.3) {\(\mu_2\)};
	\node at (1,1.3) {\(\nu_1\)}; \node at (2,1.3) {\(\nu_2\)};
	\node at (3,0) {\(=\)};
	\end{tikzpicture}
	\begin{tikzpicture}
	\draw (2.5,-.5) -- (3.5,.5); \draw (2.5,.5) -- (3.5,-.5);
	\draw (.5,-.5) -- (2.5,-.5); \draw (.5,.5) -- (2.5,.5);
	\draw (1.5,-.5) -- (1,-1); \draw (1.5,-.5) -- (2,-1);
	\draw (1.5,.5) -- (1,1);  \draw (1.5,.5) -- (2,1);
	\draw plot [smooth cycle] coordinates {(1.5,-.5) (1.3,0) (1.5,.5) (1.7,0)};
	\node at (0.3,.55) {\(\tilde{\beta}\)}; \node at (0.3,-.55) {\(\beta\)};
	\node at (3.7,.55) {\(\alpha\)}; \node at (3.7,-.55) {\(\tilde{\alpha}\)};
	\node at (1,-1.3) {\(\mu_1\)}; \node at (2,-1.3) {\(\mu_2\)};
	\node at (1,1.3) {\(\nu_1\)}; \node at (2,1.3) {\(\nu_2\)};
	\end{tikzpicture}
	\caption{Diagrammatic representation of the fundamental commutation relation.}
	\label{fig:fcr}
\end{figure}
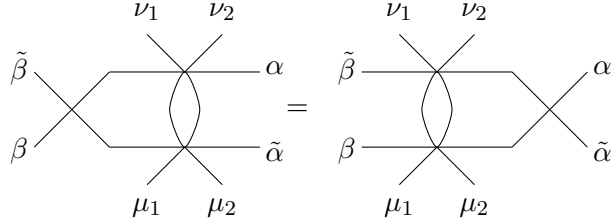

We want to close this section writing the constraints imposed by the fundamental commutation relation 
on the Boltzmann weights. We have seen that \({\cal L}\in \textrm{End}(\tilde{V}\otimes V \otimes V)\), where 
\(V \simeq \tilde{V}\simeq \mathbb{C}^2\). Moreover, as we have seen before, \({\cal L}\) is an \(8\times 8\) 
matrix, therefore, \({\cal R}\in \textrm{End}(\tilde{V} \otimes \tilde{V})\) is a \(4\times 4\) matrix, 
and its entries are \(2\times 2\) operators acting on the space \(V\). Let us write 
\be 
\label{eq:Roperator}
{\cal R}_{ab}|\alpha \tilde{\alpha}\rangle =
\sum_{\beta \tilde{\beta}}{\cal R}\left[ \begin{smallmatrix} \beta \ & \tilde{\beta}  \\ \alpha & \tilde{\alpha} \end{smallmatrix} \right]
|\beta \tilde{\beta}\rangle\; .
\ee
then, one can show that the Fundamental Commutation Relation gives the following constraints on the Boltzmann weights
\be 
\label{eq:fcr-comp}
\sum_{\gamma, \tilde{\gamma}, \lambda_1, \lambda_2} 
{\cal R}\Big[ u, v \Big| \begin{smallmatrix} \beta \ & \tilde{\beta}  \\ \gamma & \tilde{\gamma} \end{smallmatrix} \Big]
{\cal W}\Big[ u\Big| \begin{smallmatrix} \nu_1 \ & \nu_2 \\  \alpha & \gamma \\ \lambda_1 & \lambda_2  \end{smallmatrix} \Big] 
{\cal W}\Big[ v \Big| \begin{smallmatrix} \lambda_1 & \lambda_2 \\ \tilde{\alpha} & \tilde{\gamma} \\ \mu_1 & \mu2  \end{smallmatrix} \Big]
= \sum_{\gamma, \tilde{\gamma}, \lambda_1, \lambda_2} 
{\cal W}\Big[ v\Big| \begin{smallmatrix} \nu_1 \ & \nu_2 \\  \tilde{\gamma} & \tilde{\beta} \\ \lambda_1 & \lambda_2  \end{smallmatrix} \Big] 
{\cal W}\Big[ u\Big| \begin{smallmatrix} \lambda_1 & \lambda_2 \\ \gamma & \beta \\ \mu_1 & \mu2  \end{smallmatrix} \Big]
{\cal R}\Big[u,v \Big| \begin{smallmatrix} \gamma \ & \tilde{\gamma}  \\ \alpha & \tilde{\alpha} \end{smallmatrix} \Big]\; .
\ee
We represent these constraints diagrammatically as in figure~\ref{fig:fcr}.

One natural question that emerges now is if we can consider these constraints on the Boltzmann weights and 
rebuild the Kagome lattice from these two subsystems. It is not clear if this process gives a consistent lattice. 

\subsection{2-particle scattering and descendant of the vertex models}
\label{sec:vertexmodel}

The pinnacle of integrability is the factorization of an n-scattering process into a collection of 2-scattering processes, but the decomposition of the hexagons in terms of 3-particles scattering~\ref{fig:scattering} seems to be a direct violation of this condition. On the other hand, we have found the fundamental commutation relation~\ref{fig:fcr} that is the underlying algebraic structure in integrable models. In this section we would like to explain this apparent conundrum.

First, remember that in each hexagon the allowed pairs in the Y-rows are
\(|\circ\circ\rangle\), \(|\circ\bullet\rangle\) and \(|\bullet\circ\rangle\), while \(|\bullet\bullet\rangle\) decouples from the system.
In other words, physical states live in \(\mathbb{C}^3 \subset \mathbb{C}^2\times \mathbb{C}^2\).
Observe that the X-rows states form a four dimensional space, that is
\(|\circ\circ\rangle_X\), \(|\circ\bullet\rangle_X\), \(|\bullet \circ\rangle_X\) and \(|\bullet\bullet\rangle_X\) are all physical.

Consider the adjoint representation of \(sl(2, \mathbb{C})\), and we denote its states as
\be 
\label{eq:adjoint}
	|+1\rangle \equiv |\bullet  \circ \rangle  \qquad 
	|0\rangle \equiv |\circ  \circ \rangle  \qquad 
	|-1\rangle \equiv |\circ \bullet \rangle  \; .
\ee
In this representation, there are \(36\) Boltzmann weights, but using the constraints of section~\ref{sec:boltzmann}, one 
can show that half of them are trivial and we recover the \(18\) weights in figure~\ref{fig:18hvertex}. Moreover, it is 
easy to rewrite the nontrivial Boltzmann weights~\ref{fig:18hvertex} in the \(sl(2, \mathbb{C})\) adjoint representation.

We want to rewrite~(\ref{eq:laxoperator}) in a better form, that is
\bse
\be 
{\cal L}_{a,m}(u)|\alpha , M\rangle = \sum_{\beta M}
{\cal W}\big[u \big| \begin{smallmatrix} N \vspace{-2.5pt} \\ \vspace{-2.5pt} \alpha \quad \beta \\ M \end{smallmatrix}\big]
|N , \beta \rangle \; ,
\ee
where 
\be 
|\alpha, M \rangle=| \alpha \rangle\otimes \left| \mu_1 \mu_2\right\rangle\; \quad 
|N, \beta \rangle=\left| \nu_1 \nu_2\right\rangle \otimes | \beta \rangle\; ,
\ee
and \(M, N=0, \pm 1\) with \( +1 \equiv \bullet \circ\), \(0 \equiv \circ \circ\) and \( -1 \equiv  \circ \bullet\)\; .
In other words, the Boltzmann weights are the components of \({\cal L}_{a, m}(u)\)
\be 
[{\cal L}_{a, m}(u)]_{\alpha M}^{\beta N} \equiv \langle N, \beta | {\cal L}_{a , m}(u)|\alpha , M\rangle = 
{\cal W}\big[u \big| \begin{smallmatrix} N\vspace{-2.5pt} \\ \vspace{-2.5pt} \alpha \quad \beta \\ M \end{smallmatrix}\big]\; .
\ee
\ese
Now, let us rewrite equations~(\ref{eq:L-expression}) as
\begin{flalign} 
\nn 
{\cal L}_{a,m} |\circ\rangle \left|\circ \circ \right\rangle =
\left( {\cal W}\left[ \begin{smallmatrix} \circ & \circ\\  \circ & \circ \\ \circ & \circ \end{smallmatrix} \right] 
\left|\circ \circ \right\rangle + {\cal W}\left[ \begin{smallmatrix} \bullet & \circ\\  \circ & \circ \\ \circ & \circ \end{smallmatrix} \right] 
\left|\bullet \circ \right\rangle+{\cal W}\left[ \begin{smallmatrix} \circ & \bullet\\  \circ & \circ \\ \circ & \circ \end{smallmatrix} \right] 
\left|\circ \bullet \right\rangle \right) |\circ\rangle + 
\left({\cal W}\left[ \begin{smallmatrix} \circ & \circ\\  \circ & \bullet \\ \circ & \circ \end{smallmatrix} \right] 
\left|\circ \circ \right\rangle + 
{\cal W}\left[ \begin{smallmatrix} \bullet & \circ\\  \circ & \bullet \\ \circ & \circ \end{smallmatrix} \right] 
\left|\bullet \circ \right\rangle\right) |\bullet \rangle &&
\end{flalign} 
\begin{flalign}
\nn
{\cal L}_{a,m} |\circ\rangle \left|\bullet \circ \right\rangle  = 
\left(
{\cal W}\left[ \begin{smallmatrix} \circ & \circ\\  \circ & \circ \\ \bullet & \circ \end{smallmatrix} \right] 
\left|\circ \circ \right\rangle +
{\cal W}\left[ \begin{smallmatrix} \bullet & \circ\\  \circ & \circ \\ \bullet & \circ \end{smallmatrix} \right] 
\left|\bullet \circ \right\rangle
+ {\cal W}\left[ \begin{smallmatrix} \circ & \bullet\\  \circ & \circ \\ \bullet & \circ \end{smallmatrix} \right] 
\left|\circ \bullet \right\rangle  \right) |\circ\rangle  + 
\left( {\cal W}\left[ \begin{smallmatrix} \circ & \circ\\  \circ & \bullet \\ \bullet & \circ \end{smallmatrix} \right] 
\left|\circ \circ \right\rangle  + 
{\cal W}\left[ \begin{smallmatrix} \bullet & \circ\\  \circ & \bullet \\ \bullet & \circ \end{smallmatrix} \right] 
\left|\bullet \circ \right\rangle  \right) |\bullet\rangle  &&
\end{flalign} 
\begin{flalign} 
{\cal L}_{a,m} |\circ\rangle \left| \circ \bullet \right\rangle =
\left({\cal W}\left[ \begin{smallmatrix} \circ & \circ\\  \circ & \circ \\ \circ & \bullet \end{smallmatrix} \right] 
\left|\circ \circ \right\rangle +
{\cal W}\left[ \begin{smallmatrix} \bullet & \circ\\  \circ & \circ \\ \circ & \bullet \end{smallmatrix} \right] 
\left| \bullet \circ  \right\rangle
+ {\cal W}\left[ \begin{smallmatrix} \circ & \bullet\\  \circ & \circ \\ \circ & \bullet \end{smallmatrix} \right] 
\left| \circ \bullet\right\rangle\right)  |\circ \rangle &&
\end{flalign} 
\begin{flalign} 
\nn
{\cal L}_{a,m} |\bullet\rangle \left|\circ \circ\right\rangle = 
\left( {\cal W}\left[ \begin{smallmatrix} \circ & \circ\\  \bullet & \circ \\ \circ & \circ \end{smallmatrix} \right] 
\left| \circ \circ \right\rangle + 
{\cal W}\left[ \begin{smallmatrix} \circ & \bullet\\  \bullet & \circ \\ \circ & \circ \end{smallmatrix} \right] 
\left|\circ \bullet\right\rangle \right) |\circ\rangle 
+ {\cal W}\left[ \begin{smallmatrix} \circ & \circ\\  \bullet & \bullet \\ \circ & \circ \end{smallmatrix} \right] 
\left|\circ \circ \right\rangle|\bullet\rangle  &&
\end{flalign} 
\begin{flalign} 
\nn
{\cal L}_{a,m} |\bullet\rangle \left|\circ \bullet \right\rangle = 
\left( {\cal W}\left[ \begin{smallmatrix} \circ & \circ\\  \bullet & \circ \\ \circ & \bullet \end{smallmatrix} \right] 
\left|\circ \circ \right\rangle + 
{\cal W}\left[ \begin{smallmatrix} \circ & \bullet\\  \bullet & \circ \\ \circ & \bullet \end{smallmatrix} \right] 
\left|\circ \bullet \right\rangle \right) |\circ \rangle  &&
\end{flalign} 
where we have kept the tensor product implicit. The Lax operator is simply
\be 
\label{eq:Loper}
{\cal L}_{a,m} = 
\begin{pmatrix}
	{\cal L}_\circ^\circ & {\cal L}_\circ^\bullet \\ 
	{\cal L}_\bullet^\circ& {\cal L}_\bullet^\bullet
\end{pmatrix}
\ee
where \([{\cal L}_\alpha^\beta]\) are \(3\times 3\) matrices defined by
\bse 
\be 
[{\cal L}_\alpha^\beta] =
\begin{pmatrix}
	\langle \circ \circ | {\cal L}_\alpha^\beta|\circ \circ \rangle & \langle \circ \circ | {\cal L}_\alpha^\beta |\bullet \circ \rangle	& \langle \circ \circ | {\cal L}_\alpha^\beta|\circ \bullet \rangle  \\
	\langle \bullet \circ |{\cal L}_\alpha^\beta |\circ \circ \rangle & \langle \bullet \circ | {\cal L}_\alpha^\beta |\bullet \circ \rangle & \langle \bullet \circ | {\cal L}_\alpha^\beta |\circ \bullet \rangle \\
	\langle \circ \bullet |{\cal L}_\alpha^\beta |\circ \circ \rangle & \langle \circ \bullet| {\cal L}_\alpha^\beta |\bullet \circ \rangle 	& \langle \circ \bullet |{\cal L}_\alpha^\beta |\circ \bullet \rangle 
\end{pmatrix}
\ee
with 
\be 
\begin{split}
	{\cal L}_{a,m} |v\rangle\otimes |\circ \rangle & =  {\cal L}_\circ^\circ |v\rangle \otimes |\circ \rangle + {\cal L}_\circ^\bullet |v\rangle \otimes |\bullet \rangle\\
	{\cal L}_{a,m} |v\rangle\otimes |\bullet \rangle & =  {\cal L}_\bullet^\circ |v\rangle \otimes |\circ \rangle + {\cal L}_\bullet^\bullet |v\rangle \otimes |\bullet \rangle\; .
\end{split}
\ee
\ese

Therefore
\bse
\be 
\begin{split}
	{\cal L}_\circ^\circ |\circ \circ \rangle & =
	{\cal W}\left[ \begin{smallmatrix} \circ & \circ\\  \circ & \circ \\ \circ & \circ \end{smallmatrix} \right] 
	\left|\circ \circ \right\rangle + {\cal W}\left[ \begin{smallmatrix} \bullet & \circ\\  \circ & \circ \\ \circ & \circ \end{smallmatrix} \right] 
	\left|\bullet \circ \right\rangle+{\cal W}\left[ \begin{smallmatrix} \circ & \bullet\\  \circ & \circ \\ \circ & \circ \end{smallmatrix} \right] 
	\left|\circ \bullet \right\rangle \\
	{\cal L}_\circ^\circ |\bullet \circ \rangle & = 
	{\cal W}\left[ \begin{smallmatrix} \circ & \circ\\  \circ & \circ \\ \bullet & \circ \end{smallmatrix} \right] 
	\left|\circ \circ \right\rangle +
	{\cal W}\left[ \begin{smallmatrix} \bullet & \circ\\  \circ & \circ \\ \bullet & \circ \end{smallmatrix} \right] 
	\left|\bullet \circ \right\rangle
	+ {\cal W}\left[ \begin{smallmatrix} \circ & \bullet\\  \circ & \circ \\ \bullet & \circ \end{smallmatrix} \right] 
	\left|\circ \bullet \right\rangle \\
	{\cal L}_\circ^\circ |\circ \bullet \rangle & = 
	{\cal W}\left[ \begin{smallmatrix} \circ & \circ\\  \circ & \circ \\ \circ & \bullet \end{smallmatrix} \right] 
	\left|\circ \circ \right\rangle +
	{\cal W}\left[ \begin{smallmatrix} \bullet & \circ\\  \circ & \circ \\ \circ & \bullet \end{smallmatrix} \right] 
	\left| \bullet \circ  \right\rangle
	+ {\cal W}\left[ \begin{smallmatrix} \circ & \bullet\\  \circ & \circ \\ \circ & \bullet \end{smallmatrix} \right] 
	\left| \circ \bullet\right\rangle
\end{split}
\ee

\be 
\begin{split}
	{\cal L}_\circ^\bullet |\circ \circ \rangle & =
	{\cal W}\left[ \begin{smallmatrix} \circ & \circ\\  \circ & \bullet \\ \circ & \circ \end{smallmatrix} \right] 
	\left|\circ \circ \right\rangle + 
	{\cal W}\left[ \begin{smallmatrix} \bullet & \circ\\  \circ & \bullet \\ \circ & \circ \end{smallmatrix} \right] 
	\left|\bullet \circ \right\rangle\\
	{\cal L}_\circ^\bullet |\bullet \circ \rangle & = 
	{\cal W}\left[ \begin{smallmatrix} \circ & \circ\\  \circ & \bullet \\ \bullet & \circ \end{smallmatrix} \right] 
	\left|\circ \circ \right\rangle  + 
	{\cal W}\left[ \begin{smallmatrix} \bullet & \circ\\  \circ & \bullet \\ \bullet & \circ \end{smallmatrix} \right] 
	\left|\bullet \circ \right\rangle \\
	{\cal L}_\circ^\bullet | \circ \bullet \rangle & = 0
\end{split}
\ee

\be 
\begin{split}
	{\cal L}_\bullet^\circ |\circ \circ \rangle & =
	{\cal W}\left[ \begin{smallmatrix} \circ & \circ\\  \bullet & \circ \\ \circ & \circ \end{smallmatrix} \right] 
	\left| \circ \circ \right\rangle + 
	{\cal W}\left[ \begin{smallmatrix} \circ & \bullet\\  \bullet & \circ \\ \circ & \circ \end{smallmatrix} \right] 
	\left|\circ \bullet\right\rangle \\
	{\cal L}_\bullet^\circ |\bullet \circ \rangle & = 0 \\
	{\cal L}_\bullet^\circ |\circ \bullet  \rangle & =
	{\cal W}\left[ \begin{smallmatrix} \circ & \circ\\  \bullet & \circ \\ \circ & \bullet \end{smallmatrix} \right] 
	\left|\circ \circ \right\rangle + 
	{\cal W}\left[ \begin{smallmatrix} \circ & \bullet\\  \bullet & \circ \\ \circ & \bullet \end{smallmatrix} \right] 
	\left|\circ \bullet \right\rangle
\end{split}
\ee
and
\be 
\begin{split}
	{\cal L}_\bullet^\bullet |\circ \circ \rangle & =
	{\cal W}\left[ \begin{smallmatrix} \circ & \circ\\  \bullet & \bullet \\ \circ & \circ \end{smallmatrix} \right] 
	\left|\circ \circ \right\rangle \\ 
	{\cal L}_\bullet^\bullet |\bullet \circ \rangle & = 0 \\
	{\cal L}_\bullet^\bullet |\circ \bullet \rangle & = 0
\end{split}
\ee
\ese
Using these expressions one can easily find~(\ref{eq:Loper}). It is worthwhile to observe that although the physical degrees of freedom do not change, using the \(sl(2, \mathbb{C})\) adjoint representation the Lax operator is a singular \(6\times 6\) matrix, whilst in the fundamental representation we have used before, it is a \(8\times 8\) singular matrix. We know that when we select a 5D subspace where the Lax operator \({\cal L}\) has a suitable interpretation as an S-matrix.

\begin{figure}[h]
	\centering
	\begin{tikzpicture}
	\draw (-.5,-.5) -- (.5,.5); \draw (.5,-.5) -- (-.5,.5);
	\draw (.5,-.5) -- (2.5,-.5); \draw (.5,.5) -- (2.5,.5);
	\draw (1.47,-1) -- (1.47,1); \draw (1.53,-1) -- (1.53,1);
	\node at (-0.7,.55) {\(\tilde{\beta}\)}; \node at (-0.7,-.55) {\(\beta\)};
	\node at (2.7,.55) {\(\alpha\)}; \node at (2.7,-.55) {\(\tilde{\alpha}\)};
	\node at (1.5,-1.3) {\(M\)}; \node at (1.5,1.3) {\(N\)};
	\node at (3,0) {\(=\)};
	\end{tikzpicture}
	\begin{tikzpicture}
	\draw (2.5,-.5) -- (3.5,.5); \draw (2.5,.5) -- (3.5,-.5);
	\draw (.5,-.5) -- (2.5,-.5); \draw (.5,.5) -- (2.5,.5);
	\draw (1.47,-1) -- (1.47,1); \draw (1.53,-1) -- (1.53,1);
	\node at (0.3,.55) {\(\tilde{\beta}\)}; \node at (0.3,-.55) {\(\beta\)};
	\node at (3.7,.55) {\(\alpha\)}; \node at (3.7,-.55) {\(\tilde{\alpha}\)};
	\node at (1.5,-1.3) {\(M\)}; \node at (1.5,1.3) {\(N\)}; 
	\end{tikzpicture}
	\caption{Diagrammatic representation of the fundamental commutation relation.}
	\label{fig:fcr-rep}
\end{figure}
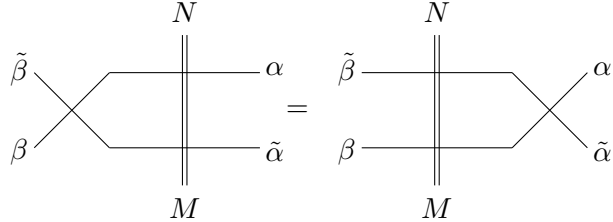

From all these definitions, one can write the fundamental commutation relation~(\ref{eq:fcr-comp}) as
\be 
\boxed{ 
\sum_{\gamma, \tilde{\gamma}, P}
{\cal R}\big[u, v \big| \begin{smallmatrix} \beta & \tilde{\beta} \\ \gamma & \tilde{\gamma} \end{smallmatrix}\big]
{\cal W}\big[u \big| \begin{smallmatrix} N\vspace{-2.5pt} \\ \vspace{-2.5pt} \alpha \quad \gamma \\ P \end{smallmatrix}\big]
{\cal W}\big[u \big| \begin{smallmatrix} P\vspace{-2.5pt} \\ \vspace{-2.5pt} \tilde{\alpha} \quad \tilde{\gamma} \\ M \end{smallmatrix}\big] = 
\sum_{\gamma, \tilde{\gamma}, P}
{\cal W}\big[v \big| \begin{smallmatrix} N\vspace{-2.5pt} \\ \vspace{-2.5pt} \tilde{\gamma} \quad \tilde{\beta} \\ P \end{smallmatrix}\big]
{\cal W}\big[u \big| \begin{smallmatrix} P\vspace{-2.5pt} \\ \vspace{-2.5pt} \gamma \quad \beta \\ M \end{smallmatrix}\big]
{\cal R}\big[u, v \big| \begin{smallmatrix} \gamma & \tilde{\gamma} \\ \alpha & \tilde{\alpha} \end{smallmatrix}\big]
}
\ee
Now we can represent this expression diagrammatically as figure~\ref{fig:fcr-rep}, where the double line denotes the 
adjoint representation of \(sl(2, \mathbb{C})\).

The definitions above characterized the even and odd subsystems as descendant of the vertex model~\cite{Gomez:1996az}. More specifically, it is often assumed that the auxiliary and quantum spaces are isomorphic, but in our case these two vector spaces are clearly different. As a relevant example, the 6-vertex model is defined from the fundamental representation of \(su(2)\), and the Lax operator is simply 
\({\cal L}^{6v}\equiv {\cal L}^{(\frac{1}{2}, \frac{1}{2})}  \in \textrm{End}( V_a \times V_q)\), where \(V_a\simeq V_q \simeq \mathbb{C}^2\). Therefore
\be 
{\cal L}^{(\frac{1}{2}, \frac{1}{2})} :=\vcenter{\hbox{
\begin{tikzpicture}
\draw (0,-1) -- (0,1); \draw (-1,0) -- (1,0);
\node at (0,-1.3) {\(V_q\)}; \node at (0,1.3) {\(V_q\)};
\node at (-1.3,0) {\(V_a\)}; \node at (1.3,0) {\(V_a\)};
\end{tikzpicture}}}
\ee

The subsystem defined in this section, which culminated in the fundamental commutation relation~\ref{fig:fcr-rep}, is defined by a Lax operator of the form \({\cal L}\equiv {\cal L}^{(\frac{1}{2}, 1)}  \in \textrm{End}( V_a \times \tilde{V}_q)\), where \(\tilde{V}_q\simeq \mathbb{C}^3\), so that
\be 
{\cal L}^{(\frac{1}{2}, 1)} :=\vcenter{\hbox{
\begin{tikzpicture}
\draw (-1,0) -- (1,0); 
\draw (-0.07,-1) -- (-0.07,1); \draw (0.03,-1) -- (0.03,1);
\node at (0,-1.3) {\(\tilde{V}_q\)}; \node at (0,1.3) {\(\tilde{V}_q\)};
\node at (-1.3,0) {\(V_a\)}; \node at (1.3,0) {\(V_a\)};
\end{tikzpicture}}}\; ,
\ee
where the double line denotes the adjoint representation of \(sl(2, \mathbb{C})\). A clear exposition of these descendant models, including the Bethe ansatz analysis, can be found in section \(\S 2.4\) of~\cite{Gomez:1996az}.

%%%%%%%%%%%%%%%%%%%%%%%%%%%%%%%%%%%%%
%%%%%%%%%%%%%%%%%%%%%%%%%%%%%%%%%%%%%
%%%%%%%%%%%%%%%%%%%%%%%%%%%%%%%%%%%%%
%%%%%%%%%%%%%%%%%%%%%%%%%%%%%%%%%%%%%

\section{Conclusions, discussions and perspectives}

In this paper we have studies the Kagome lattice introduced in~\cite{Araujo:2020opn}. The most appealing aspect of this system is the plane partition grading of the Hilbert space. We expect that the new features we studied in the present work will eventually shed some light on the quantum crystal melting integrability.

In section 2, we wrote the Kagome lattice system as a stack of two types of spin chains, we call them \(X\)- and \(Y\)-rows, which are respectively defined in terms of two fermionic fields with components \(\psi_m\) and \(\theta_{m+1/2}\), for \(m\in \mathbb{Z}\). This description brings the discussion to a formalism closer to the one considered in~\cite{Dijkgraaf:2008ua}. Physical states, which are classified by plane partitions, were considered in section 3, where we have also found explicit expressions for the growth operators, and for states corresponding to the 0-, 1- and 2-boxes partitions. 

Moreover, it is an important exercise to compare the results of this section with the 2D case~\cite{Dijkgraaf:2008ua} where all integer partitions could be easily written once the growth operators and a reference state, which is defined by the filled Fermi sea, were properly determined. In the present case, we need to define two families of reference states (because there are two types of rows, the \(X\) and \(Y\)-chains), and these reference states are generally harder than the filled Fermi sea. Quite remarkably, we have shown that these fiducial states are manageable, and the construction of plane partitions states is almost immediate once the 0-box configuration is completely defined.

Furthermore, in section 4 we have also written the crystal melting Hamiltonian using the formalism we develop in section 2. In spite all the evidences collected in~\cite{Dijkgraaf:2008ua, Araujo:2020opn} and in the present paper, the exact solvability of the 3D problem has not been completely settled, and it should naturally be contrasted with the 1D and 2D cases~\cite{Dijkgraaf:2008ua}. 

In section 5 we associate, to the quantum Kagome lattice system, a classical statistical model which is defined when we assign a Boltzmann weight to each allowed hexagon configurations in the Kagome lattice. In other words, we have shown that the plane partition states can be completely defined using 18 different local configurations, and if we assign a Boltzmann weight to each of these states, we immediately find a statistical problem that, as we seen, is not very different from a vertex model. The original quantum problem is completely established when we find the correct Boltzmann weights, which when appropriately combined give the quantum Hamiltonian eigenenergies~(\ref{eq:Hamiltonian}).

The classical integrability of a system is one of the first signs for the integrability of its quantum version. As we have mentioned before, evidences on the integrability of the quantum Hamiltonian~(\ref{eq:Hamiltonian}) have been collected in~\cite{Dijkgraaf:2008ua, Araujo:2020opn}, but since a definite answer has not been achieved yet, we start analyzing the integrability of the classical system just mentioned.

In order to address the integrability of this classical model, we try to follow the most straightforward approach available, namely, we try to organize the Boltzmann weights in a matrix, and we impose the Yang-Baxter equation. If we can find a solution to this system, it means that the classical problem is integrable. The absence of solutions, on the other hand, does not necessarily imply the non-integrability of the system, it can simply denote that our choices have not been appropriate.

With these principles in mind, section \(5\) initiates a program to address the exact solvability of the associate classical system. We have defined a Lax operator, and we describe a partially unsuccessful attempt to define a row-to-row transfer matrix. The partial failure in defining an appropriate row-to-row matrix prevents us from writing the conditions on the weights imposed by the Yang-Baxter equation. At this point, it is fundamental to know if one can find a more convenient row-to-row (or rows-to-rows) transfer matrix, or if there are obstructions for the correct definition of these objects. This particular point is currently under investigation.

Section 6 concludes this paper with a curious analysis. We have shown that after a judicious choice for the rules connecting local hexagon configurations, one can define a descendant of the 6-vertex model, and it explains why we said that the row-to-row matrix was just partially unsuccessful. More precisely, notwithstanding the noncommutativity \([{\bm t}(u), \tilde{\bm t}(v)]\neq 0\), the transfer matrices \({\bm t}\) and \(\tilde{\bm t}\) do commute among themselves, that is \([{\bm t}(u), {\bm t}(v)] = 0\) and \([\tilde{\bm t}(u), \tilde{\bm t}(v)] = 0\), and these objects can be used to define two completely equivalent integrable subsystems. 

These models descend from the 6 vertex model, and it is assumed that the states in the \(X\)-spin chains fulfill a fundamental representation of the \(su(2)\) while the pair of states in the \(Y\)-spin chains are associated to the \(sl(2, \mathbb{C})\) adjoint representation. The role of this soluble subsystem is an important open problem, and in particular, it is fundamental to determine if explicit solutions of these descendant models can be used as building blocks for plane partitions.

There are many possible directions of research, here is a short (and biased) list of interesting problems. The system we defined in section 5 seems to be a rich statistical model, and given the results we found, the most immediate question is to what extent the Yang-Baxter subsystems  (the descendant of the 6-vertex model) can be used as building blocks for plane partitions. In other words, given any solution to the descendant vertex models, can we use its Boltzmann weights to address the crystal melting problem? In fact, the classical statistical system above has a much bigger physical space, so it is important to develop techniques to separate the macroscopic states with corresponding plane partitions analogues from those states without such a description.

In this paper, and in~\cite{Araujo:2020opn}, we have studied just one of the two conjectures of~\cite{Dijkgraaf:2008ua}, namely, the one which states the integrability of the 3D problem. We have not touched, yet, the second (and perhaps more interesting)  conjecture on the mass gap of this system. One natural question is if the ideas we developed here and in~\cite{Araujo:2020opn} can help to prove (or disprove) the mass gap conjecture of~\cite{Dijkgraaf:2008ua}. As we have mentioned before, the ground state of the Hamiltonian~(\ref{eq:Hamiltonian}) is the sum of states labeled by plane partitions weighted by a factor \(q^{\# Boxes/2}\), and norm squared given by the famous MacMahon function. It is reasonable to assume that the first excited state can be similarly written as a sum over partitions weighted by different functions \(f_{\Lambda}(q)\), where \(\Lambda\) is a given plane partition. The explicit form of these functions is an interesting problem to be further analyzed.

Finally, we should also point that the original motivation for our work was to shed some light on the variety of relations between plane partitions, \(N=2\) supersymmetric theories, topological strings, the AGT and the AdS\({}_3\)/Higher spin correspondence \cite{maulik2012, Prochazka:2015deb, Gaberdiel:2017dbk}; and we have not addressed any of these connections yet. This particular problem is currently under investigation and we hope to provide some answers in a future publication. Let us hope nature does not disappoint us.

\paragraph{Acknowledgments}

I am supported by the Swiss National Science Foundation under grant number \textsc{PP00P2\_183718/1}. I would also like to thank Susanne Reffert and Domenico Orlando for many discussions and collaboration on related projects. I am also grateful to the anonymous referee for many valuable suggestions.

\addcontentsline{toc}{section}{References}
\bibliographystyle{utphys}
\bibliography{library.bib}

\providecommand{\href}[2]{#2}\begingroup\raggedright\begin{thebibliography}{10}

\bibitem{Niss2008}
M.~Niss, ``{History of the Lenz–Ising Model 1950–1965: from irrelevance to
  relevance},'' \href{http://dx.doi.org/10.1007/s00407-008-0039-5}{{\em Archive
  for History of Exact Sciences} {\bf 63} (2008)  243}.
  \url{https://doi.org/10.1007/s00407-008-0039-5}.

\bibitem{Hitchin1999}
N.~Hitchin, R.~Nigel J.~Hitchin, N.~Hitchin, S.~Hitchin, G.~Segal,
  N.~Woodhouse, R.~Ward, L.~Segal, O.~Conference~on Integrable Systems~(1997,
  O.~Conference~on Integrable Systems.~1997, {\em et al.}, {\em Integrable
  Systems: Twistors, Loop Groups, and Riemann Surfaces}.
\newblock Oxford Graduate Texts in Mathematics. Clarendon Press, 1999.
\newblock \url{https://books.google.ch/books?id=SSwSDAAAQBAJ}.

\bibitem{Miwa2000}
T.~Miwa, M.~Jinbo, M.~Jimbo, E.~Date, M.~Reid, W.~Fulton, A.~Katok, F.~Kirwan,
  B.~Bollobas, P.~Sarnak, {\em et al.}, {\em Solitons: Differential Equations,
  Symmetries and Infinite Dimensional Algebras}.
\newblock Cambridge Tracts in Mathematics. Cambridge University Press, 2000.
\newblock \url{https://books.google.ch/books?id=kQDw1ZcqLjUC}.

\bibitem{Okounkov:2003sp}
A.~Okounkov, N.~Reshetikhin, and C.~Vafa, ``{Quantum Calabi-Yau and classical
  crystals},'' \href{http://dx.doi.org/10.1007/0-8176-4467-9_16}{{\em Prog.
  Math.} {\bf 244} (2006)  597},
\href{http://arxiv.org/abs/hep-th/0309208}{{\tt arXiv:hep-th/0309208
  [hep-th]}}.
%%CITATION = HEP-TH/0309208;%%.

\bibitem{Nekrasov:2003rj}
N.~Nekrasov and A.~Okounkov, ``{Seiberg-Witten theory and random partitions},''
  \href{http://arxiv.org/abs/hep-th/0306238}{{\tt arXiv:hep-th/0306238}}.

\bibitem{Heckman:2006sk}
J.~J. Heckman and C.~Vafa, ``{Crystal Melting and Black Holes},''
  \href{http://dx.doi.org/10.1088/1126-6708/2007/09/011}{{\em JHEP} {\bf 09}
  (2007)  011}, \href{http://arxiv.org/abs/hep-th/0610005}{{\tt
  arXiv:hep-th/0610005}}.

\bibitem{Okounkov:2006}
A.~Okounkov, ``The uses of random partitions,''
  \href{http://arxiv.org/abs/math-ph/0309015}{{\tt arXiv:math-ph/0309015
  [math-ph]}}.

\bibitem{Gaberdiel:2017dbk}
M.~R. Gaberdiel, R.~Gopakumar, W.~Li, and C.~Peng, ``{Higher Spins and Yangian
  Symmetries},'' \href{http://dx.doi.org/10.1007/JHEP04(2017)152}{{\em JHEP}
  {\bf 04} (2017)  152},
\href{http://arxiv.org/abs/1702.05100}{{\tt arXiv:1702.05100 [hep-th]}}.
%%CITATION = ARXIV:1702.05100;%%.

\bibitem{macmahon1915}
P.~MacMahon, {\em Combinatory Analysis}.
\newblock No.~v. 1 in Combinatory Analysis. The University Press, 1915.
\newblock \url{https://books.google.ch/books?id=qvLuAAAAMAAJ}.

\bibitem{macmahon1916}
P.~MacMahon, {\em Combinatory Analysis}.
\newblock No.~v. 2 in Combinatory Analysis. The University Press, 1916.
\newblock \url{https://books.google.ch/books?id=A\_PuAAAAMAAJ}.

\bibitem{Dijkgraaf:2008ua}
R.~Dijkgraaf, D.~Orlando, and S.~Reffert, ``{Quantum Crystals and Spin
  Chains},'' \href{http://dx.doi.org/10.1016/j.nuclphysb.2008.11.027}{{\em
  Nucl. Phys.} {\bf B811} (2009)  463--490},
\href{http://arxiv.org/abs/0803.1927}{{\tt arXiv:0803.1927
  [cond-mat.stat-mech]}}.
%%CITATION = ARXIV:0803.1927;%%.

\bibitem{Kenyon2003}
R.~Kenyon, ``An introduction to the dimer model,''
  \href{http://arxiv.org/abs/math/0310326}{{\tt arXiv:math/0310326 [math.CO]}}.

\bibitem{Orlando:2009kd}
D.~Orlando, S.~Reffert, and N.~Reshetikhin, ``{On domain wall boundary
  conditions for the XXZ spin Hamiltonian},''
\href{http://arxiv.org/abs/0912.0348}{{\tt arXiv:0912.0348 [math-ph]}}.
%%CITATION = ARXIV:0912.0348;%%.

\bibitem{Araujo:2020opn}
T.~Araujo, D.~Orlando, and S.~Reffert, ``{Quantum crystals, Kagome lattice and
  plane partitions fermion-boson duality},''
  \href{http://arxiv.org/abs/2005.09103}{{\tt arXiv:2005.09103 [hep-th]}}.

\bibitem{maulik2012}
D.~Maulik and A.~Okounkov, {\em Quantum Groups and Quantum Cohomology}.
\newblock Soci{\'e}t{\'e} Math{\'e}matique de France, 2019.
\newblock \href{http://arxiv.org/abs/1211.1287}{{\tt arXiv:1211.1287
  [math.AG]}}.
\newblock \url{https://books.google.ch/books?id=9lTUxQEACAAJ}.

\bibitem{Prochazka:2015deb}
T.~Prochazka, ``{$ \mathcal{W} $ -symmetry, topological vertex and affine
  Yangian},'' \href{http://dx.doi.org/10.1007/JHEP10(2016)077}{{\em JHEP} {\bf
  10} (2016)  077},
\href{http://arxiv.org/abs/1512.07178}{{\tt arXiv:1512.07178 [hep-th]}}.
%%CITATION = ARXIV:1512.07178;%%.

\bibitem{Gomez:1996az}
C.~Gomez, G.~Sierra, and M.~Ruiz-Altaba,
  \href{http://dx.doi.org/10.1017/CBO9780511628825}{{\em {Quantum groups in
  two-dimensional physics}}}.
\newblock Cambridge Monographs on Mathematical Physics. Cambridge University
  Press,
2011.
\newblock
%%CITATION = INSPIRE-428978;%%.

\bibitem{Ogura2020}
M.~Ogura, Y.~Imamura, N.~Kameyama, K.~Minami, and M.~Sato, ``{Geometric
  Criterion for Solvability of Lattice Spin Systems},''
  \href{http://arxiv.org/abs/2003.13264}{{\tt arXiv:2003.13264
  [cond-mat.stat-mech]}}.

\bibitem{Kazuhiko2016}
K.~Minami, ``Solvable hamiltonians and fermionization transformations obtained
  from operators satisfying specific commutation relations,''
  \href{http://dx.doi.org/10.7566/JPSJ.85.024003}{{\em Journal of the Physical
  Society of Japan} {\bf 85} (2016) no.~2, 024003},
  \href{http://arxiv.org/abs/https://doi.org/10.7566/JPSJ.85.024003}{{\tt
  https://doi.org/10.7566/JPSJ.85.024003}}.
  \url{https://doi.org/10.7566/JPSJ.85.024003}.

\bibitem{Minami:2017lld}
K.~Minami, ``{Infinite number of solvable generalizations of XY-chain, with
  cluster state, and with central charge c = m /2},''
  \href{http://dx.doi.org/10.1016/j.nuclphysb.2017.10.004}{{\em Nucl. Phys. B}
  {\bf 925} (2017)  144--160}, \href{http://arxiv.org/abs/1710.01851}{{\tt
  arXiv:1710.01851 [cond-mat.stat-mech]}}.

\bibitem{Minami:2019bry}
K.~Minami, ``{Honeycomb lattice Kitaev model with Wen\textendash{}Toric-code
  interactions, and anyon excitations},''
  \href{http://dx.doi.org/10.1016/j.nuclphysb.2018.12.029}{{\em Nucl. Phys. B}
  {\bf 939} (2019)  465--484}, \href{http://arxiv.org/abs/1901.04117}{{\tt
  arXiv:1901.04117 [cond-mat.stat-mech]}}.

\end{thebibliography}\endgroup
 
\end{document}